\documentclass[11pt,oneside]{article}

\def\Ndot{{%
    \setbox0\hbox{$\N$}%
    \rlap{\hbox to \wd0{\hss \raisebox{4pt}{\scriptsize {$\centerdot$}}\hss}}\box0
}}
\def\Nsp{{%
    \setbox0\hbox{$\N$}%
    \rlap{\hbox to \wd0{\hss \raisebox{3pt}{\scriptsize {\textbf s}}\hss}}\box0
}}

\usepackage{jheppub} %Declare other packages before it, and option modifications after it.
\usepackage{aj-definitions}
\renewcommand\binom[2]{
	{\textstyle {#1\choose #2}}
}
\usepackage{tablefootnote}

\title{Null Fluids -- A New Viewpoint of Galilean Fluids}

\author[a]{Nabamita Banerjee,}
\author[b]{Suvankar Dutta}
\author[c]{and Akash Jain.}

\affiliation[a]{Dept. of Physics, Indian Institute of Science Education and Research (IISER), Pune,
  India,} \affiliation[b]{Dept. of Physics, Indian Institute of Science Education and Research
  (IISER), Bhopal, India,} \affiliation[c]{Centre for Particle Theory \& Dept. of Mathematical
  Sciences, Durham University, UK.}

\emailAdd{nabamita@iiserpune.ac.in}
\emailAdd{suvankar@iiserb.ac.in}
\emailAdd{akash.jain@durham.ac.uk}

\abstract{This article is a detailed version of our short letter `On equilibrium partition function
  for non-relativistic fluid' \cite{Banerjee:2015uta} extended to include an anomalous $U(1)$
  symmetry.  We construct a relativistic system, which we call {\it null fluid} and show that it is
  in one-to-one correspondence with a Galilean fluid living in one lower dimension. The
  correspondence is based on light cone reduction, which is known to reduce the Poincar\'e symmetry
  of a theory to Galilean in one lower dimension. We show that the proposed {\it null fluid} and the
  corresponding Galilean fluid have exactly same symmetries, thermodynamics, constitutive relations,
  and equilibrium partition to all orders in derivative expansion. We also devise a mechanism to
  introduce  $U(1)$ anomaly in even dimensional Galilean theories using light cone reduction, and
  study its effect on the constitutive relations of a Galilean Fluid.}

\begin{document}

\maketitle

%\setcounter{tocdepth}{3}
%\tableofcontents

\newpage

\section{Introduction and Summary}

Non-relativistic systems enjoy an active interest in the physics community primarily for two main
reasons.
% or Galilean\footnote{Non-relativistic theories are low velocity limit of
%   relativistic theories, and are only approximately Galilean. In this work however, we would only talk about
%   theories with } systems, in particular, are
% interesting for two main reasons. 
First, they are expected to be realised in the low energy physics experiments. Second and more
fundamentally grounded reason is that a non-relativistic system can be thought of as an effective
low energy description of an underlying relativistic theory. Hence, it is natural to expect that the
constitutive relations of a non-relativistic fluid, obtained as an effective description of a
relativistic theory, may contain new terms which are not considered in the coarse grained
description of hydrodynamics \cite{landau1959fluid}. For example, if the system breaks parity
symmetry at the microscopic level, that will enforce us to add parity-odd terms in the constitutive
relations. The goal of this paper is to revisit the paradigm of non-relativistic charged
hydrodynamics. We devise a consistent mechanism to derive the parity-even and odd terms in the
constitutive relations of a non-relativistic fluid up to leading order in derivatives, starting from a
relativistic theory.

Galilean fluids\footnote{A non-relativistic system is defined by $c\ra \infty$ limit of a
  relativistic system, while a Galilean system is one whose isometry group is Galilean. The two are
  only approximately the same. In this paper however, we only talk about Galilean theories, as they
  are much easier to handle.} have been an interesting and active topic of research in recent years
\cite{Son:2008ye, Son:2013rqa, Jensen:2014aia, Jensen:2014ama, Jensen:2014wha, Banerjee:2014pya,
  Banerjee:2014nja, Banerjee:2015tga, Geracie:2015xfa}. \cite{Geracie:2015xfa} worked out a
consistent way to write Galilean fluid constitutive relations in Newton-Cartan covariant formalism,
and used the second law of thermodynamics to constrain the hydrodynamic transport in two spatial
dimensions. It is known that Newton-Cartan geometry with Galilean isometry follows from light cone
reduction of a relativistic geometry in one higher dimension \cite{Duval:1984cj,Julia:1994bs}. The
idea behind this is that the Poincar\'e algebra in $(d+2)$-dim has a $(d+1)$-dim Galilean subalgebra
embedded into it. As suggested in \cite{Jensen:2014aia}, this approach can be used to construct
Galilean covariant tensors in Newton-Cartan formalism, which is otherwise a non-trivial
task. Similar ideas were also used in \cite{Geracie:2015xfa} where authors constructed an extended
representation of Galilean group by embedding it in one higher dimension, and used it to present the
Galilean fluid dynamics in a manifestly covariant manner.

In this work, we take this approach a step ahead and ask if we can construct a relativistic fluid,
whose symmetry algebra when restricted to the Galilean subalgebra, is equivalent to a Galilean fluid
in one lower dimension. This idea has been explored in the past, starting with
\cite{Rangamani:2008gi} which showed that dynamics of a relativistic fluid reduces to that of a
Galilean fluid under light cone reduction. However in \cite{Banerjee:2014mka}, we observed that this
na{\"\i}ve approach runs into some troubles -- the thermodynamics that the reduced Galilean fluid
follows is restricted (mass chemical potential is not an independent variable; look
\cref{faulty_thermo}). We also found that the parity-odd sector only survives if the fluid is
incompressible and is kept in a constant magnetic field. It strongly hints that to get the most
generic Galilean fluid via light cone reduction, we need to start with a modified relativistic
system.

More precisely, we start with a flat background (metric and gauge field),
\begin{equation}
	\df s^2_{flat} = -2 \df x^- \df t + \sum_{i=1}^d (\df x^{i})^2,\qquad
	\bcA_{flat} = 0,
\end{equation}
which has $(d+2)$-dim Poincar\'e invariance. $(d+1)$-dim Galilean algebra sits inside Poincar\'e --
all generators which commute with $P_- = \dow_-$ (c.f. \cite{Son:2008ye}). Hence a theory on this
background which respects $x^-$ independent isometries $x^M \ra x^M + \xi^M(t,\vec x)$,
$x^M = \{x^-,t,x^i\}$ enjoys Galilean invariance. Compactifying the $x^-$ direction, we can recover
a $(d+1)$-dim flat Galilean background on which non-relativistic theories can be defined. This is
known as \emph{light cone reduction (LCR)}. We turn on $x^-$ independent fluctuations around the
flat background,
\begin{equation} \label{E:first-metric}
	\df s^2 = 
	- 2 e^{-\Phi} (\df t + a_i \df x^i) (\df x^- - \cB_t \df t -\cB_i \df x^i) + g_{ij}\df x^i\df x^j, \qquad
	\bcA = \cA_t \df t + \cA_i \df x^i.
\end{equation}
Galilean theories can then be described by a partition function
$\cZ[\cB_t,\cB_i,\Phi,a_i,g_{ij}, \cA_t, \cA_i]$. Treating these fluctuations as sources, we can
define the following observables, evaluated in absence of sources (i.e. on flat background),
\begin{equation}\nn
	\r = \frac{\d W}{\d \cB_t} \bigg\vert_{flat}, \quad
	j_{\r}^i = \frac{\d W}{\d \cB_i} \bigg\vert_{flat}, \quad
	\e = \frac{\d W}{\d \Phi} \bigg\vert_{flat}, \quad
	j^i_{\e} = \frac{\d W}{\d a_i} \bigg\vert_{flat}, \quad
	t^{ij} = 2\frac{\d W}{\d g_{ij}} \bigg\vert_{flat},
\end{equation}
\begin{equation}\label{E:expvar}
	q = \frac{\d W}{\d \cA_t} \bigg\vert_{flat}, \quad
	j_{q}^i = \frac{\d W}{\d \cA_i} \bigg\vert_{flat}.
\end{equation}
Here $W = \ln\cZ$, and $\rho$, $j_{\r}^i$, $\e$, $j^i_{\e}$, $t^{ij}$, $q$, $j_{q}^i$ are mass
density, mass current, energy density, energy current, stress tensor, charge density and charge
current respectively of the Galilean theory. Invariance of partition function under $x^-$
independent diffeomorphisms will imply the following conservation equations,
\begin{equation}
	\dow_t \r + \dow_i j_\r^i = 0, \qquad
	\dow_t \e + \dow_i j_\e^i = 0, \qquad
	\dow_t j_\r^i + \dow_j t^{ji} = 0, \qquad
	\dow_t q + \dow_i j_q^i = 0.
\end{equation}
These are exactly what we expect for a Galilean system, if we identify $t$ with Galilean time, as
suggested by the notation.

Above procedure can be made manifestly covariant as proposed in \cite{Duval:1984cj} and later
developed by \cite{Julia:1994bs,Hassaine:1999hn,Christensen:2013rfa} and many others. Consider a
curved background, a metric $\rmG_{MN}$ and a gauge field $\cA_M$ with a null Killing vector $V^M$
normalized as $V^M V_M = 0$ and $V^M \cA_M = 0$; we call this background \emph{null
  background}\footnote{Backgrounds admitting a covariantly constant null Killing vector are termed
  as Bargmann structures in \cite{Duval:1984cj}. Null backgrounds are special cases of Bargmann
  structures on which gauge fields and connections' component along the null Killing vector are
  fixed.}. Theories on null background, which we call \emph{null theories}, are demanded to be
invariant under $V^M$ preserving diffeomorphisms and gauge transformations. The background given in
\cref{E:first-metric} with respective Galilean symmetry then just follows by a choice of basis
$x^M = \{ x^-, t, x^i \}$ such that $V = \dow_-$ ($t$ is not necessarily null). It suggests that
null theories are entirely equivalent to Galilean theories, and are related by merely this choice of
basis. More formally, null theories exhibit Galilean invariance upon null reduction, i.e. getting
rid of $V$ direction through compactification.

We can now study a \emph{null fluid} on this null background, with the hope to get the most generic
Galilean fluid after reduction. Unlike `usual' relativistic fluids, in this case isometry $V$ is
also a background field and hence must be considered while writing the respective constitutive
relations. This simple consideration happens to resolve all the issues we enlisted before. In fact
it does much more that that; even before LCR, $(d+2)$-dim null fluid is essentially equivalent to a
$(d+1)$-dim Galilean fluid, as they have same symmetries. As we shall show, their constitutive
relations, conservation equations, thermodynamics etc. match exactly to all orders in the derivative
expansion.

Another motivation to study null backgrounds is Galilean anomalies\footnote{We will only be talking
  about global t'Hooft anomalies appearing in Galilean theories as described by \cite{Jensen:2014hqa}. Our working definition of anomaly  shall be that the respective
  conservation laws are violated by certain terms purely dependent on the background sources. We do not
  dwell in the microscopic interpretation of these anomalies.}. The thumb rule for anomalies
tells us that they can only exist in even dimensions. But since light cone reduction reduces
dimension of the theory by one, even if we start with an even-dimensional anomalous relativistic
theory, the reduced Galilean theory is odd-dimensional and hence all anomalous terms should
vanish\footnote{This is in contrast with the results of \cite{Jensen:2014hqa}, where author found
  that the relativistic anomalies survive the light cone reduction and show up as
  gauge/gravitational and Milne anomalies in the Galilean theory. We observe that this is because of
  the presence of extra scalar sources in the Galilean theory that are reminiscent of reduction and
  must be switched off in a physically realizable theory. We present a detailed analysis on these
  issues in a companion paper \cite{akash}.}. However this argument about dimensionality can be
bypassed by working in null backgrounds. Since there is an extra vector field $V$ in the theory, the
tensor structure allows for anomalies only in odd dimensions. In fact, one can reconstruct anomalies
in an even dimensional Galilean theory by starting with an odd dimensional anomalous null theory. We
would like to use the null fluid construction to see how these anomalies affect the Galilean
hydrodynamic transport.

It is known that constitutive relations of a relativistic fluid at local thermodynamic equilibrium
can be obtained from an equilibrium partition function up to some undetermined `transport
coefficients' \cite{Banerjee:2012iz,Jensen:2012jh}. These coefficients can be determined either from
experiments or through a microscopic calculation. If we think of Galilean fluid as a limit of an
underlying relativistic theory, we would expect that its constitutive relations will also follow
from such an equilibrium partition function, which has been discussed in \cite{Jensen:2014ama}. We
expect that a similar partition function can also be achieved via light cone reduction by setting
the theory on background \cref{E:first-metric} to be independent of $t$ direction. In this
configuration symmetries of the theory break down to $\text{diff}\times U(1)^3$ (spatial
diffeomorphisms, Kaluza-Klein transformations, mass transformation and gauge transformation), and
one can easily write down the equilibrium partition function invariant under these symmetries as a
gauge invariant scalar made out of the background fields. In null background picture, same story
follows by introducing another isometry $K^M$, and choosing a basis such that $K= \dow_t$.

Hence the refined goal of this paper is to \emph{set up a consistent theory of hydrodynamics on null
backgrounds. We want to find the most generic constitutive relations for a fluid on null backgrounds
constrained by the second law of thermodynamics and requirement of an equilibrium partition
function. Later employing light cone reduction, we interpret these null fluid constitutive relations as constitutive relations of the most generic Galilean fluid}.

Organization of this paper is as follows. In \cref{construction} we review the construction of
torsionless null backgrounds, and construct an equilibrium partition function for null
theories. Then in \cref{null_fluids} we study hydrodynamics on these null backgrounds, and put
constraints on its dynamics by equilibrium partition function and second law of themodynamics. We
devote \cref{LCR} to review the procedure to obtain Galilean theories from null theories by light
cone reduction, and use it to study Galilean hydrodynamics in \cref{nonrel_hydro}. Finally in
\cref{anomalies} we extend this entire construction to anomalous fluids. In \cref{MinTor} we extend
the entropy current calculation in presence of minimal compatible torsion, which is required to get
agreement between equilibrium partition function and entropy current constraints. In \cref{non_cov}
we express all these results in conventional non-covariant basis. In \cref{Geracie:2015xfa} we
provide a comparison of our results with those of \cite{Geracie:2015xfa}. At the end, in
\cref{forms} we mention notations and conventions of differential forms used throughout this paper.

\section{Construction of Null Backgrounds} \label{construction}

We start our discussion by formally setting up null backgrounds, which will prove to be a natural
`embedding' of Galilean (Newton-Cartan) backgrounds into a spacetime of one higher dimension. These
kind of backgrounds were first considered in \cite{Duval:1984cj} and further explored by
\cite{Hassaine:1999hn,Julia:1994bs,Christensen:2013rfa} where authors recovered Newton-Cartan gravity by light cone reduction of
general relativity. We will refine the approach by constraining the background field content so that
it exactly matches that of a non-relativistic theory, hence letting us study physically realizable
Galilean fluids later.

Let us consider a manifold $\cM_{(d+2)}$ equipped with a metric $\df s^2 = \rmG_{MN} \df x^M \df x^N$ and a $U(1)$ gauge field $\bcA = \cA_M \df x^M$  together referred as \emph{background fields/sources}. $\cM_{(d+2)}$ is also provided with the Levi-Civita connection,
\bee{
	\G^R_{\ MS} = \half \rmG^{RN} \lb \dow_M \rmG_{NS} + \dow_S \rmG_{NM} - \dow_N \rmG_{MS} \rb,
}
and a covariant derivative $\N_M$ associated with $\G^R_{\ MS}$ and $\cA_M$. We demand that physical theories on $\cM_{(d+2)}$ are left invariant by diffeomorphisms and gauge transformations parametrized by infinitesimal parameters $\p_\xi = \lbr \xi = \xi^M \dow_M, \L_{(\xi)} \rbr$ which we call \emph{symmetry data}. Action of $\p_\xi$ (denoted by $\d_\xi$) on various background fields is given as,
\bee{
	\d_\xi \rmG_{MN} = \lie_\xi \rmG_{MN} = 2\N_{(M} \xi_{N)}, \qquad
	\d_\xi \cA_M = \dow_M \lb\L_{(\xi)} + \xi^N\cA_N\rb + \xi^N\cF_{NM}, \label{E:gauge_invariance}
}
where $\lie_\xi$ denotes Lie derivative along $\xi$ and $\cF_{MN}$ is the field strength of $\cA_M$. One can check that symmetry data $\p_\xi$ form an algebra with commutator defined by,
\bee{
	\p_{[\xi_1,\xi_2]} \equiv [\p_{\xi_1},\p_{\xi_2}] = \d_{\xi_1} \p_{\xi_2} = - \d_{\xi_2} \p_{\xi_1} 
	= \lbr \lie_{\xi_1} \xi_2 = - \lie_{\xi_2} \xi_1, \lie_{\xi_1} \L_{(\xi_2)} - \lie_{\xi_2} \L_{(\xi_1)} \rbr.
}
Correspondingly their action on a general field (suppressing all the indices) $\vf$ also forms an algebra with commutator given by $[\d_{\xi_1}, \d_{\xi_2}] \vf = \d_{[\xi_1,\xi_2]} \vf$. Physical theories on $\cM_{(d+2)}$ can be described by a generating functional\footnote{Actually $W = \ln \cZ$ where $\cZ$ is the QFT generating functional.} $W[\rmG_{MN},\cA_M]$ which is seen as a functional of the background sources. Under infinitesimal variation of these sources linear response of $W$ is captured by,
\bee{\label{E:rel(d+2)PF_full}
	\d W = \int \lbr \df x^M \rbr \sqrt{-\rmG} \lB 
		\half T^{MN} \d \rmG_{MN} 
		+ J^M \d\cA_M
	\rB.
}
$T^{MN}$, $J^M$ are called energy-momentum tensor/current and charge current respectively. Demanding partition function to be invariant under the action of $\p_\xi$ given in \cref{E:gauge_invariance}, we can obtain a set of \emph{Ward identities} these currents must follow,
\bee{\label{E:relcons_full}
	\N_M T^{MN} = \cF^{NM} J_M, \qquad
	\N_M J^M = 0.
}
These are the energy-momentum and charge conservation laws of a relativistic theory. It is not mandatory for a physical theory to admit a Lagrangian description, in which case the theory itself can be characterized in terms of conserved currents $T^{MN},J^M$ with dynamics provided by \emph{equations of motion} (\ref{E:relcons_full}).

\subsection{Compatible Null Isometry}

So far whatever we have said applies to any relativistic theory. We now specialize to our case of interest -- `null backgrounds' by introducing a null Killing vector.
More formally, a symmetry data $\p_V = \lbr V = V^M \dow_M, \L_{(V)} \rbr$ will be said to generate a \emph{compatible null isometry} on $\cM_{(d+2)}$ if it follows,
\begin{enumerate}
	\item Action of $\p_V$ is an isometry, $\d_V \rmG_{MN} = \d_V\cA_M =  0$,
	\item $V$ is null, $V^M V_M = 0$,
	\item $V$ is preserved under covariant transport, $\N_M V^N = 0$, and,
	\item Component of gauge field $\cA$ along $V$ is fixed to: $V^M \cA_M = - \L_{(V)}$.
\end{enumerate}

We will call backgrounds admitting a compatible null isometry to be \emph{null backgrounds}.
%\footnote{In the introduction we had chosen $\L_V = 0$ for brevity.}.
Since we are working with torsionless manifolds, one can check that above conditions imply that $\cH_{MN} = \dow_M V_N - \dow_N V_M = 0$. This is a dynamic constraint and can be violated by quantum fluctuations off-shell -- a fact that will become important when we write equilibrium partition function for fluids on null backgrounds in \cref{sec:EQBPF_con}.

Null backgrounds possess some nice features, first one being: $V^M \N_M \vf = \d_V \vf$ for any contra-co-variant tensor $\vf$ (all indices suppressed), transforming in appropriate representation of the gauge group. Further if $\vf$ is entirely made up of $\rmG_{MN},\cA_M$, by first consistency condition, $V^M \N_M \vf = \d_V \vf = 0$. These consistency conditions also imply,
\bee{
%	\hat\square_{(M} V_{M)} = 0, \qquad 
	V^N\cF_{NM} = V^N\cR_{NMRS} = \cR_{NMRS} V^R = 0,
}
where $\cR_{MNRS}$ is the Riemann curvature tensor. We term physical theories on null backgrounds (with compatible null isometry $\p_V$) as \emph{null theories}, and demand them to be invariant under $\p_V$ preserving symmetry transformations i.e. $[\p_\xi,\p_V] = 0$.  This will break down the Poincar\'e symmetry algebra to Galilean, and give null theories a Galilean interpretation. Algebraic relations (2) and (4) in the definition of compatible null isometry, will imply,
\bea{\label{E:PF_redefinitions}
	\d (V^M V_M) = 0 &\qquad\Ra\qquad V^M V^N \d \rmG_{MN} = - 2V_M \d V^M, \nn\\
	\d (V^M \cA_M + \L_{(V)}) = 0 &\qquad\Ra\qquad  V^M \d \cA_M = - \d \L_{(V)} - \cA_M \d V^M.
}
%Relations (1) and (3) are dynamic in nature, and hence will be imposed on-shell. 
It immediately follows that under a variation of background sources restricted by $\d\p_V = 0$, linear response of partition function \cref{E:rel(d+2)PF_full} is still completely characterized by $T^{MN}$, $J^M$, with an added ambiguity in currents,
\bee{\label{E:constitutive_redefinitions}
	T^{MN} \ra T^{MN} + \q_1 V^M V^N, \qquad
	J^M \ra J^M + \q_2 V^M,
} 
where $\q$'s are some arbitrary scalars. One can check that they leave the conservation equations (\ref{E:relcons_full}) invariant, provided $\d_V \q_1 = \d_V \q_2 = 0$. This point onwards whenever we talk about variation, we implicitly assume it to follow $\d\p_V = 0$. Also for the following analysis, we partially fix $\p_V$ by choosing $\L_{(V)} = 0$ for convenience.

\subsection{Equilibrium}\label{relEQB}

%In a covariant theory, there is no preferred notion of time. Different observers can perceive the theory with their own definition of time $T = T^M \dow/\dow x^M$ ($T_M T^M$ < 0). 
For our later discussion on equilibrium partition function of hydrodynamics, it will be helpful to define a notion of equilibrium on null backgrounds.
A system is said to be in equilibrium if it admits a time-like isometry generated by $\p_K = \lbr K = K^M\dow_M, \L_{(K)} \rbr$, i.e. $K^M K_M < 0$ and $\d_K \rmG_{MN}= \d_K\cA_M = 0$. Using $K$ we can define a null field, 
\bee{
	\bar V^M_{(K)} = \E{\F} \lb  K^M + \cB_t V^M \rb, \qquad 
	\E{\F} = - \frac{1}{K^M V_M}, \qquad
	\cB_t = - \frac{K_M K^M}{2 K_N V^N},
}
which is orthonormal to $V$, i.e. $\bar V^M_{(K)} V_M = -1$, $\bar V^M_{(K)} \bar V_{(K)M} = 0$. We define \emph{spatial slice} $\cM_{(d)}$ as the spacetime transverse to $V$ and $\bar V_{(K)}$ with projection operator, 
\bee{
	P^{MN}_{(K)} = \rmG^{MN} - 2 V^{(M}\bar V^{N)}_{(K)}.
}
Using diffeomorphism and gauge invariance of $\cM_{(d+2)}$ we pick up coordinates $x^M = \{x^-, t, x^i\}$ such that, 
\bee{
	\p_V = \lbr V = \dow_-, \L_{(V)} = 0 \rbr, \qquad
	\p_K = \lbr K = \dow_t, \L_{(K)} = 0 \rbr,
}	
and coordinates $\vec x = \{x^i\}$ span $\cM^{K}_{(d)}$. In this basis, background fields can be decomposed as, 
\bea{
	\df s^2_{(d+2)} &= \rmG_{MN} \df x^M \df x^N %&= - 2 V_M \bar V_{(K)N} \df x^M \df x^N + g_{ij} \df x^i \df x^j \nn \\
	= -2\E{-\F} \lb \df t + a_i \df x^i \rb \lb \df x^- - \cB_t \df t - \cB_i \df x^i \rb + g_{ij} \df x^i \df x^j, \nn\\
	\bcA &= \cA_M \df x^M = \cA_t \df t + \cA_i \df x^i.
}
Indices on $\cM^K_{(d)}$ can be raised and lowered by $g^{ij}$ and its inverse $g_{ij}$. Decomposition of other derived fields follow trivially from here,
\bee{\nn
%	V^M = \begin{pmatrix} 1 \\ 0 \\ 0 \end{pmatrix}, \quad
	V_M = \begin{pmatrix} 0 \\ - \E{-\F} \\ - \E{-\F} a_i \end{pmatrix}, \qquad
	\bar V^M_{(K)} = \begin{pmatrix} \E{\F} \cB_t \\ \E{\F} \\ 0 \end{pmatrix}, \qquad
	\bar V_{(K)M} = \begin{pmatrix} -1 \\ \cB_t \\ \cB_i \end{pmatrix}, \quad
}
\bee{
	P_{(K)MN} = \begin{pmatrix}
		0 & 0 & 0 \\
		0 & 0 & 0 \\
		0 & 0 & g_{ij}
	\end{pmatrix}, \qquad
	P^{MN}_{(K)} = \begin{pmatrix}
		B_k B^k & - a_k B^k & B^j \\
		- a_k B^k & a^k a_k & - a^j \\
		B^i & - a^i & g^{ij}
	\end{pmatrix},
}
where we have defined $B_i = \cB_i - a_i \cB_t$, $A_i = \cA_i - a_i \cA_t$. Under this choice of basis, one can check that residual symmetry transformations are parametrized by $\vec x$ dependent symmetry data $\p_{\xi} = \{ \xi^t, \xi^-, \vec \xi, \L_{(\xi)} \}$, which acts on reduced set of background fields as,
%Under the residual The reduced set of background fields $\F$, $a_i$, $\cB_+$, $B_i$, $g_{ij}$, $\cA_+$ and $A_i$ transform nicely at equilibrium under symmetries restricted to the choice of basis,
\bee{\nn
	\d_\xi \F = \lie_{\vec\xi}\ \F, \quad
	\d_\xi a_i = \dow_i \xi^t + \lie_{\vec\xi}\ a_i, \quad
	\d_\xi \cB_t = \lie_{\vec\xi}\ \cB_t, \quad
	\d_\xi B_i = -\dow_i \xi^- + \lie_{\vec\xi}\ B_i, \quad
	\d_\xi g_{ij} =\lie_{\vec\xi}\ g_{ij},
}
\bee{\nn
	\d_\xi \cA_t = \lie_{\vec\xi}\ \cA_t, \quad
	\d_\xi A_i = \lie_{\vec\xi}\ A_i + \dow_i \L_{(\x)},
}
where $\lie_{\vec\xi}$ denotes lie derivative with respect to $\vec\xi$. Its trivial to see that $a_i, B_i, A_i$ transform as $U(1)$ vector gauge fields, $\F, \cB_t, \cA_t$ transform as scalars, and $g_{ij}$ transform as rank 2 tensor. The response of partition function \cref{E:rel(d+2)PF_full} in equilibrium under infinitesimal variation of these sources can be worked out to be,
\bem{\label{E:deqbreducedPF}
	\d W^{eqb} = \int \lbr \df x^i \rbr \sqrt{g} \lB 
		\E{\F}\lb T_{t-} + T_{--} \cB_t \rb \frac{1}{\vq_o^2} \d \vq_o
		+ \frac{1}{\vq_o} \lB T^i{}_{t} + J^i \cA_t \rB \d a_i \dbrk
		+ \frac{1}{2\vq_o} T^{ij} \d g_{ij}
		+ \lb T_{--} \d \vp_o - \frac{1}{\vq_o} T^{i}{}_{-} \d B_i \rb
		- \lb J_- \d \nu_o - \frac{1}{\vq_o} J^i \d A_i \rb
	\rB,
}
where we have defined:
\bee{
	\vq_o = \tilde \vq \E{\F}, \qquad
	\vp_o = \frac{1}{\tilde \vq}\cB_t , \qquad 
	\nu_o = \frac{1}{\tilde \vq}\cA_t.
}
$\tilde \vq = 1/(\tilde\b\tilde R)$ where $\tilde\b$ is the radius of the euclidean time $\t = \i t$ and $\tilde R$ is the radius of compactified $x^-$. We define a connection on $\cM_{(d)}$ as,
\bee{
	\g^k_{\ ij} = \half g^{kl} \lb \dow_i g_{jl} + \dow_k g_{ij} - \dow_l g_{ij} \rb,
}
and $\Nsp_i$ as its associated covariant derivative. We call the associated Riemann curvature tensor $R_{ijkl}$. Note that condition for torsion-less manifolds, $\cH_{MN} = 0$ implies that in equilibrium,
\bee{
	f_{ij} = \dow_{i} a_{j} - \dow_j a_i = 0, \qquad
	\dow_i \vq_o = 0. 
}
Again, these conditions can be violated offshell, which will be important in next subsection when we start construction the equilibrium partition functions.

\subsubsection{Constructing Equilibrium Partition Function} \label{sec:EQBPF_con}

Motivated by applications in hydrodynamics, we want to write the most generic form of equilibrium partition function allowed by symmetries arranged in a derivative expansion of the background sources. Partition function is generally written as integration of scalar densities. While the partition function is itself invariant under symmetries, such statement cannot be made for the integrand. In fact, terms can be added to it whose variation is gauge invariant only upto some boundary terms. 
We can hence decompose $W^{eqb}$ into\footnote{
Usage of subscripts $\rmH_S$, $\rmH_V$ is motivated from eightfold classification of relativistic transport in \cite{Haehl:2014zda}. It is yet not clear if such a classification is also applicable to null backgrounds, so for us this usage is purely notational.
},
\bee{
	W^{eqb} = W_{\rmH_S}^{eqb} + W_{\rmH_V}^{eqb}, \qquad 
	W_{\rmH_S}^{eqb} = \int \lbr \df x^i\rbr \sqrt{g} \frac{1}{\vq_o} P_{\rmH_S}, \qquad
	W_{\rmH_V}^{eqb} = - \int_{\cM_{(d)}} \bI_{CS}^{(d)},
}
where $P_{\rmH_S}$ is a scalar, and $\bI_{CS}^{(d)}$ is the $d$-dimensional `\emph{Chern-Simons}' form. $\bI_{CS}^{(d)}$ is defined such that $\df \bI_{CS}^{(d)} = \bcP^{(d+1)}_{CS}$ is a `polynomial' made out of field strengths $\df \bm a$, $\df \bB$, $\df \bA$ and curvature 2-form $\bR^i{}_j = R_{kl}{}^i{}_j \df x^k \wedge \df x^l$.
% Here $\bcP^{(d+1)}_{CS}$ is any $(d+1)$-form invariant under symmetries and satisfying $\df\bcP^{(d+1)}_{CS} = 0$.
%Exploiting Bianchi identities one can see that the most generic form of $\bcP^{(d+1)}_{CS}$ is combination of various field strengths, and 
It is known that $\bcP^{(d+1)}_{CS}$ can only be written in odd spatial dimensions $(d = 2n-1)$, and upto first non-trivial order in derivatives is given as,
\bee{
	\bcP^{(2n)}_{CS} = \sum_{r=0}^{n} \binom{n}{r} \sum_{s=0}^{n-r} \binom{n-r}{s} C_{(r,s)}(\df \bA)^{\wedge r} \wedge (\df \bB)^{\wedge s} \wedge (\tilde\vq\df \bm a)^{\wedge (n-r-s)},
}
where $C_{(r,s)}$ are constants and $\binom{n}{r} = \frac{n!}{r!(n-r)!}$ is the binomial coefficient introduced for later convenience. Note that the torsionlessness condition $\bcH = 0$ would imply $\bm f = \df \bm a = 0$ on-shell, but since partition functions are to be written off-shell, we include these terms. 
From here we can find the on-shell variation of $W_{\rmH_V}^{eqb}$ ignoring some boundary terms,
\bee{
	\d W_{\rmH_V}^{eqb} = - \int_{\cM_{(d)}} n \sum_{r=0}^{n-1} \binom{n-1}{r} (\df \bA)^{\wedge r} \wedge (\df \bB)^{\wedge (n-r-1)} 
	 \wedge \lb C_{1,(r)} \tilde\vq \d \bm a + C_{2,(r+1)} \d \bA + C_{2,(r)} \d \bB \rb,
}
where $C_{1,(r)} = C_{(r,n-r-1)}$ and $C_{2,(r)} = C_{(r,n-r)}$. From here we can trivially read out the contribution of $W_{\rmH_V}^{eqb}$ to currents in equilibrium; we will come back to it in \cref{const_PF}. Coming back to $W_{\rmH_S}^{eqb}$, it is now just integration of the most generic scalar $P_{\rmH_S}$ made out of background sources arranged in a derivative expansion. 
%This is generally done in a derivative expansion, about a preferred background. 
At ideal order (no derivatives) $P_{\rmH_S,ideal} = P_o$ is defined as a gauge invariant function of $\vq_o,\vp_o,\nu_o$,
\bee{\label{nulPF_ideal}
	W_{\rmH_S,ideal}^{eqb} = \int \lbr \df x^i \rbr \sqrt{g} \frac{1}{\vq_o} P_o (\vq_o,\vp_o,\nu_o).
}
We take this opportunity to define near equilibrium thermodynamics on null backgrounds. It is known that euclidean partition function $\cZ = \E{W}$ can be identified with grand canonical partition function of statistical mechanics. It follows that $P$ can be identified as \emph{pressure density}, $\vq$ as \emph{temperature}, and $\nu$, $\vp$ as \emph{chemical potentials} scaled with temperature, which boil down to their $_o$ values at equilibrium. Differential of $P(\vq,\vp,\nu)$ can be expanded as,
\bee{\label{dP}
	\df P = S \df \vq + R \df (\vp\vq) + Q \df (\nu\vq),
}
where we identify $S$ as \emph{entropy density}, $R$ as `\emph{mass density}' and $Q$ as \emph{charge density}. We can also define an \emph{energy density} E by invoking Gibbs-Duhem relation,
\bee{
	E = S\vq + \vq R \vp + \vq Q \nu - P.
}
Taking a derivative of this relation and using \cref{dP} we can find the first law of thermodynamics,
\bee{
	\df E = \vq \df S + \vq\vp\df R + \vq\nu\df Q.
}
Existence of a mass density makes this thermodynamic system already look Galilean and we take it as first hint that (at least) near equilibrium theories on null backgrounds are secretly Galilean\footnote{\label{faulty_thermo}
	It is interesting to see that a null theory satisfies different thermodynamics than a relativistic theory. It was noted in \cite{Banerjee:2014mka} that if we start with a relativistic fluid (following relativistic thermodynamics), thermodynamics of Galilean fluid after reduction gets restricted. In our setting this restriction manifests itself as $E + P + R\vq\vp = 0$. After a non-trivial redefinition of thermodynamic functions,
\bee{\nn
	E_{rel} = 2E+P, \quad
	P_{rel} = P, \quad
	S_{rel} = \frac{1}{a} S, \quad
	Q_{rel} = \frac{1}{a} Q, \quad
	\vq_{rel} = a\vq, \quad
	\mu_{rel} = \vq\nu, \quad 
	\text{where} \quad a = \frac{1}{\sqrt{-2\vq\vp}},
}
this restricted thermodynamics is equivalent to relativistic thermodynamics,
\bee{
	\df E_{rel} = \vq_{rel} \df S_{rel} + \mu_{rel} \df Q_{rel}, \qquad
	E_{rel} = S_{rel}\vq_{rel} + Q_{rel} \mu_{rel} - P_{rel}.
}
Interestingly this map between relativistic and restricted null thermodynamics is exactly the same as the map between relativistic and restricted Galilean thermodynamics found in \cite{Banerjee:2014mka} by null reduction with $a = ``u^+"$ in their language. It motivates us to propose that thermodynamic systems on null backgrounds are equivalent to thermodynamic systems on Galilean backgrounds.
}. 
Coming back to equilibrium, we can vary partition function \cref{nulPF_ideal} and use \cref{E:deqbreducedPF} to read out components of currents in equilibrium at ideal order,
\bee{\label{E:ideal_constraints}
	(T^{ij})_{o,ideal} = P_o g^{ij}, \quad
	(T_{--})_{o,ideal} = R_o, \quad
	\E{\F}(T_{t-} + T_{--}\cB_t)_{o,ideal} = E_o, \quad
	- (J_{-})_{o,ideal} = Q_o,
}
and rest all spatial currents zero. Using $V^M$, $\bar V^M_{(K)}$, $P_{(K)}^{MN}$ we can recompile these into covariant language,
\bee{\label{E:ideal_eqb_currents}
	T^{MN}_{o,ideal} = R_o \bar V^M_{(K)} \bar V^N_{(K)} + 2 E_o V^{(M} \bar V^{N)}_{(K)} + P_o P_{(K)}^{MN}, \qquad
	J^{M}_{o,ideal} = Q_o \bar V^M_{(K)}.
}
These look like some sort of ideal fluid constitutive relations, but are quite different from a relativistic fluid. We will make notion of this fluid on null backgrounds -- \emph{null fluids} more precise in next section.

%Here, considering $\vq_o,\vp_o,\nu_o$ as thermodynamic potentials at equilibrium, we have defined the thermodynamics as following:
%\bee{\label{E:thermo}
%	\df P = \frac{E+P}{\vq} \df \vq + \vq R \df \vp + \vq Q \df \nu, \qquad
%	E + P = S\vq + \vq R \vp + \vq Q \nu.
%}
%This finishes our general study of a theory in null background. Next, we specialize to study hydrodynamics in this background.
%\paragraph*{CPT Invariance:} We might want the theories of interest to be invariant under CPT. The relativistic metric $G_{MN}$ (and %hence all its components) are CPT invariant. On the other hand $\cA_M$ (and hence its components) are odd under CPT. Demanding %invariance of equilibrium partition function under CPT would imply:
%\begin{enumerate}
%	\item Even $n$: $C_{(2r,s)} = 0 \quad \forall \ r\geq 0$.
%	\item Odd $n$: $C_{(2r+1,s)} = 0 \quad \forall \ r\geq 0$.
%\end{enumerate}

\section{Hydrodynamics on Null Backgrounds} \label{null_fluids}

Having already developed some intuition in last section, we proceed to formally construct hydrodynamics on null backgrounds in its full generality. Any quantum field theory in near equilibrium regime can be described by hydrodynamics. Systems having a hydrodynamic description, \emph{fluids} are in local thermodynamic equilibrium, i.e. variations away from equilibrium are on scales much much larger than the characteristic scale of the system. This essentially means that variables defining the fluid are much much larger than their space-time derivatives. One can therefore express observables (currents) of the theory as a derivative expansion of symmetry covariant data made out of fluid variables.

Note that conservation laws (\ref{E:relcons_full}) are $(d+3)$ independent equations, so any system with $(d+3)$ variables would be exactly solvable. We choose to describe our system by a fluid with null velocity $u$ normalized as $u^M u_M = 0, u^M V_M = -1$, which will give us $(d)$ degrees of freedom, and three thermodynamic variables: temperature $\vq$, mass chemical potential $\vq\vp$, and charge chemical potential $\vq\nu$. We are interested in configurations which respect the isometry generated by $\p_V$. Hence, all the fluid variables as well as constitutive relations are annihilated by action of $\p_V$.

Hydrodynamics (due to dissipation) is not described by a partition function; rather it is characterized by the most generic form of currents $T^{MN},J^M$ in terms of background fields $\rmG_{MN},\cA_M,\p_V$ and fluid variables $u^M,\vq,\vp,\nu$ known as `\emph{constitutive relations}' of the fluid. Dynamics of these currents is given by Ward identities \cref{E:relcons_full} imposed as equations of motion. The constitutive relations are further constrained by certain physicality arguments, like second law of thermodynamics or existence of an equilibrium configuration. Using $u^M, V^M$ and $P^{MN} = \rmG^{MN} + 2V^{(M}u^{N)}$ we can decompose constitutive relations as,
\bea{\label{E:RelCons}
	T^{MN} &= \cR u^M u^N + 2 \cE u^{(M}V^{N)} + \cP P^{MN} + 2 \bbR^{(M}u^{N)} + 2 \bbE^{(M}V^{N)} + \bbT^{MN}, \nn \\
	J^M &= \cQ u^M + \bbJ^M,
}
where we have used redefinitions \cref{E:constitutive_redefinitions} to get rid of some terms. $\cR,\cE,\cP,\cQ$ are some arbitrary functions of $\vq,\vp,\nu$. The tensors $\bbR^M, \bbE^M, \bbT^{MN}, \bbJ^M$ contains derivative corrections and are transverse to $u^M$ and $V^M$, and $\bbT^{MN}$ is traceless. Comparing constitutive relations \cref{E:RelCons} to \cref{E:ideal_eqb_currents} we can refer that, at ideal order $\cR,\cE,\cP,\cQ$ boil down to respective thermodynamic variables $R,E,P,Q$. In the presence of dissipation however, these functions can deviate from their thermodynamic values.

\subsection{Hydrodynamic Frames}

Note that fluid variables $u^M,\vq,\vp,\nu$ are some arbitrary dynamical fields introduced to describe the near equilibrium quantum system. Like any field theory, these fields can be subjected to arbitrary field redefinition, called the \emph{hydrodynamic redefinition freedom}. Some of this freedom is already fixed by the ideal order equilibrium partition function, requiring that these fields  boil down to $\bar V^M_{(K)},\vq_o,\vp_o,\nu_o$ in equilibrium configuration at ideal order. Away from equilibrium however we are free to perturb these variables the way we like as long as the mentioned restriction holds,
\bee{\label{field_trans}
	u^M \ra u^M + \d u^M, \qquad
	\vq \ra \vq + \d \vq, \qquad
	\vp \ra \vp + \d \vp, \qquad
	\nu \ra \nu + \d \nu,
}
where the variations are some arbitrary functions of fluid variables and background fields, subjected to velocity normalization conditions $u_M\d u^M = V_M \d u^M =0$. Note that near equilibrium assumption requires these variations should contain at least one derivative. For the physics to remain invariant under these transformations, we require that the functional form of $T^{MN},J^M$ remain unmodified. Hence to first non-trivial order in derivatives we will get,
\bea{
	T^{MN} &\ra
	(\cR + \d R) u^M u^N 
	+ 2 (\cE+\d E) u^{(M}V^{N)}
	+ (\cP + \d P) P^{MN} \nn \\ &\qquad
	+ 2 \lb \bbR^{(M} + R \d u^{(M} \rb u^{N)} 
	+ 2 \lb \bbE^{(M} + (E+P) \d u^{(M} \rb V^{N)}
	+ \bbT^{MN}, \nn \\
	J^M &\ra (\cQ + \d Q) u^M + Q \d u^M + \bbJ^M,
}
from where we can obtain \emph{hydrodynamic frame transformations},
%from where we can read out,
\bee{\nn
	\cR \ra \cR + \d R, \qquad
	\cE \ra \cE + \d E, \qquad
	\cP \ra \cP + \d P, \qquad
	\cQ \ra \cQ + \d Q, \qquad
	\bbT^{MN} \ra \bbT^{MN},
}
\bee{\label{consti_redef}
	\bbJ^M \ra \bbJ^M + Q \d u^M, \qquad
	\bbR^M \ra \bbR^M + R \d u^{M}, \qquad
	\bbE^M \ra \bbE^M + (E+P) \d u^{M}.
}
%with $\d P$ determined by thermodynamics. There is slight subtlety here that is often overlooked; e.g. $\bbJ^M$ will not transform under transformations (\ref{field_trans}) upto leading order in derivatives, as it contains at-least one derivative, but \cref{consti_redef} tells us it transforms as $\bbJ^M \ra \bbJ^M + Q \d u^M$. Transformations \cref{consti_redef} are hence to be seen independent of \cref{field_trans} which tells us how the functional form of constitutive relations change upon a field redefinition. These transformations we call \emph{hydrodynamic frame transformations}.
Out of these we can construct three hydrodynamic frame invariants, i.e. quantities that do not transform under hydrodynamic frame transformations,
\bea{
	\Pi^{MN} &= \bbT^{MN} + P^{MN}\lB(\cP - P) - (\cE-E) \frac{\dow}{\dow E} P - (\cR - R) \frac{\dow}{\dow R} P - (\cQ - Q) \frac{\dow}{\dow Q} P \rB, \nn\\
%	\Pi^{MN} &= \bbT^{MN} + P^{MN}\lB \cP - \cE \frac{\dow}{\dow E} P - \cR \frac{\dow}{\dow R} P - \cQ \frac{\dow}{\dow Q} P \rB, \nn\\
	\U^M &= \bbJ^M - \frac{Q}{R} \bbR^M, \qquad 
	\cE^M = \bbE^M - \frac{E+P}{R} \bbR^M.
}
All the physical information about fluid constitutive relations is encoded in these invariants. It is sometimes convenient to fix a \emph{hydrodynamic frame} to be able to talk about the physical constitutive relations directly. Most popular choices involve identifying $\cE,\cR,\cQ$ with $E,R,Q$, and dumping all the dissipation into $\cP$. This fixes the ambiguity in $\vq,\vp,\nu$. 
%There exists a magnetovortical frame for parity-odd fluids which changes this convention to include magnetic field and vorticity in the definitions.

For fixing the velocity redefinition, in spirit with the `usual' relativistic fluids we can use, `Eckart Frame' in which $\bbJ^M$ is chosen to be zero, or `Landau Frame' in which $\bbE^M$ is zero. A more natural\footnote{Upon reduction this will imply that mass current does not have any dissipation, i.e. we associate the fluid velocity with the flow of mass.} frame in this case is the `Mass Frame' where $\bbR^M$ is chosen to be zero, which aligns velocity along $R$ flow and all dissipation transverse to $V^M$. We will mainly work in mass frame for which constitutive relations are given as,
\bea{\label{E:MassFrame}
	T^{MN} &= R u^M u^N + 2 E u^{(M}V^{N)} + P P^{MN} + 2 \cE^{(M}V^{N)} + \Pi^{MN}, \nn \\
	J^M &= Q u^M + \U^M.
}
Another helpful frame for our work is to choose all the fluid variables to be equal to their value at equilibrium\footnote{This frame choice however does not completely fix the hydrodynamic ambiguity. You can still shift fluid variables with terms that vanish in equilibrium. However for equilibrium partition function calculations, it is good enough.} exactly, not just at ideal order. We call this `Equilibrium Frame'. This has the advantage that equilibrium partition function naturally gives constitutive relations in this frame. To be precise, in equilibrium configuration setting $\{u^M,\vq,\vp,\nu\} = \{\bar V^M_{(K)},\vq_o,\vp_o,\nu_o\}$ in the constitutive relations \cref{E:RelCons}, and putting them into equilibrium partition function variation \cref{E:deqbreducedPF}, we can deduce that,
\bee{\nn
	\cR_o = \frac{\d W^{eqb}_{(d)}}{\d \vp_o}, \quad
	\bbR_o^i = \vq_o\frac{\d W^{eqb}_{(d)}}{\d B_i}, \quad
	\cQ_o = \frac{\d W^{eqb}_{(d)}}{\d \nu_o}, \quad
	\bbJ_o^i = \vq_o\frac{\d W^{eqb}_{(d)}}{\d A_i},
}
\bee{
	\cE_{o} = \vq_o^2\frac{\d W^{eqb}_{(d)}}{\d \vq_o}, \quad
	\bbE^i_{o} - \vp_o\vq_o \bbR_{o}^i - \nu_o\vq_o \bbJ_{o}^i  = - \vq_o\E{\F}\frac{\d W^{eqb}_{(d)}}{\d a_i}, \quad
	\cP_o g^{ij} + \bbT_o^{ij} = 2\vq_o\frac{\d W^{eqb}_{(d)}}{\d g_{ij}}.
}
Switching back and forth between frames is a non-trivial task, and has to be done order by order in derivatives. We shall see in the subsequent sections that different physical aspects of our theory of interest are better understood in different frames. We have to switch between frames accordingly.
%one does not have to correct fluid variables at every order, and as a price, velocity and thermodynamic variables' definitions are no longer tied with the constitutive relations. 
%One is however free to make any frame transformation, so long as the invariants remain unchanged.

%and sometimes in equilibrium frame,
%\bea{\label{E:EqbFrame}
%	T^{MN} &= \cR_o \bar V_{(K)}^M \bar V_{(K)}^N + 2 \cE_o \bar V_{(K)}^{(M}V^{N)} + \cP_o P^{MN}_{(K)} + 2 \bbR_o^{(M}\bar V_{(K)}^{N)} + 2 \bbE_o^{(M}V^{N)} + \bbT^{MN}_o, \nn \\
%	J^M &= \cQ_o \bar V_{(K)}^M + \bbJ_o^M.
%}
%In equilibrium configuration, on choosing the basis introduced in \cref{relEQB}, we 
%Since $u_o^i = 0$, one can trivially see that these are the quantities (time-independent\footnote{Equilibrium frame as such allows time-derivative dissipation as well, as this is just a hydrodynamic frame. However, partition function only generates time independent configurations.}) that will be generated by equilibrium partition function \cref{E:deqbreducedPF},

\subsection{Entropy Current}

Since hydrodynamics is an effective field theory, we start by writing down
all possible expressions, compatible with symmetry, that can contribute to  $\bbR^M, \bbE^M, \bbT^{MN}, \bbJ^M$.
In addition since we are dealing with a thermodynamic system, we must ensure that
the second law of thermodynamics is satisfied, i.e., there must exist an entropy current $J_{s}^M$, whose divergence is positive semi-definite,
\bee{
	\N_M J_{s}^M \geq 0.
}
We can construct the most generic entropy current for the fluid as,
\bee{\label{E:entropycurrent}
	J_{s}^M = J_{s,can}^M + \U^M_s, \qquad
	J_{s,can}^M = \frac{1}{\vq} P u^M - \frac{1}{\vq} T^{MN} u_N + \vp T^{MN} V_N - \nu J^M,
}
which is just $S u^M$ at ideal order. $J_{s,can}^M$ is called the canonical entropy current, and is given purely in terms of constitutive relations. $\U^M_s$ on the other hand, are arbitrary derivative corrections to the entropy current. Note that $\U^M_s$, unlike $\U^M$, is not required to be transverse to $u^M$ and $V^M$.
%Since $J_{s}^M$ is a physical quantity, it must be hydrodynamic frame invariant. Note that $J_{s(can)}^M$ is invariant under hydrodynamic frame transformations, therefore so must be $\U^M_s$. 
Using first order equations of motion we can obtain,
\bee{\nn
	u^M \dow_M E = -(E+P) \Q, \qquad
	u^M \dow_M R = -R \Q, \qquad
	u^M \dow_M Q = -Q \Q,
}
\bee{\label{E:firstEOM}
	P^{MN} \lB 
		R \lb \O_{NR} u^R - \vq\dow_N \vp \rb 
		- (E+P) \frac{1}{\vq} \dow_N\vq 
		+ Q \lb \cF_{NR} u^R - \vq \dow_N \nu \rb \rB = 0,
}
where we have defined,
\bee{
	\O_{MN} = \dow_M u_N - \dow_N u_M, \qquad
	\Q = \N_M u^M.
}
Using these, divergence of canonical entropy current can be computed to be,
%\bem{
%	\N_M J_{s(can)}^M 
%	=
%	- \frac{1}{\vq^2} (\cE - E) u^M \dow_M \vq
%	- (\cR - R) u^M \dow_M \vp
%	- (\cQ- Q) u^M \dow_M \nu
%	- \frac{1}{\vq}( \cP - P) \N_M u^M \\
%	- \frac{1}{\vq^2} \bbE^{M} \N_M \vq
%	+ \frac{1}{\vq} \bbR^{M} \lb
%		U_{MN} u^{N}
%		- \vq\N_M \vp
%	\rb
%	+ \frac{1}{\vq} \bbJ^M \lb
%		\cF_{MN}u^N
%		- \vq\N_M \nu
%	\rb
%	- \frac{1}{\vq} \bbT^{MN} \N_M u_N.
%}
%Here $U_{MN} = 2\dow_{[M}u_{N]}$. It is convenient to express it in Mass Frame:
\bee{\label{E:entropycurrent_div}
	\vq\N_M J_{s(can)}^M 
	=
	- \Pi^{MN} \N_M u_N
	- \frac{1}{\vq}  \cE^{M}\dow_M \vq
	+ \U^M \lb
		\cF_{MN}u^N
		- \vq\dow_M \nu
	\rb,
}
which will come in handy later. Note that each term in above expression is product of derivatives (called composites). This heavily constraints the form of $\U^M_s$. Its divergence must not contain any pure derivative terms (terms which are not composites), otherwise total entropy current cannot be ensured positive semi-definite.

% Consequently, $\U^M_{s}$ can either have composite terms as they will give composites on differentiation by parts, or it can have exact terms with vanishing divergence.

In next subsection, we write the most generic constitutive relations of a null fluid upto leading derivative order\footnote{By leading order we mean the first derivative corrections appearing in constitutive relations. In parity-even sector it happens at one derivative order itself. In parity-odd sector however it depends on the number of dimensions -- in odd dimensions $(d=2n-1)$ the first correction appear at $(n-1)$ derivative order, and in even dimensions $(d=2n)$ in appears at $(n)$ derivative order.} in parity-odd and even sectors. We further impose constraints on these constitutive relations by imposing second law of thermodynamics and requirement of an equilibrium partition function independently, and compare the results from both the approaches. Readers who are more interested in Galilean fluid results, can skip this computation and directly proceed to subsection \cref{recap} where the final results for null fluid have been summarized. These results can be directly used to read off the constitutive relations of a Galilean fluid, which has been done in \cref{LCR,nonrel_hydro}.
\subsection{Leading Order Hydrodynamics}

In \cite{Banerjee:2015vxa} we discussed in detail the procedure to count various independent data that appear in constitutive relations of usual relativistic fluids. This can be easily extended to null fluid. However in this work we are only interested in leading order null fluid, so we can write the required data by hand without going into the technicalities of \cite{Banerjee:2015vxa}. All possible scalars, vectors and symmetric traceless tensors made out of background fields and fluid variables has been enlisted in \cref{tab:data}; data marked with $*$ can be eliminated by using first order equations of motion (\ref{E:firstEOM}).
{\renewcommand{\arraystretch}{1.5}
\mktbl{t}{tab:data}{Leading Derivative Order Data for Null Fluid}{} {r|l|l|} { 
	
	\cline{2-3}
	& \textbf{Data} & \textbf{Value at Equilibrium} \\
	\cline{2-3}\cline{2-3}
	& \multicolumn{2}{|c|}{\textbf{Parity Even}} \\ \cline{2-3}
	
	& $\Q \equiv \N_M u^M$ & 0 \\ \cline{2-3}
	* & $u^M \dow_M \vq$, $u^M \dow_M \vp$, $u^M \dow_M \nu$ & 0, 0, 0 \\ \cline{2-3}
	& $P^{MN}\dow_N \vq$, $P^{MN}\dow_N \vp$, $P^{MN}\dow_N \nu$ & 0, $\Nsp^i \vp$, $\Nsp^i \nu$ \\ \cline{2-3}
	* & $P^{MN}\lb \O_{NR} u^R - \vq \dow_N \vp \rb$ & 0 \\ \cline{2-3}
	& $P^{MN}\lb \cF_{NR} u^R - \vq \dow_N \nu \rb$ & 0 \\ \cline{2-3}
	& $\s^{MN} \equiv 2 P^{MR} P^{NS} \N_{(R} u_{S)} - \frac{2}{d} P^{MN} \Q $ & 0 \\ \cline{2-3}\cline{2-3}
	
	& \multicolumn{2}{|c|}{\textbf{Parity Odd -- Odd Dimensions $(d=2n-1)$}} \\ \cline{2-3}
	& $l^M_{(r)} \big\vert_{r=0}^{n-1} \equiv \star \lB \bV \wedge \bm u \wedge \bcF^{\wedge r} \wedge \bm\O^{\wedge (n-r-1)} \rB^M$ & 
	$l^i_{o(r)} \big\vert_{r=0}^{n-1} \equiv \ast \lB (\df \bA)^{\wedge r} \wedge (\df \bB)^{\wedge (n-r-1)} \rB^i$ \\ \cline{2-3}\cline{2-3}
	
	& \multicolumn{2}{|c|}{\textbf{Parity Odd -- Even Dimensions $(d=2n)$}} \\ \cline{2-3}
	& $l_{(r)} \big\vert_{r=0}^{n} \equiv \star \lB \bV \wedge \bm u \wedge \bcF^{\wedge r} \wedge \bm\O^{\wedge (n-r)} \rB$ & 
	$l_{o(r)} \big\vert_{r=0}^{n} \equiv \star \lB (\df \bA)^{\wedge r} \wedge (\df \bB)^{\wedge (n-r)} \rB$ \\ \cline{2-3}
	& $l_{(r)}^{MN}\dow_N \vq$, $l_{(r)}^{MN}\dow_N \vp$, $l_{(r)}^{MN}\dow_N \nu$ & 0, $l^{ij}_{(r)} \dow_j \vp$, $l^{ij}_{(r)} \dow_j \nu$ \\ \cline{2-3}
	* & $l_{(r)}^{MN}\lb \O_{NR} u^R - \vq \dow_N \vp \rb$ & 0 \\ \cline{2-3}
	& $l_{(r)}^{MN}\lb \cF_{NR} u^R - \vq \dow_N \nu \rb$ & 0 \\
\cline{2-3}
	& $l_{(r)}^{R(M} \s_{R}^{N)}$ & 0 \\ \cline{2-3}
	& \multicolumn{2}{l|}{where,} \\ 
	& $l^{MN}_{(r)} \big\vert_{r=0}^{n-1} \equiv \star \lB \bV \wedge \bm u \wedge \bcF^{\wedge r} \wedge \bm\O^{\wedge (n-r-1)} \rB^{MN}$ & 
	$l^{ij}_{o(r)} \big\vert_{r=0}^{n-1} \equiv \star \lB (\df \bA)^{\wedge r} \wedge (\df \bB)^{\wedge (n-r-1)} \rB^{ij}$ \\ \cline{2-3}
}
}

Using data in \cref{tab:data} we can now write the most generic form of leading order constitutive relations. For parity-even sector we will get,
\bea{\label{leading_cons}
	\Pi_{(1)}^{MN} &= - \eta \s^{MN} - P^{MN}\z \Q, \nn\\
	\cE_{(1)}^{M} &= P^{MN} \lB \l_{\e\vp} \dow_N \vp + \l_{\e\nu} \dow_N \nu + \k_{\e} \dow_N \vq + \s_{\e} \lb \cF_{NR}u^R - \vq \dow_N \nu \rb \rB, \nn\\
	\U_{(1)}^{M} &= P^{MN} \lB \l_{q\vp} \dow_N \vp + \l_{q\nu} \dow_N \nu + \k_{q} \dow_N \vq + \s_{q} \lb \cF_{NR}u^R - \vq \dow_N \nu \rb \rB.
}
In parity-odd sector however, in odd number of dimensions $(d=2n-1)$ we will get,
\bee{\label{leading_cons_odd_1}
	\tilde\Pi_{(n-1)}^{MN} = 0, \qquad
	\tilde\cE_{(n-1)}^{M} = \sum_{r=0}^{n-1} \binom{n-1}{r}  \tilde\o_{\e(r)} l^M_{(r)}, \qquad
	\tilde\U_{(n-1)}^{M} = \sum_{r=0}^{n-1} \binom{n-1}{r}  \tilde\o_{q(r)} l^M_{(r)},
}
and in even number of dimensions $(d=2n)$,
\bea{\label{leading_cons_odd_2}
	\tilde\Pi_{(n)}^{MN} &= 
	- P^{MN} \sum_{r=0}^{n} \binom{n}{r} \tilde\z_{(r)} l_{(r)}
	- \sum_{r=0}^{n-1} \binom{n-1}{r} \tilde\eta_{(r)} l_{(r)}^{R(M} \s^{N)}{}_R, \nn\\
	\tilde\cE_{(n)}^{M} &= \sum_{r=0}^{n-1} \binom{n-1}{r}  l^{MN}_{(r)} 
		\lB \tilde\l_{\e\vp(r)} \dow_N \vp + \tilde\l_{\e\nu(r)} \dow_N \nu + \tilde\k_{\e(r)} \dow_N \vq + \tilde\s_{\e(r)} \lb \cF_{NR}u^R - \vq \dow_N \nu \rb \rB, \nn\\
	\tilde\U_{(n)}^{M} &= \sum_{r=0}^{n-1} \binom{n-1}{r}  l^{MN}_{(r)} 
		\lB \tilde\l_{q\vp(r)} \dow_N \vp + \tilde\l_{q\nu(r)} \dow_N \nu + \tilde\k_{q(r)} \dow_N \vq + \tilde\s_{q(r)} \lb \cF_{NR}u^R - \vq \dow_N \nu \rb \rB.
}
%At equilibrium their spatial components reduce to $A,B$ and $F_{(A)},F_{(B)}$ respectively. 
%Consequently $l^M_{(r)}$ reduces to $l^i_{o(r)} \equiv l^i_{o(r,n-r-1)}$ at equilibrium using $\star[V \wedge \bar V_{(K)}]$ as the area element on $\cM_{(2n-1)}$; which also implies:
%\bee{
%	\e^{i_1\ldots i_{d}} = \e^{MN i_1\ldots i_{d}} V_M \bar V_{(K)N} = \E{\F}\e_{-+}^{\ \ \ \ i_1\ldots i_{d}}.
%}
Similarly we can work out constitutive relations to arbitrary high derivative orders, but in this work we will not be interested in those.

\subsubsection{Constraints through Equilibrium Partition Function} \label{const_PF}

The constitutive relations in described above are constrained by the requirement of existence of an equilibrium partition function. The statement is that at equilibrium, any null theory must be determined by the most generic partition function made out of background fields discussed in \cref{relEQB}. We have already seen that at ideal order, equilibrium partition function gives thermodynamic meaning to various functions. Even at further order in derivatives, equilibrium partition function turns out to be very useful to (partially) determine the constitutive relations. It gives constraints on various transport coefficients, and tells us which of them are physical.
%It turns out that constraints we get from equilibrium partition function is a subset of those gained by entropy current positivity; in particular we only get equality type constraints (may not be all). 
We along with many people in past have used this approach to find transport of a relativistic fluid. Here we attempt to outline a similar procedure for null fluid up to leading order in derivatives.

\paragraph{Leading Order Parity Even Sector:}
Leading order parity even sector contains one derivative corrections to ideal fluid dynamics.
Using \cref{tab:data}, we see that at equilibrium only terms coupling to $\l$'s survive in frame invariants \cref{leading_cons},
\bee{
	\Pi_{o(1)}^{ij} = 0, \qquad
	\cE_{o(1)}^{i} = \l_{o\e\vp} \N^i \vp_o + \l_{o\e\nu} \N^i \nu_o,  \qquad
	\U_{o(1)}^{i} = \l_{oq\vp} \N^i \vp_o + \l_{oq\nu} \N^i \nu_o.
}
On the other hand there are no one-derivative scalars at equilibrium to construct partition function. Hence all the coefficients appearing above must vanish,
\beebox{\label{E:constraints_PF_leading_even}
	\l_{\e\vp} = \l_{\e\nu} = \l_{q\vp} = \l_{q\nu} = 0.
}
Since equilibrium partition function is identically zero, none of the fluid variables get order one even correction out of equilibrium (in mass frame).

\paragraph{Leading Order Parity Odd Sector (for $d=2n-1$):}

In odd dimensions, $d=2n-1$, the first parity odd contributions show up at $(n-1)$-derivative order.
At equilibrium all the parity-odd terms survive in constitutive relations \cref{leading_cons_odd_1}. On the other hand there are no gauge invariant scalars to construct equilibrium partition function, and it gets contributions only from the Chern-Simons piece (cf. \cref{sec:EQBPF_con}). Consequently we get constitutive relations in equilibrium frame,
%\bee{
%	l^M_{(r,s)} = \star \lB V \wedge u \wedge (\cF + \vq\nu \cH)^{\wedge r} \wedge (U + \vq\vp \cH)^{\wedge s} \wedge (-\vq \cH)^{\wedge (n-r-s-1)} \rB^M,
%}
\bea{\label{E:eqbdefns_GI}
	\tilde\bbE_{o(n-1)}^i
	&= \vq_o^2 n \sum_{r=0}^{n-1}\binom{n-1}{r} l^i_{o(r)} \lb
		C_{1,(r)}
		- \vp_o C_{2,(r)}
		- \nu_o C_{2,(r+1)}
	\rb , \nn\\
	\tilde\bbR^i_{o(n-1)} &= - \vq_o n \sum_{r=0}^{n-1} \binom{n-1}{r} l^i_{o(r)} C_{2,(r)}, \nn\\
	\tilde\bbJ^i_{o(n-1)} &= - \vq_o n \sum_{r=0}^{n-1} \binom{n-1}{r} l^i_{o(r)} C_{2,(r+1)},
}
and rest all zero. Here $C_{1,(r)}$, $C_{2,(r)}$ are constants introduced in \cref{sec:EQBPF_con}. Performing a hydrodynamic frame transformation, we can get the transport coefficients introduced in frame invariants \cref{leading_cons_odd_1} as, 
\beabox{\label{E:constraints_PF_leading_odd}
	\tilde\o_{\e(r)}
	&=
	\vq n \lb
		\vq C_{1,(r)}
		+ \frac{E + P - \vq\vp R}{R} C_{2,(r)}
		- \vq\nu C_{2,(r+1)}
	\rb, \nn \\
	\tilde\o_{q(r)}
	&= 
	\vq n \lb
		\frac{Q}{R} C_{2,(r)}
		- C_{2,(r+1)}
	\rb.
}
We see that both (set of) transport coefficients are completely determined upto some constants. Outside equilibrium, fluid velocity gets a correction (in mass frame) given by:
\bee{
	\tilde\D^{(n-1)} u^{i}
	= - \frac{\vq_o n}{R_o} \sum_{r=0}^{n-1} \binom{n-1}{r} l^i_{o(r)} C_{2,(r)}.
}
Corrections to other components of velocity can be determined by this using normalization conditions. Other fluid variables do not get any leading order odd correction.

\paragraph{Leading  Order Parity Odd Sector (for $d=2n$):}
 Contrary to the last case studied, in even dimensions $d=2n$, the first parity odd contributions show up at $n$-derivative order. 
 In even number of dimensions, only terms coupling to $\tilde\l$'s and $\tilde\z$ survive in frame invariants \cref{leading_cons_odd_2},
\bea{\label{leading_cons_odd_2_eqb}
	\tilde\Pi_{o(n)}^{ij} &= - g^{ij} \sum_{r=0}^{n} \binom{n}{r} \tilde\z_{o(r)} l_{o(r)}, \nn\\
	\tilde\cE_{o(n)}^{i} &= \sum_{r=0}^{n-1} \binom{n-1}{r}  l^{ij}_{o(r)} 
		\lB \tilde\l_{o\e\vp(r)} \dow_j \vp + \tilde\l_{o\e\nu(r)} \dow_j \nu \rB, \nn\\
	\tilde\U_{o(n)}^{i} &= \sum_{r=0}^{n-1} \binom{n-1}{r}  l^{ij}_{o(r)} 
		\lB \tilde\l_{oq\vp(r)} \dow_j \vp + \tilde\l_{oq\nu(r)} \dow_j \nu \rB.
}
On the other hand using data in \cref{tab:data}, we can write the equilibrium partition function as,
\bee{
	W^{eqb} = - \int_{\cM_{(d)}} \sum_{r=0}^{n} \binom{n}{r} \sum_{s=0}^{n-r} \binom{n-r}{s} \cS_{o(r,s)} (\df \bA)^{\wedge r} \wedge (\df \bB)^{\wedge s} \wedge (\tilde\vq \df \bm a)^{\wedge (n-r-s)}.
}
Varying this partition function on-shell, we can compute the constitutive relations in equilibrium frame,
\bea{\nn
	\cP_o &= 0, \qquad 
	\bbT_o^{ij} = 0, \nn\\
	\cE_o &= - \vq_o^2 \sum_{r=0}^{n} \binom{n}{r} \frac{\dow}{\dow\vq_o} \cS_{o2,(r)} l_{o(r)}, \quad
	\cR_o = - \sum_{r=0}^{n} \binom{n}{r} \frac{\dow}{\dow\vp_o} \cS_{o2,(r)} l_{o(r)}, \quad
	\cQ_o = - \sum_{r=0}^{n} \binom{n}{r} \frac{\dow}{\dow\nu_o} \cS_{o2,(r)} l_{o(r)}, \nn\\
	\bbE^i_{o}  &= \vq^2_o n\sum_{r=0}^{n-1} \binom{n-1}{r} l^{ij}_{o(r)} 
		\lb \dow_j\cS_{o1,(r)} - \vp_o \dow_j\cS_{o2,(r)} - \nu_o \dow_j\cS_{o2,(r+1)} \rb \nn\\
	\bbR^i_o &= - \vq_o n\sum_{r=0}^{n-1} \binom{n-1}{r} l^{ij}_{o(r)} \dow_j\cS_{o2,(r)} \qquad
	\bbJ^i_o = - \vq_o n\sum_{r=0}^{n-1} \binom{n-1}{r} l^{ij}_{o(r)} \dow_j\cS_{o2,(r+1)},
}
where $\cS_{1,(r)} = \cS_{(r,n-r-2)}$, $\cS_{2,(r)} = \cS_{(r,n-r-1)}$. Transforming these to mass frame, one can compute the transport coefficients appearing in \cref{leading_cons_odd_2_eqb},
\beabox{\label{E:constraints_PF_leading_odd2}
	\tilde\z_{(r)} &= - \lB\vq^2 \frac{\dow P}{\dow E} \frac{\dow}{\dow\vq} + \frac{\dow P}{\dow R} \frac{\dow}{\dow\vp} + \frac{\dow P}{\dow Q} \frac{\dow}{\dow\nu} \rB \cS_{2,(r)}, \nn\\
	\tilde\l_{\e\vp(r)} &= \vq n\lB \vq\frac{\dow}{\dow\vp}\cS_{1,(r)} + \frac{E + P - \vq\vp R}{R} \frac{\dow}{\dow\vp}\cS_{2,(r)} - \vq\nu \frac{\dow}{\dow\vp}\cS_{2,(r+1)} \rB, \nn\\
	\tilde\l_{\e\nu(r)} &= \vq n\lB \vq \frac{\dow}{\dow\nu} \cS_{1,(r)} + \frac{E + P - \vq\vp R}{R} \frac{\dow}{\dow\nu} \cS_{2,(r)} - \vq\nu \frac{\dow}{\dow\nu} \cS_{2,(r+1)} \rB, \nn\\
	\tilde\l_{q\vp(r)} &= \vq n \lB \frac{Q}{R} \frac{\dow}{\dow\vp}\cS_{2,(r)} - \frac{\dow}{\dow\vp}\cS_{2,(r+1)} \rB, \nn\\
	\tilde\l_{q\nu(r)} &= \vq n \lB \frac{Q}{R} \frac{\dow}{\dow\nu}\cS_{2,(r)} - \frac{\dow}{\dow\nu}\cS_{2,(r+1)} \rB.
}
%We need to do hydrodynamic frame transformation here.
We see that $5$ (set of) transport coefficients $\tilde\z_{(r)}$, $\tilde\l_{\e\vp(r)}$, $\tilde\l_{\e\nu(r)}$, $\tilde\l_{q\vp(r)}$, $\tilde\l_{q\nu(r)}$ are determined in terms of $2$ (set of) functions $\cS_{1,(r)}$, $\cS_{2,(r)}$. Corrections to fluid variables outside equilibrium in mass frame are given as,
\bea{
	\tilde\D^{(n)} \vq &= - \sum_{r=0}^{n} \binom{n}{r} l_{o(r)} \lB\vq_o^2 \frac{\dow \vq_o}{\dow E_o} \frac{\dow}{\dow\vq_o} + \frac{\dow \vq_o}{\dow R_o} \frac{\dow}{\dow\vp_o} + \frac{\dow \vq_o}{\dow Q_o} \frac{\dow}{\dow\nu_o} \rB \cS_{o2,(r)}, \nn\\
	\tilde\D^{(n)} \vp &= - \sum_{r=0}^{n} \binom{n}{r} l_{o(r)} \lB\vq_o^2 \frac{\dow \vp_o}{\dow E_o} \frac{\dow}{\dow\vq_o} + \frac{\dow \vp_o}{\dow R_o} \frac{\dow}{\dow\vp_o} + \frac{\dow \vp_o}{\dow Q_o} \frac{\dow}{\dow\nu_o} \rB \cS_{o2,(r)}, \nn\\
	\tilde\D^{(n)} \nu &= - \sum_{r=0}^{n} \binom{n}{r} l_{o(r)} \lB\vq_o^2 \frac{\dow \nu_o}{\dow E_o} \frac{\dow}{\dow\vq_o} + \frac{\dow \nu_o}{\dow R_o} \frac{\dow}{\dow\vp_o} + \frac{\dow \nu_o}{\dow Q_o} \frac{\dow}{\dow\nu_o} \rB \cS_{o2,(r)}, \nn\\
	\tilde\D^{(n)} u^i &= - \frac{\vq_o}{R_o} n\sum_{r=0}^{n-1} \binom{n-1}{r} l^{ij}_{o(r)} \dow_j\cS_{o2,(r)}.
}

\subsubsection{Constraints through Entropy Current} \label{const_EC}

As we have said, the second law of thermodynamics for null fluid  implies the existence of an entropy current with non-negative divergence. From our experience of usual relativistic fluids, we expect second law requirement to give all the constraints we found through equilibrium partition function, and more. However as we shall see, we will not get all the partition function constraints through entropy current. This can be accounted to the fact that in this computation we will miss constraints coupling to $\bcH = \df \bV$, which is set to zero by requirement of manifold being torsionless. Since this condition can be violated off-shell, equilibrium partition function can however `see' these constraints. In \cref{MinTor} we will turn on minimal amount of torsion to allow non-zero $\bcH$, and will verify that we get all the partition function constraints through entropy current analysis as well. Here we perform torsionless computation to leading derivative order.

\paragraph{Leading Order Parity Even Sector:}

At leading even order, no terms can be introduced in $\U^M_{s}$ without having pure derivative terms
in the divergence, hence, $ J_{s}^M =  J_{s(can)}^M$ whose divergence using \cref{E:entropycurrent_div,leading_cons} is given as,
\bem{
	\vq\N_M J_{s}^M 
	=
	- \frac{1}{\vq}  P^{MN} \lb \l_{\e\vp} \dow_N \vp + \l_{\e\nu} \dow_N \nu \rb \dow_M \vq 
	+ P^{MN} \lb\l_{\e\vp} \dow_N \vp + \l_{\e\nu} \dow_N \nu \rb \lb \cF_{MR}u^R - \vq\dow_M \nu \rb \\
	- \frac{1}{\vq} (\s_\e - \vq\k_q) \lb \cF_{NR}u^R - \vq\dow_N \nu\rb P^{NM} \dow_M \vq  \\
	- \frac{1}{\vq} \k_\e P^{MN} \dow_M \vq \dow_N \vq
	+ \s_q P^{MN} \lb \cF_{MR}u^R - \vq\dow_M \nu \rb \lb \cF_{NR}u^R - \vq\dow_N \nu \rb
	+ \half \eta \s^{MN}\s_{MN}
	+ \z \Q^2.
}

Demanding $\N_M J_{s}^M \geq 0$, from the first line we get all the  equilibrium partition function constraints \cref{E:constraints_PF_leading_even}, and in addition from last two lines,
\beebox{\label{E:leading_ineq_cons}
	\eta, \zeta, \s_q \geq 0, \qquad \k_\e \leq 0, \qquad \s_\e = \vq\k_q.
}

\paragraph{Leading Order Parity Odd Sector (for $d=2n-1$):}
In parity odd sector however, there are terms we can write in $\U^M_{s}$ which have composite divergence. We first consider the odd dimensional case for which we will get\footnote{The $C$'s in this expression are arbitrary, and a priori have no connection to the $C$'s introduced in previous sections. However, as is suggested by the notation, both will eventually turn out to be the same quantities in constitutive relations. Also, the entropy current need not be gauge invariant.},
\bee{\label{E:odd_dim_EC}
	\U^M_s = \sum_{r=0}^{n-1} \lB 
		n \binom{n-1}{r} \tilde\o_{s(r)} l^M_{(r)} 
		- \binom{n}{r+1} C_{2,(r+1)}\star \lB \bm{{\hat\cA}} \wedge \bm u \wedge \bm{{\hat\cF}}^{\wedge r} \wedge \bm{{\hat\O}}^{\wedge (n-r-1)} \rB^M
	\rB.
}
Here $\hat\cA_M = \cA_M + \vq\nu V_M$, $\hat u_M = u_M + \vq\vp V_M$ and $\hat\cF_{MN}$, $\hat\O_{MN}$ are their field strengths. $C_{2(r)}$ is a constant. One can check that no other terms are allowed. Note that since entropy current is not a direct observable, only its divergence is, we are allowed to include gauge non-invariant terms in $\U^M_s$ as long as the divergence is gauge invariant. Computing divergence of \cref{E:odd_dim_EC} we can obtain,
\bem{
	- \N_M \U^M_s 
	= 
	l^M_{(0)}  n\lB 
		- \dow_M \tilde\o_{s(0)} + C_{2,(1)} \nu \dow_M \vq
		- C_{2,(1)} \lb \cF_{MN} u^N - \vq \dow_M \nu \rb
	\rB \\
	+ \sum_{r=1}^{n-1} \binom{n-1}{r} l^M_{(r)}
	n \lB 
		- \N_M \tilde\o_{s(r)} - \lb \frac{E+P - \vq\vp R}{R} C_{2,(r)} - \vq\nu C_{2,(r+1)} \rb \frac{1}{\vq} \dow_M \vq \dbrk
		+ \lb \frac{Q}{R} C_{2,(r)} - C_{2,(r+1)} \rb \lb \cF_{MN} u^N - \vq \dow_M \nu \rb
	\rB.
}
On the other hand divergence of canonical entropy current through \cref{E:entropycurrent_div,leading_cons_odd_1} is given as,
\bee{
	\vq\N_M J_{s(can)}^M 
	=
	\sum_{r=0}^{n-1} \binom{n-1}{r} l^M_{(r)} 
	\lB
	- \tilde\o_{\e(r)} \frac{1}{\vq} \dow_M \vq
	+ \tilde\o_{q(r)} \lb
		\cF_{MN}u^N
		- \vq\dow_M \nu
	\rb
	\rB.
}
Combining the two pieces and demanding $\N_M J_{s}^M \geq 0$, we find a consistency condition in entropy current that $\tilde\o_{s(r)} = \tilde\o_{s(r)}(\vq)$ must not be a function of $\vp,\nu$. From here parity odd transport coefficients in \cref{leading_cons} are determined to be,
\bee{
	\tilde\o_{\e(0)} = \vq n \lb \vq C_{1,(0)} - \vq\nu C_{2,(1)} \rb, \qquad
	\tilde\o_{q(0)} = - \vq n C_{2,(1)},
}
and for $r\neq 0$,
\bea{
	\tilde\o_{\e(r)} &= \vq n \lb \vq C_{1,(r)} + \frac{E+P - \vq\vp R}{R} C_{2,(r)} - \vq\nu C_{2,(r+1)} \rb, \nn\\
	\tilde\o_{q(r)} &= \vq n \lb \frac{Q}{R} C_{2,(r)} - C_{2,(r+1)} \rb,
}
where $C_{1,(r)} = \frac{\df}{\df \vq}\tilde\o_{s(r)}(\vq)$ is an arbitrary function of $\vq$. Compared to equilibrium partition function constraints \cref{E:constraints_PF_leading_odd}, we have one additional constraint,
\beebox{
	C_{2,(0)} = 0,
}
and one less constraint: $C_{1,(r)}$ is not a constant but a function of $\vq$. In \cref{MinTor} we show that on introducing torsion, entropy current positivity will indeed set $C_{1,(r)}$ to be a constant.

\paragraph{Leading Order Parity Odd Sector (for $d=2n$):}
Now we perform a similar analysis for even dimensional parity odd sector. Similar to odd dimensional case, here also we can have terms in $\U^M_{s}$ whose divergence does not have any pure derivative terms\footnote{The $\cS$'s in this expression are arbitrary, and a priori have no connection to the $\cS$'s introduced in previous sections. However, as is suggested by the notation, both will eventually turn out to be the same quantities in constitutive relations.},
\bem{\label{E:EC_d=2n}
	\U^M_s = \sum_{r=0}^{n} \binom{n}{r} \cS_{2,(r)} \star\lB \bm u \wedge \bm{{\hat\cF}}^{\wedge r} \wedge \bm{{\hat\O}}^{\wedge (n-r)} \rB^M \\
	+ \sum_{r=0}^{n-1} \binom{n-1}{r}  l^{MN}_{(r)}
		\lB \tilde\l_{s\vp(r)} \dow_N \vp + \tilde\l_{s\nu(r)} \dow_N \nu + \tilde\k_{s(r)} \dow_N \vq \rB.
}

One can check that any other term if included, will give pure derivative terms in divergence. Divergence of this object can be computed fairly easily to be,
\bem{
	\N_M\U^M_s = \Q\sum_{r=0}^{n} \binom{n}{r} l_{(r)} 
	\lB\vq \frac{\dow P}{\dow E} \frac{\dow}{\dow\vq} + \frac{1}{\vq} \frac{\dow P}{\dow R} \frac{\dow}{\dow\vp} + \frac{1}{\vq} \frac{\dow P}{\dow Q} \frac{\dow}{\dow\nu} \rB \cS_{2,(r)} \\
	+ \sum_{r=0}^{n-1} \binom{n-1}{r}  l^{MN}_{(r)} 
	\lB 
		\lb
			\frac{\dow}{\dow \vq} \tilde\l_{s\vp(r)} - \frac{\dow}{\dow \vp} \tilde\k_{s(r)}
			+ n \frac{E+P - \vq\vp R}{\vq R} \frac{\dow}{\dow \vp} \cS_{2,(r)}
			- n \nu \frac{\dow}{\dow \vp} \cS_{2,(r+1)}
		\rb \dow_M \vq \dow_N \vp \dbrk
		+ \lb
			\frac{\dow}{\dow \vq} \tilde\l_{s\nu(r)} 
			- \frac{\dow}{\dow \nu} \tilde\k_{s(r)}
			+ n \frac{E+P- \vq\vp R}{\vq R} \frac{\dow}{\dow \nu} \cS_{2,(r)}
			- n \nu \frac{\dow}{\dow \nu} \cS_{2,(r+1)}
		\rb \dow_M \vq \dow_N \nu \dbrk
		+ \lb
			\frac{\dow}{\dow \nu} \tilde\l_{s\vp(r)} 
			- \frac{\dow}{\dow \vp} \tilde\l_{s\nu(r)}
		\rb \dow_M \nu \dow_N \vp
		+ n \lb
			\frac{Q}{R} \dow_M \cS_{2,(r)}
			- \dow_M \cS_{2,(r+1)}
		\rb \lb \cF_{NR} u^R - \vq \dow_N \nu \rb
	\rB.
}
On the other hand divergence of canonical entropy current through \cref{E:entropycurrent_div,leading_cons_odd_2} is,
\bem{
	- \vq\N_M J_{s(can)}^M 
	=
	- \Q \sum_{r=0}^{n} \binom{n}{r} \tilde\z_{(r)} l_{(r)}
	+ \sum_{r=0}^{n-1} \binom{n-1}{r}  l^{MN}_{(r)} 
	\lB 
		\tilde\l_{\e\vp(r)} \frac{1}{\vq} \dow_M \vq \dow_N \vp 
		+ \tilde\l_{\e\nu(r)} \frac{1}{\vq} \dow_M \vq \dow_N \nu 
	\rB \\
	+ \sum_{r=0}^{n-1} \binom{n-1}{r}  l^{MN}_{(r)} 
	\lB 
		\tilde\l_{q\vp(r)} \dow_M \vp 
		+ \tilde\l_{q\nu(r)} \dow_M \nu 
		+ \lb \vq\tilde\k_{q(r)} + \tilde\s_{\e(r)} \rb \frac{1}{\vq} \dow_M \vq
	\rB 
	\lb
		\cF_{NR}u^R
		- \vq\dow_N \nu
	\rb.
}
Combining the two pieces and demanding $\N_M J_{s}^M \geq 0$, we get a consistency condition on entropy current,
%\bee{
%	\frac{\dow}{\dow \nu} \tilde\l_{s\vp(r)} = \frac{\dow}{\dow \vp} \tilde\l_{s\nu(r)}.
%}
\bee{
	\frac{\dow}{\dow \nu} \tilde\l_{s\vp(r)} = \frac{\dow}{\dow \vp} \tilde\l_{s\nu(r)},
}
whose most generic solution is,
\bee{
	\l_{s\vp(r,s)} = \frac{\dow}{\dow\vp} f_1(\vq,\vp,\nu), \qquad
	\l_{s\nu(r,s)} = \frac{\dow}{\dow\nu} f_1(\vq,\vp,\nu) + \frac{\dow}{\dow\nu} f_2(\vq,\nu),
}
for some functions $f_1(\vq,\vp,\nu)$, $f_2(\vq,\nu)$.  We define,
\bee{
	n\cS_{1,(r)} = - \tilde\k_{s(r)} + \frac{\dow}{\dow \vq} f_1.
}
Expressed in these variables, one can check that entropy current positivity gives all the partition function constraints \cref{E:constraints_PF_leading_odd2}, except the expression for $\tilde\l_{\e\nu(r)}$ modifies to,
\bee{
	\tilde\l_{\e\nu(r)} = \vq n\lB 
	\vq \frac{\dow^2}{\dow\nu\dow\vq} f_2(\vq,\nu)
	+ \vq \frac{\dow}{\dow\nu} \cS_{1,(r)} + \frac{E + P - \vq\vp R}{R} \frac{\dow}{\dow\nu} \cS_{2,(r)} - \vq\nu \frac{\dow}{\dow\nu} \cS_{2,(r+1)} \rB,
}
and in addition we get,
\beebox{
	\tilde\k_{q(r)} + \frac{1}{\vq} \tilde\s_{\e(r)}
	= \vq n \lB
		\frac{Q}{R} \frac{\dow}{\dow \vq} \cS_{2,(r)}
		- \frac{\dow}{\dow \vq} \cS_{2,(r+1)}
	\rB.
}
Like even dimensional case, we again see that we get an additional constraint through entropy current, but one constraint turns out to be weaker. Equilibrium partition function sets $\frac{\dow^2}{\dow\nu\dow\vq} f_2(\vq,\nu) = 0$ which entropy current fails to do. In \cref{MinTor} we will show that introducing torsion remedies this situation.

\subsection{Recap} \label{recap}

In this section we summarize the results for leading derivative order null fluid in mass frame, taking into account constraints from equilibrium partition function and second law of thermodynamics. The constitutive relations for null fluid are given in terms of fluid variables $\vq$, $\vp$, $\nu$, $u^M$,
\bea{
	T^{MN} &= R u^M u^N + 2 E u^{(M}V^{N)} + P P^{MN} + 2 \cE^{(M}V^{N)} + \Pi^{MN}, \nn \\
	J^M &= Q u^M + \U^M,
}
where $P,R,E,Q$ are thermodynamic pressure, mass density, energy density and charge density expressed as functions of $\vq$, $\vp$, $\nu$. These constitutive relations follow the conservation laws,
\bee{
	\N_M T^{MN} = \cF^{NM}J_M, \qquad
	\N_M J^M = 0.
}
In odd number of dimensions $(d=2n-1)$, the form of hydrodynamic frame invariant corrections $\Pi^{MN}$, $\cE^M$, $\U^M$ to leading order in derivatives are given as, 
\bea{
	\Pi^{MN} &= - \eta \s^{MN} - P^{MN}\z \Q, \nn\\
	\cE^{M} &= \k_{\e} P^{MN} \dow_N \vq + \vq \k_q P^{MN} \lb \cF_{NR}u^R - \vq \dow_N \nu \rb 
	+ \sum_{r=0}^{n-1} \binom{n-1}{r}  \tilde\o_{\e(r)} l^M_{(r)}, \nn\\
	\U^{M} &= \k_{q} P^{MN} \dow_N \vq + \s_{q} P^{MN} \lb \cF_{NR}u^R - \vq \dow_N \nu \rb 
	+ \sum_{r=0}^{n-1} \binom{n-1}{r}  \tilde\o_{q(r)} l^M_{(r)},
}
where transport coefficients $\eta$ (shear viscosity), $\z$ (bulk viscosity), $\s_q$ (electric conductivity) are some non-negative, $\k_\e$ (thermal conductivity) is a non-positive and $\k_q$ (thermo-electric coefficient) is an arbitrary function of $\vq$, $\vp$, $\nu$. Parity-odd transport coefficients (Hall conductivities) are however completely determined upto some constants as,
\bea{\label{E:recap_cons_odd_1}
	\tilde\o_{\e(r)}
	&=
	\vq n \lb
		\vq C_{1,(r)}
		+ \frac{E + P - \vq\vp R}{R} C_{2,(r)}
		- \vq\nu C_{2,(r+1)}
	\rb, \nn \\
	\tilde\o_{q(r)}
	&= 
	\vq n \lb
		\frac{Q}{R} C_{2,(r)}
		- C_{2,(r+1)}
	\rb.
}
where $C$'s are some arbitrary constants, and $C_{2,(0)} = 0$. In even number of dimensions $(d = 2n)$ however the corrections are given as,
\bea{\label{E:d=2n_recap}
	\Pi^{MN} &= - \eta \s^{MN} - \sum_{r=0}^{n-1} \binom{n-1}{r} \tilde\eta_{(r)} l_{(r)}^{R(M} \s^{N)}{}_R
	- P^{MN} \lb \z \Q + \sum_{r=0}^{n} \binom{n}{r} \tilde\z_{(r)} l_{(r)} \rb, \nn\\
	\cE^{M} &= 
	\lb
		P^{MN} \k_{\e}
		+ \sum_{r=0}^{n-1} \binom{n-1}{r}  l^{MN}_{(r)} \tilde\k_{\e(r)}
	\rb \dow_N \vq 
	+ \vq \lb
		P^{MN} \k_q
		- \sum_{r=0}^{n-1} \binom{n-1}{r}  l^{MN}_{(r)} \tilde\k_{q(r)} 
	\rb \lb \cF_{NR}u^R - \vq \dow_N \nu \rb\nn\\
	&\qquad + \vq n \sum_{r=0}^{n-1} \binom{n-1}{r}  l^{MN}_{(r)} 
		\lb \vq \dow_N \cS_{1,(r)} + \frac{E + P - \vq\vp R}{R} \dow_N \cS_{2,(r)} - \vq\nu \dow_N \cS_{2,(r+1)} \rb, \nn\\
	\U^{M} &= 
	\lb
		P^{MN} \k_{q}
		+ \sum_{r=0}^{n-1} \binom{n-1}{r}  l^{MN}_{(r)} \tilde\k_{q(r)}
	\rb \dow_N \vq 
	+ \lb
		\s_{q} P^{MN}
		+ \sum_{r=0}^{n-1} \binom{n-1}{r}  l^{MN}_{(r)} \tilde\s_{q(r)} 
	\rb \lb \cF_{NR}u^R - \vq \dow_N \nu \rb \nn\\
	&\qquad + \vq n\sum_{r=0}^{n-1} \binom{n-1}{r}  l^{MN}_{(r)} \lb \frac{Q}{R} \dow_N \cS_{2,(r)} - \dow_N \cS_{2,(r+1)}\rb,
}
where we have made following redefinitions with respect to \cref{leading_cons_odd_2},
\bea{
	\tilde\k_{\e(r)} &\ra \tilde\k_{\e(r)} + \vq n\lB \vq \frac{\dow}{\dow\vq} \cS_{1,(r)} + \frac{E + P - \vq\vp R}{R} \frac{\dow}{\dow\vq} \cS_{2,(r)} - \vq\nu \frac{\dow}{\dow\vq} \cS_{2,(r+1)} \rB, \nn\\
	\tilde\k_{q(r)} &\ra \tilde\k_{q(r)} + \vq n \lB
		\frac{Q}{R} \frac{\dow}{\dow \vq} \cS_{2,(r)}
		- \frac{\dow}{\dow \vq} \cS_{2,(r+1)}
	\rB.
}
The transport coefficients in parity even sector are same as before; however parity-odd transport coefficients $\tilde\eta_{(r)}$ (Hall viscosity), $\tilde\k_{\e(r)}$ (thermal Hall conductivity), $\tilde\k_{q(r)}$ (thermo-electric Hall coefficient), $\tilde\s_{q(r)}$ (electric Hall conductivity), $\cS_{1,(r)}$ and $\cS_{2,(r)}$ are some arbitrary functions of $\vq,\vp,\nu$. Finally $\tilde\z_{(r)}$ is determined as,
\bee{
	\tilde\z_{(r)} = - \lB\vq^2 \frac{\dow P}{\dow E} \frac{\dow}{\dow\vq} + \frac{\dow P}{\dow R} \frac{\dow}{\dow\vp} + \frac{\dow P}{\dow Q} \frac{\dow}{\dow\nu} \rB \cS_{2,(r)}.
}
All the constitutive relations satisfy the physical requirements of existence of an equilibrium partition function and entropy current. To leading order in derivatives they are given as, in odd number of dimensions $(d = 2n-1)$,
\bea{
	W^{eqb}
	&= \int \lbr \df x^i\rbr \sqrt{g} \lB \frac{1}{\vq_o} P_o - \sum_{r=0}^{n-1} l_{o(r)}^i \lbr n \binom{n-1}{r} C_{1,(r)} \tilde\vq a_i + \binom{n}{r+1} C_{2,(r+1)} A_i \rbr \rB, \nn\\
	J^M_{s} &= J_{s(can)}^M + \sum_{r=0}^{n-1} \star \lB \lbr
		n \binom{n-1}{r} C_{1,(r)} \vq \bV 
		- \binom{n}{r+1} C_{2,(r+1)} \bm{{\hat\cA}}
	\rbr \wedge \bm u \wedge \bm{{\hat\cF}}^{\wedge r} \wedge \bm{{\hat \O}}^{\wedge (n-r-1)} \rB^M,
}
and in even number of dimensions $(d = 2n)$,
\bea{
	W^{eqb}
	&= \int \lbr \df x^i\rbr \sqrt{g} \lB 
		\frac{1}{\vq_o} P_o 
		- n \sum_{r=0}^{n-1} \binom{n-1}{r} l^{ij}_{o(r)} \cS_{o1,(r)} \tilde\vq \dow_i a_j
		- \sum_{r=0}^{n} \binom{n}{r} l_{o(r)} \cS_{o2(r)} 
	\rB, \nn\\
	J^M_{s} &= J_{s(can)}^M + \star\lB \lbr
		n \sum_{r=0}^{n-1} \binom{n-1}{r} \cS_{1,(r)}  \bV \wedge \df \vq 
		+ \sum_{r=0}^{n} \binom{n}{r} \cS_{2,(r)} \bm{{\hat \O}}
	\rbr
	\wedge \bm u \wedge \bm{{\hat\cF}}^{\wedge r} \wedge \bm{{\hat \O}}^{\wedge (n-r-1)} \rB^{M}.
}
While writing entropy current from \cref{E:EC_d=2n}, some total derivative terms have been dropped, as they will have zero divergence. 
We have included the additional constraints coming from entropy current analysis while writing the partition function and vice versa. This finishes the discussion of null fluid upto leading order in derivatives in arbitrary number of dimensions.
Next we turn on to study the light cone reduction and how to get Galilean fluids via reduction of a null fluid.

\section{Light Cone Reduction} \label{LCR}

We want to study a Galilean system in $(d+1)$ dimensions. So we essentially want to compactify the $V$ direction as $\cM_{(d+2)} = S^1_V \times \cM_{(d+1)}$. But, $V$ is null, and thus is transverse to itself, so it is not possible make such decomposition uniquely. It is therefore convenient to introduce another vector field (we call \emph{time field}) $T = T^M \dow_M$, which can be used to define a unique decomposition $\cM_{(d+2)} = S^1_V \times \bbR^1_T \times \cM^T_{(d)}$ where $\cM_{(d)}$ is the set of vectors transverse to both $V$ and $T$. The time field $T$ provides a \emph{reference frame} for compactified Galilean theory. We formally define \emph{light cone reduction} as this choice of frame and subsequent compactification.

%\begin{shaded}
%We will work here in the `extended-space representation' of Galilean physics motivated from \cite{Geracie:2015xfa}. 
%It is essentially decomposition of null theory on $\cM_{(d+2)} = S^1_V \times \bbR^1_T \times \cM^T_{(d)}$ using null isometry $V^M$, a conjugate null field defined from $T$ ($\bar V^M_{(T)} \bar V_{(T)M} = 0$, $\bar V^M_{(T)} V_M = -1$),
%\bee{\label{E:barV_defn}
%	\bar V^M_{(T)} = \frac{1}{\a} \lb  T^M + \frac{\o}{2\a} V^M \rb, \qquad \o = T_M T^M, \qquad \a = - T^M V_M,
%}
%and a projection operator,
%\bee{
%	P_{(T)}^{MN} = G^{MN} + 2 \bar V^{(M}_{(T)} V^{N)}, \qquad
%	\cM^T_{(d)} = \lbr P^{MN}_{(T)} \p_N : \p^N \in \cM_{(d+2)} \rbr.
%}
%\end{shaded}

Using $T$ we can define another null field orthonormal to $V$,
\bee{\label{E:barV_defn}
	\bar V^M_{(T)} = - \frac{1}{T^N V_N} \lb  T^M - \frac{T_R T^R}{2 T^S V_S} V^M \rb,
}
which satisfies $\bar V^M_{(T)} \bar V_{(T)M} = 0$, $\bar V^M_{(T)} V_M = -1$, and a projection operator transverse to $V$ and $\bar V$,
\bee{
	P_{(T)}^{MN} = G^{MN} + 2 \bar V^{(M}_{(T)} V^{N)}, \qquad
	\cM^T_{(d)} = \lbr P^{MN}_{(T)} \p_N : \p^N \in \cM_{(d+2)} \rbr.
}

Since the choice of $T$ is arbitrary and does not have any physical significance, null theories are invariant under an arbitrary redefinition of $T \ra T'$, which we parametrize as,
\bee{\label{E:milne_defn}
	T^M \ra T'^M = a\lB T^M - T^N V_N \p^M \rB,
}
%\bee{\label{E:milne_defn}
%	T^M \ra T'^M = a\lB T^M - V_A T^A \lb P^{MN}_{(T)} \p_N + \half P^{AB}_{(T)} \p_A\p_B V^M + b V^M \rb \rB,
%}
where $\p^M V_M = 0 $ and $a\in \bbR$. 
%We will call this Milne Boosts\footnote{Upon reduction this will give the Milne Boosts of Galilean theories.}. 
One can check that inverse transformation is simply $a\ra 1/a$, and $ \p^M \ra - \p^M$. Our funny parametrization has a benefit that under $T$ redefinition, transformation of $\bar V^M_{(T)}$ and $P^{MN}_{(T)}$ only depends on $\bar\p^M = P^{MN}_{(T)}\p_N$,
\bee{
	\bar V^M_{(T)} \ra \bar V^M_{(T)} + \bar\p^M + \half \bar\p^2 V^M, \qquad
	P^{MN}_{(T)} \ra P^{MN}_{(T)} + 2 V^{(M} \bar\p^{N)} + \bar\p^2 V^M V^N,
%	= \lb P_{(T)}^{MR} + V^M \bar\p^R  \rb \lb P_{(T)R}^{N} + \bar\p_R V^N  \rb,
}
where $\bar\p^2 = P^{MN}_{(T)}\p_M\p_N$.  So our light cone reduced theory is described on a compactified null background $\cM_{(d+2)}$ with isometry $\p_V = \lbr V^M ,\L_{(V)} \rbr$, and a time field $T^M$, modded by diffeomorphisms, gauge transformation and $T$ redefinition. In light cone reduction approach however, we need not worry too much about $T$ redefinition. Since the original theory on null background did not depend on $T$, so the reduced theory will also be invariant under its redefinition automatically.

\subsection{Newton-Cartan Backgrounds by Light Cone Reduction}
It is easy to see how the Newton-Cartan structure comes out by light cone reduction.
We identify $\cM^{\text{NC}}_{(d+1)} = \bbR^1_T \times \cM^T_{(d)}$ as the degenerate Newton-Cartan (NC) manifold. Without loss of generality we can choose a basis $x^M = \{x^-,x^\mu \}$ in the original manifold $\cM_{(d+2)}$ such that $\p_V = \{V = \dow_-, \L_{(V)} = 0 \}$. $x^\mu$ will then provide a basis on NC manifold $\cM^{\text{NC}}_{(d+1)}$.
This mechanism to generate NC manifold via null reduction was first found in \cite{Duval:1984cj} and has been further developed in \cite{Julia:1994bs,Jensen:2014aia}.

\paragraph*{Reduction of Background Fields:}

We can decompose background fields according to this choice of basis as,
\bee{
	V^M = \begin{pmatrix} 1 \\ 0 \end{pmatrix}, \qquad
	V_M = \begin{pmatrix} 0 \\ - n_\mu \end{pmatrix}, \qquad
	\bar V^M_{(T)} = \begin{pmatrix} v^\mu \cB_\mu \\ v^\mu \end{pmatrix}, \qquad
	\bar V_{(T)M} = \begin{pmatrix} -1 \\ \cB_\mu \end{pmatrix},
}
\bee{
	P_{(T)MN} = \begin{pmatrix}
		0 & 0 \\
		0 & p_{\mu\nu}
	\end{pmatrix}, \qquad
	P^{MN}_{(T)} = \begin{pmatrix}
		p^{\nu\r}\cB_\nu\cB_\r & p^{\mu\nu}\cB_\nu \\
		p^{\mu\nu}\cB_\nu & p^{\mu\nu}
	\end{pmatrix},
}
such that 
\bee{
	n_\mu v^\mu = 1, \qquad 
	v^\mu p_{\mu\nu} = 0, \qquad
	n_\mu p^{\mu\nu} = 0, \qquad
	p_{\mu\r}p^{\r\nu} + n_\mu v^\nu = \d_{\mu}^{\ \nu}.
}
This is the well known Newton-Cartan structure.
%In this basis metric and gauge field will decompose as:
%\bee{
%	\df s^2 = - n_\mu \df x^\mu (\df x^- - \cB_\nu \df x^\nu) + p_{\mu\nu} \df x^\mu \df x^\nu, \qquad
%	\cA = \cA_\mu \df x^\mu.
%}
The $T$-redefinition invariance introduced above becomes Milne boosts on NC manifold,
\bee{
	v^\mu \ra v^\mu + \bar\p^\mu, \qquad
	\cB_\mu \ra \cB_\mu + \bar\p_{\mu} - \half n_\mu \bar\p^2, \qquad
	p_{\mu\nu} \ra p_{\mu\nu} - 2n_{(\mu} \bar\p_{\nu)} + n_\mu n_\nu \bar\p^2. 
}
from where we can identify $v^\mu$ as NC frame velocity. Similarly the Levi-Civita connection $\G^{R}_{\ MS}$ can be reduced in NC basis as,
%$\cM_{(d+2)} = S^1_V \times \bbR^1_T \times \cM^T_{(d)}$ as,
%\bem{\label{E:reduced_connection}
%	\G^{R}_{\ MS}
%	= 
%	- V^R \lb \dow_{(M} \bar V_{S)} - \N_{(M}\bar V_{S)} \rb
%	- \bar V^R \dow_{(M} V_{S)}
%	+ \half P^{RN} \lb \dow_M P_{NS} + \dow_S P_{NM} - \dow_N P_{MS} \rb \\
%	- V_{(M} \O_{S)}^{ \ \ R},
%}
%where we have chosen $\N_{M}\bar V^{N}$ to be an independent field and defined $\O_{MN} = 2\N_{[M}\bar V_{N]} = \dow_M \bar V_{N} - \dow_N \bar V_{M}$. In NC basis only surviving components of connection are,
\bea{\label{NC_Connection}
	\G^{\l}_{\ \mu\nu}
	&= 
	v^\l \dow_{(\mu} n_{\nu)}
	+ \half p^{\l\r} \lb \dow_\mu p_{\r\nu} + \dow_\nu p_{\r\mu} - \dow_\r p_{\mu\nu} \rb
	+ \O_{(T)\s(\nu} n_{\nu)} p^{\s\l}, \nn\\
	\G^{-}_{\ \mu\nu}
	&= 
	p_{\l(\mu} \Ndot_{\nu)}v^\l
	- \Ndot_{(\mu} \cB_{\nu)}.
}
and all other components zero. Here we have identified $\G^{\l}_{\ \mu\nu}$ as Newton-Cartan connection and denoted respective covariant derivative by $\Ndot_\mu$. In \cref{NC_Connection} we have treated derivative of frame velocity $\Ndot_{\mu}v^\nu$ as an independent variable, and have defined frame vorticity in terms of it as $\O_{(T)\mu\nu} = 2p_{\s[\nu}\Ndot_{\mu]}v^\s$. 
The covariant derivative can be checked to act on NC structure appropriately,
\bee{
	\Ndot_\mu n_\nu = 0, \qquad
	\Ndot_\mu p^{\r\s} = 0, \qquad
	\Ndot_\mu p_{\nu\r} = - 2 n_{(\nu} p_{\r)\s} \Ndot_\mu v^\s.
}
We define the spatial volume element on NC manifold as,
\bee{
	\ve^{\mu\nu\ldots} = \e^{MN\mu\nu\ldots}V_M \bar V_{(T)N} = - \e^{-\r\mu\nu\ldots} n_\r, \qquad
	\ve_{\mu\nu\ldots} = p_{\mu\r}p_{\nu\s}\ldots \ve^{\r\s\ldots}.
%	\ve^{\mu\nu\ldots} &= \e^{-+\mu\nu\ldots} = \e^{MN\mu\nu\ldots}V_M \bar V_{(T)N}
}
and $*$ as the Hodge duality operation associated with it. The notations and conventions on differential forms can be found in \cref{forms}. Finally the only surviving components of gauge field strength are $\cF_{\mu\nu}$, which can be identified as NC gauge field strength. We can further decompose $\cF_{\mu\nu}$ and $\O_{(T)\mu\nu}$ into,
\bee{
	\cF_{\mu\nu} = 2 e_{[\mu} n_{\nu]}  + \b_{\mu\nu}, \qquad 
	\O_{(T)\mu\nu} = - 2 \a_{(T)[\mu} n_{\nu]} + \o_{(T)\mu\nu}.
}
All the introduced tensors are transverse to $v^\mu$. Here $e_\mu$ is the electric field while $\b_{\mu\nu}$ is the dual magnetic field defined with respect to frame $T$. Similarly $\a_{(T)\mu}$ is frame acceleration and $\o_{(T)\mu\nu}$ is spatial frame vorticity. We can similarly define the fluid acceleration and vorticity as well, which will be used later in \cref{nonrel_hydro}.
%\begin{shaded}
%
%The extended-space representation is covariant and is endowed with a non-degenerate metric, making the tensorial manipulations easier. One can however get the well known Newton-Cartan covariant Galilean dynamics from here by choosing a basis $x^M = \{x^-,x^\mu \}$ such that $V = \dow/\dow x^-$ \cite{Duval:1984cj}. This will effectively curl away the $x^-$ direction into a very small circle upon compactification, and effective large coordinates will be $x^\mu$ on $\cM_{(d+1)} = \bbR^1_T \times \cM^T_{(d)}$ which will have Newton-Cartan structure. In particular $T$-redefinition will give rise to Milne boost in Newton-Cartan language. We have discussed it in detail in \cref{NewtonCartan}. 
%We can also go to conventional non-covariant Galilean formalism by making a further basis choice $x^M = \{x^-,x^+, x^i \}$ such that $V = \dow/\dow x^-$ and $T = \dow/\dow x^+$. This is equivalent to viewing non-relativistic system from local-rest of frame $T$ with $x^+$ being the respective time. Details of this has been worked out in \cref{sec:inertial}.
%
%\end{shaded}
%Next we perform reduction of various relativistic quantities, and mention their transformation properties under $T$ redefinition. We will drop the subscript $(T)$ in this section, and denote $\bar V = \bar V_{(T)}$, $P = P_{(T)}$, $\bar V' = \bar V_{(T')}$ and $P' = P_{(T')}$.
%\end{shaded}

%\bee{
%	Q^{MN}_{(T)} = P^{MN}_{(T)} - \bar V^M_{(T)} V^N
%}

\paragraph*{Reduction of Currents:}

We decompose the currents of null theory on $\cM_{(d+2)} = S^1_V \times \bbR^1_T \times \cM^T_{(d)}$ as,
\bea{\label{E:RelCons_canonical}
	T^{MN} &= \r \bar V^M_{(T)} \bar V^N_{(T)} + 2 \e_{tot} V^{(M}\bar V^{N)}_{(T)} + 2 j_\r^{(M}\bar V^{N)}_{(T)} + 2 j_\e^{(M}V^{N)} + t^{MN} + \q_1 V^M V^N, \nn \\
	J^M &= q \bar V^M_{(T)} + j_q^M + \q_2 V^M.
}
All the introduced tensors are projected along $P^{MN}$.
Note that redefinitions \cref{E:constitutive_redefinitions} can be used to get rid of $\q$ terms. Choosing Newton-Cartan basis we identify, $t^{\mu\nu}$ as stress-energy tensor, $\r, j^\mu_\r$ as mass density and current, $\e_{tot}$, $j_\e^\mu$ as energy density and current, and $q, j^\mu_q$ as charge density and current, as seen by frame $T$. Under a finite $T$ redefinition \cref{E:milne_defn} they transform as,
\bee{\nn
	\r \ra \r, \qquad
	j^\mu_\r \ra j^\mu_\r - \r \bar\p^\mu, \qquad
	t^{\mu\nu}  \ra t^{\mu\nu} - 2 j^{(\mu}_\r \bar\p^{\nu)} + \r \bar\p^\mu \bar\p^\nu,
}
\bee{\nn
	\e_{tot} \ra \e_{tot} - j^\mu_\r \bar\p_\mu + \half \r \bar\p^2, \qquad
	j^\mu_\e \ra
		\lb j^\mu_\e - \e_{tot} \bar \p^\mu \rb
		- \lb t^{\mu\nu} - \bar\p^\mu j^\nu_\r \rb \bar\p_\nu
		+ \half \bar\p^2 \lb j^\mu_\r - \r \bar \p^\mu \rb,
}
\bee{
	q \ra q, \qquad
	j^\mu_q \ra j^\mu_q - q \bar\p^\mu.
}
%Therefore these relativistic currents are $T$ redefinition invariant only upto redefinitions \cref{E:constitutive_redefinitions}. 
Note that we have used the same time field (reference frame) to decompose currents/densities as well as background fields. It is sometimes required to define background in one reference frame (e.g. lab frame) but currents and densities in some other reference frame (e.g. co-moving frame). One can merely perform a $T$ redefinition on various quantities noted above and gain the desired result.

\paragraph*{Reduction of Ward Identities:} %\label{sec:PF_redn}

On the decomposition $\cM_{(d+2)} = S^1_V \times \bbR^1_T \times \cM^T_{(d)}$, background field content is $V_M$, $\bar V_{(T)M}$, $P_{(T)MN}$ and $\cA_M$, so any physical theory should be described by a partition function $W[V_M,\bar V_{(T)M}, P_{(T)MN}, A_M]$.
 Using the current redefinitions \cref{E:PF_redefinitions} for null backgrounds, we can parametrize the variation of partition function as,
\bem{\label{E:reducedPF}
	\d W = \int \lbr \df x^M \rbr \sqrt{-G} \lB 
		\lb \e_{tot} \bar V^M_{(T)} + j_\e^M \rb \d V_{M}
		+ \lb \r \bar V^M_{(T)} + j_\r^M \rb \d \bar V_{(T)M} \dbrk
		+ \lb p^{M} \bar V^N_{(T)} + \half t^{MN} \rb \d P_{(T)MN}
		+ \lb q \bar V^M_{(T)} + j_q^M \rb \d \cA_M
	\rB.
}
The same partition function should also be gained by directly reducing the relativistic partition function \cref{E:rel(d+2)PF_full}. This will render the quantities in above partition function to be same as the ones defined in \cref{E:RelCons_canonical}, and in addition $p^M = j_\r^M$. The latter constraint is Ward identity of $T$ redefinition, i.e. can be gained by demanding partition function \cref{E:reducedPF} to be invariant under $T$ redefinition. 

Choosing NC basis the partition function variation \cref{E:reducedPF} can be decomposed to,
\bem{\label{NC_PF}
	\d W = \int \lbr \df x^M \rbr \sqrt{\det(p_{\mu\nu} + n_\mu n_\nu)} \bigg[
		- \lb \e_{tot} v^\mu + j_\e^\mu \rb \d n_\mu
		+ \lb \r v^\mu + j_\r^\mu \rb \d \cB_\mu \\
		+ \lb p^{\mu} v^\nu + \half t^{\mu\nu} \rb \d p_{\mu\nu}
		+ \lb q v^\mu + j_q^\mu \rb \d \cA_\mu
	\bigg].
}
Symmetry data of the light cone reduced theory is,
\bee{
	\p^{\text{NC}}_\xi = \lbr \xi^-, \xi_- = - n_\mu \xi^\mu,  \bar\xi^\mu = p^{\mu}_{\ \nu} \xi^\nu, \L_{(\xi)}, \bar\p^\mu = p^{\mu}_{\ \nu} \p^\nu \rbr,
}
where we identify $\xi^-$ as mass parameter, $\xi_-$ as time translation parameter, $\bar\xi^\mu$ as space translation parameter, $\L_{(\xi)}$ as gauge parameter, and $\bar\p^\mu$ as Milne boost parameter. The respective Ward identities can be found using \cref{NC_PF} or directly reducing the null Ward identities \cref{E:relcons_full},
\bea{\label{E:nonrelcons_full_NC}
	\text{Mass Conservation:}&\qquad {\Ndot}_\mu \lb \r v^\mu + j_{\r}^\mu \rb = 0, \nn\\
	\text{Energy Conservation:}&\qquad {\Ndot}_\mu \lb \e_{tot} v^\mu + j_{\e}^\mu \rb
	=
	e_\mu j^\mu_q
	- \lb v^{\mu} p_\nu + t^{\mu}_{\ \nu} \rb \Ndot_\mu v^\nu, \nn\\
	\text{Momentum Conservation:}&\qquad {\Ndot}_\mu \lb v^\mu p^{\nu} + t^{\mu\nu} \rb
	=
	\lB q e^\nu + \b^\nu_{\ \mu} j^\mu_q \rB
	- \lb \r v^\mu + j_{\r}^\mu \rb \Ndot_\mu v^\nu, \nn\\
	\text{Charge Conservation:}&\qquad {\Ndot}_\mu \lb q v^\mu + j_{q}^\mu \rb = 0, \nn\\
	\text{Milne Identity:}&\qquad j_\r^\mu = p^\mu.
}
First terms in the RHS of energy and momentum conservation equations are work done and Lorentz force due to electromagnetic fields, while the last terms are pseudo-energy and pseudo-force due to spacetime dependence of the frame velocity. As we already mentioned, the Milne identity is trivial in theories obtained by reduction.

%\begin{shaded}
%$\xi\cdot\bar V$, $\xi\cdot V$ and $\bar\xi^M$ generate mass, energy and momentum conservation respectively, while $\L_\xi$ generates charge conservation:
%\bea{\label{E:nonrelcons_EM}
%	\N_M \lb \hat\r \bar V^M + j_{\r}^M \rb &= 0, \nn\\
%%
%	\N_M \lb \hat\e_{tot} \bar V^M + j_{\e}^M \rb
%	&=
%	e_{M} j_q^M
%	- \lb \bar V^{M} \hat p^{N} + t^{MN} \rb \N_M \bar V_{N}, \nn\\
%%
%	P_{NR} \N_M \lb \bar V^{M} \hat p^{R} + t^{MR} \rb
%	&=
%	\lb\hat q e_{N} + \b_{NM} j^{M}_{q}\rb
%	- \lb \hat\r \bar V^M + j_{\r}^M \rb \N_M \bar V_{N}, \nn\\
%	\N_M \lb \hat q \bar V^M + j_{q}^M \rb &= 0.
%}
%First terms in the RHS of energy and momentum conservation equations are work done and Lorentz force due to electromagnetic fields, while the last terms are pseudo-energy and pseudo-force due to acceleration and vorticity of frame. 
%These equations can also be gained by reduction of relativistic energy-momentum conservation equations \cref{E:relcons_full}. $\bar \p^M$ as we have already stated, implies the equality of mass current and momentum density, which is an identity on null backgrounds.
%\end{shaded}

All these results here have been mentioned in Newton-Cartan notation, which is a nice covariant formalism for Galilean physics. However for familiarity and to build intuition, we have given all these results in conventional non-covariant notation as well in \cref{non_cov}.

\subsection{Equilibrium on Newton-Cartan Backgrounds}

From the perspective of Galilean theories, equilibrium is defined by a preferred reference frame (or time field) $K$ with respect to which system does not evolve in time. This can be achieved by reducing null theories at equilibrium, and identify the timelike isometry $\p_K$ with preferred reference frame in the Galilean theory.
Hence the variation of eqb. partition function in local rest of reference frame $K$ is essentially same as the null fluid \cref{E:deqbreducedPF} written in terms of Galilean quantities, 
\bem{\label{E:PF_eqb_LCR}
	\d W^{eqb} = \int \lbr \df x^i \rbr \sqrt{g} \lB 
		\e_{o} \frac{1}{\vq_o^2} \d \vq_o
		+ \lb 
			- \frac{1}{\vq_o} j_{o\e}^{i}
			+ \vp_o j_{o\r}^{i}
			+ \nu_o j_{oq}^{i}
		\rb \frac{1}{\vq_o} \tilde\vq \d a_i \dbrk
		+ \frac{1}{2\vq_o} t_o^{ij} \d g_{ij}
		+ \lb \r \d \vp_o + \frac{1}{\vq_o} j_{o\r}^i \d B_i \rb
		+ \lb q \d \nu_o + \frac{1}{\vq_o} j_{oq}^i \d A_i \rb
	\rB,
}
and hence,
\bee{\nn
	\r_o = \frac{\d W^{eqb}}{\d \vp_o}, \quad
	j_{o\r}^i = \vq_o\frac{\d W^{eqb}}{\d B_i}, \quad
	t_o^{ij} = 2\vq_o\frac{\d W^{eqb}}{\d g_{ij}}, \quad
	q_o = \frac{\d W^{eqb}}{\d \nu_o}, \quad
	j_{oq}^i = \vq_o\frac{\d W^{eqb}}{\d A_i},
}
\bee{
	\e_{o} = \vq_o^2\frac{\d W^{eqb}}{\d \vq_o}, \quad
	j^i_{o\e} - \vp_o\vq_o j_{o\r}^i - \nu_o\vq_o j_{oq}^i  = - \vq_o\E{\F}\frac{\d W^{eqb}}{\d a_i}.
}
Here all the observables are defined as seen by reference frame $K$, and are denoted by a subscript $_o$. These will reduce to the expected relations \cref{E:expvar} in flat space, i.e.,
\bee{
	\vq_o = 1, \qquad g_{ij} = \d_{ij}, \qquad \vp_o = \nu_o = A_i = B_i = a_i = 0.
}
Since we have fixed that $T$-redefinition symmetry by choosing a preferred reference frame $K$, the corresponding EOM does not show up. Consequently momentum current $p^\mu$ does not appear in the partition function, and can be found by using the missed EOM. In equilibrium configuration, null fluid and Galilean fluid have same field content and symmetries, so we expect the eqb. partition function to also be the same. To ideal order \cref{nulPF_ideal} it will identify $\r, \e, q$ with thermodynamic functions $R,E,Q$, and hence will give physical interpretation to thermodynamics of null theories in terms of Galilean physics. In hydrodynamic description, at further derivative orders also, it will give physical interpretation to various transport coefficients and constraints of null fluid.

\section{Galilean Hydrodynamics} \label{nonrel_hydro}

Having discussed the light cone reduction of generic null theories in last section, we can straight away perform light cone reduction of null fluids in \cref{null_fluids} and hope to get Galilean fluids. If we look at \cref{E:RelCons} closely, it is already nicely organized in local rest frame of the fluid (defined by $\bar V^M = u^M$). One just needs to apply a $T$-redefinition to it with $\p^M = - u^M$ to get densities and currents in a generic reference frame,
\bee{\nn
	j_\r^\mu = p^\mu = \r \bar u^\mu + \vs_\r^\mu, \qquad
	t^{\mu\nu} = \r \bar u^\mu \bar u^\nu + P p^{\mu\nu} + \pi^{\mu\nu} +  2 \bar u^{(\mu} \vs_\r^{\nu)}, \qquad
	j_q^\mu = q \bar u^\mu + \vs_q^\mu,
}
\bee{
	\e_{tot} = \e + \half \r \bar u^2 + \vs_\r^\mu \bar u_\mu, \qquad
	j_\e^\mu = \lb \e_{tot} + P \rb \bar u^\mu + \vs_\e^\mu + \pi^{\mu\nu}\bar u_\nu + \half \vs_\r^\mu \bar u^2,
}
where we have identified,
\bee{\nn
	\r = \cR, \qquad
	q = \cQ, \qquad
	\e = \cE, \qquad
	\bar u^\mu = u^\nu p_{\nu}^{\ \mu},
}
\bee{\label{E:nonrel-hydro-identifications} 
	\vs_\r^\mu = \bbR^\mu, \qquad
	\vs_q^\mu = \bbJ^\mu, \qquad
	\vs_\e^\mu = \bbE^{\mu}, \qquad
	\pi^{\mu\nu} = (\cP - P) p^{\mu\nu} + \bbT^{\mu\nu},
}
and $\bar u^2 = \bar u^\mu \bar u_\mu$. Similarly entropy current of the Galilean fluid can be found to be,
\bee{
	s = \frac{\e + P}{\vq} - \vp\r - \nu q - \U_{s-}, \qquad
	j_{s}^\mu = s \bar u^\mu + \frac{1}{\vq} \vs_\e^{\mu} - \vp \vs_\r^\mu - \nu \vs_q^\mu + \U^\nu_s p_{\nu}^{\ \mu},
}
which follows second law of thermodynamics 
\bee{
	\Ndot_\mu \lb s v^\mu + j_{s}^\mu \rb \geq 0.
}
Choosing `mass frame' in the null fluid, which is the most natural frame from a non-relativistic perspective, will switch off $\vs_\r^\mu = \bbR^\mu$, and hence Galilean mass current will not undergo any dissipation.  The identifications for mass frame are given by \cref{E:nonrel-hydro-identifications} can be read out in terms of frame invariants as,
\bee{\nn
	\r = R, \qquad
	q = Q, \qquad
	\e = E, \qquad
	s = S - \U_{s-}, \qquad
	\bar u^\mu = u^\nu p_{\nu}^{\ \mu},
}
\bee{\label{E:nonrel-hydro-identifications_invariants} 
	\vs_\r^\mu = 0, \qquad
	\vs_q^\mu = \U^\mu, \qquad
	\vs_\e^\mu = \cE^{\mu}, \qquad
	\pi^{\mu\nu} = \Pi^{\mu\nu},
}
and in turn constitutive relations become,
\bee{\nn
	j_\r^\mu = p^\mu = R \bar u^\mu, \qquad
	t^{\mu\nu} = R \bar u^\mu \bar u^\nu + P p^{\mu\nu} + \pi^{\mu\nu}, \qquad
	j_q^\mu = Q \bar u^\mu + \vs_q^\mu,
}
\bee{\nn
	\hat\e_{tot} = E + \half R \bar u^2, \qquad
	j_\e^\mu = \lb E + P + \half R \bar u^2 \rb \bar u^\mu + \vs_\e^\mu + \pi^{\mu\nu}\bar u_\nu,
}
\bee{
	\hat s = S - \U_{s-}, \qquad
	j_{s}^\mu = S \bar u^\mu + \frac{1}{\vq} \vs_\e^{\mu} - \nu \vs_q^\mu + \U^\nu_s p_{\nu}^{\ \mu} - \U_{s-} \bar u^\mu.
}
These are the standard Galilean constitutive relations, written in Newton-Cartan basis. We will present all these expressions in conventional non-covariant basis in \cref{non_cov} for the benefit of readers not comfortable with Newton-Cartan formalism.

{\renewcommand{\arraystretch}{1.3}
\mktbl{t}{tab:dataredn}{Leading Derivative Order Data for Galilean Fluid}{} {|l|l|} { 
	
	\hline
	\textbf{Null Fluid Data} & \textbf{Newton-Cartan Data} \\
	\hline\hline
	\multicolumn{2}{|c|}{\textbf{Parity Even}} \\ \hline
	
	$\Q \equiv \N_M u^M$ & $\Q \equiv \Ndot_\mu u^\mu$ \\ \hline
	$P^{MN}\dow_N \vq$, $P^{MN}\dow_N \vp$, $P^{MN}\dow_N \nu$ & $p^{\mu\nu}\dow_\nu \vq$, $p^{\mu\nu}\dow_\nu \vp$, $p^{\mu\nu}\dow_\nu \nu$ \\ \hline
	$P^{MN}\lb \cF_{NR} u^R - \vq \dow_N \nu \rb$ & $p^{\mu\nu}\lb \cF_{\nu\r} u^\r - \vq \dow_\nu \nu \rb$ \\ \hline
	$\s^{MN} \equiv 2 P^{MR} P^{NS} \N_{(R} u_{S)} - \frac{2}{d} P^{MN} \Q $ & $\s^{\mu\nu} \equiv 2 p^{\s(\mu} \Ndot_{\s} u^{\nu)} - \frac{2}{d} p^{\mu\nu} \Q $ \\ \hline\hline
	
	\multicolumn{2}{|c|}{\textbf{Parity Odd -- Odd Dimensions $(d=2n-1)$}} \\ \hline
	$l^M_{(r)} \big\vert_{r=0}^{n-1} \equiv \star \lB \bV \wedge \bm u \wedge \bcF^{\wedge r} \wedge \bm\O^{\wedge (n-r-1)} \rB^M$ & 
	$l^\mu_{(r)} \big\vert_{r=0}^{n-1} \equiv \ast \lB \bm\b^{\wedge r} \wedge \bm\o^{\wedge (n-r-1)} \rB^\mu$ \\ \hline\hline
	
	\multicolumn{2}{|c|}{\textbf{Parity Odd -- Even Dimensions $(d=2n)$}} \\ \hline
	$l_{(r)} \big\vert_{r=0}^{n} \equiv \star \lB \bV \wedge \bm u \wedge \bcF^{\wedge r} \wedge \bm\O^{\wedge (n-r)} \rB$ & 
	$l_{(r)} \big\vert_{r=0}^{n} \equiv \ast \lB \bm\b^{\wedge r} \wedge \bm\o^{\wedge (n-r)} \rB$ \\ \hline
	$l_{(r)}^{MN}\dow_N \vq$, $l_{(r)}^{MN}\dow_N \vp$, $l_{(r)}^{MN}\dow_N \nu$ & $l_{(r)}^{\mu\nu}\dow_\nu \vq$, $l_{(r)}^{\mu\nu}\dow_\nu \vp$, $l_{(r)}^{\mu\nu}\dow_\nu \nu$ \\ \hline
	$l_{(r)}^{MN}\lb \cF_{NR} u^R - \vq \dow_N \nu \rb$ & $l_{(r)}^{\mu\nu}\lb \cF_{\nu\r} u^\nu - \vq \dow_\nu \nu \rb$ \\
\hline
	$l_{(r)}^{R(M} \s_{R}^{N)}$ & $l_{(r)}^{\r(\mu} \s^{\nu)\s}p_{\r\s}$ \\ \hline
	\multicolumn{2}{|l|}{where,} \\ 
	$l^{MN}_{(r)} \big\vert_{r=0}^{n-1} \equiv \star \lB \bV \wedge \bm u \wedge \bcF^{\wedge r} \wedge \bm\O^{\wedge (n-r-1)} \rB^{MN}$ & 
	$l^{\mu\nu}_{(r)} \big\vert_{r=0}^{n-1} \equiv \ast \lB \bm\b^{\wedge r} \wedge \bm\o^{\wedge (n-r-1)} \rB^{\mu\nu}$ \\ \hline
}
}

Having obtained the general picture, we can now deduce constitutive relations for a Galilean fluid upto leading order in derivative expansion, using the corresponding null fluid results in \cref{recap}. In \cref{tab:dataredn} we have mentioned light cone reduction of all the leading order data to get Newton-Cartan data. Having done so, rest of the algebra is essentially trivial. In the following we will work in mass frame exclusively. 

\paragraph{Even Dimensional Galilean Fluids:}
Using reduction of data enlisted in \cref{tab:dataredn}, we can read out the even dimensional ($d=2n-1$) constitutive relations from \cref{recap},
\bea{
	\pi^{\mu\nu} &= - \eta \s^{\mu\nu} - p^{\mu\nu}\z \Q, \nn\\
	\vs_q^\mu &= \k_{q} p^{\mu\nu}\dow_\nu \vq + \s_{q} p^{\mu\nu}\lb \cF_{\nu\r}u^\r - \vq \dow_\nu \nu \rb 
	+ \sum_{r=0}^{n-1} \binom{n-1}{r}  \tilde\o_{q(r)} l^\mu_{(r)}, \nn\\
	\vs_\e^\mu &=
	\k_{\e} p^{\mu\nu}\dow_\nu \vq + \vq \k_q p^{\mu\nu}\lb \cF_{\nu\r}u^\r - \vq \dow_\nu \nu \rb
	+ \sum_{r=0}^{n-1} \binom{n-1}{r}  \tilde\o_{\e(r)} l^\mu_{(r)},
}
%\bea{
%%	j_\r^\mu &= \hat p^\mu = R v^\mu, \nn\\
%	t^{\mu\nu} &= R v^\mu v^\nu + P p^{\mu\nu} - \eta \s^{\mu\nu} - p^{\mu\nu}\z \Q, \nn\\
%	j_q^\mu &= Q v^\mu + \k_{q} p^{\mu\nu}\dow_\nu \vq + \s_{q} p^{\mu\nu}\lb \cF_{\nu\r}u^\r - \vq \dow_\nu \nu \rb 
%	+ \sum_{r=0}^{n-1} \binom{n-1}{r}  \tilde\o_{q(r)} l^\mu_{(r)}, \nn\\
%%	\hat\e_{tot} &= E + \half R v^\mu v_\mu, \nn\\
%	j_\e^\mu &= \lb E + P + \half R v^2 \rb v^\mu 
%	+ \k_{\e} p^{\mu\nu}\dow_\nu \vq + \vq \k_q p^{\mu\nu}\lb \cF_{\nu\r}u^\r - \vq \dow_\nu \nu \rb
%	- \eta \s^{\mu\nu} v_\nu - v^\mu \z \Q  \nn\\
%	&\qquad
%	+ \sum_{r=0}^{n-1} \binom{n-1}{r}  \tilde\o_{\e(r)} l^\mu_{(r)},
%}
where transport coefficients $\eta$ (shear viscosity), $\z$ (bulk viscosity), $\s_q$ (electric conductivity) are some non-negative, $\k_\e$ (thermal conductivity) is a non-positive and $\k_q$ (thermo-electric coefficient) is an arbitrary function of $\vq$, $\vp$, $\nu$. Parity-odd transport coefficients (Hall conductivities) are however completely determined upto some constants as,
\bea{\label{E:recap_cons_odd_1_NC}
	\tilde\o_{\e(r)}
	&=
	\vq n \lb
		\vq C_{1,(r)}
		+ \frac{E + P - \vq\vp R}{R} C_{2,(r)}
		- \vq\nu C_{2,(r+1)}
	\rb, \nn \\
	\tilde\o_{q(r)}
	&= 
	\vq n \lb
		\frac{Q}{R} C_{2,(r)}
		- C_{2,(r+1)}
	\rb,
}
where $C$'s are some arbitrary constants, and $C_{2,(0)} = 0$. As a special case one can obtain the 4 dimensional ($d=3$) results which will only affect the parity odd sector,
\bea{
	\pi^{\mu\nu} &= - \eta \s^{\mu\nu} - p^{\mu\nu}\z \Q, \nn\\
	\vs_q^\mu &= \k_{q} p^{\mu\nu}\dow_\nu \vq + \s_{q} p^{\mu\nu}\lb \cF_{\nu\r}u^\r - \vq \dow_\nu \nu \rb
	+ \tilde\o_{q(0)} \o^\mu
	+ \tilde\o_{q(1)} B^\mu, \nn\\
%	\hat\e_{tot} &= E + \half R v^\mu v_\mu, \nn\\
	\vs_\e^\mu &= 
	\k_{\e} p^{\mu\nu}\dow_\nu \vq + \vq \k_q p^{\mu\nu}\lb \cF_{\nu\r}u^\r - \vq \dow_\nu \nu \rb
	+ \tilde\o_{\e(0)} \o^\mu
	+ \tilde\o_{\e(1)} B^\mu,
}
where,
\bee{
	\o^\mu = l^\mu_{(0)} = \half \ve^{\nu\r\mu} \o_{\nu\r}, \qquad
	B^\mu = l^\mu_{(1)} = \half \ve^{\nu\r\mu} \b_{\nu\r},
}
are vorticity and gauge magnetic fields respectively, and,
\bea{
	\tilde\o_{\e(0)}&=2\vq \lb\vq C_{1,(0)}- \vq\nu C_{2,(1)}\rb, \qquad
	\tilde\o_{\e(1)} = 2\vq \lb \vq C_{1,(1)} + \frac{E + P - \vq\vp R}{R} C_{2,(1)}- \vq\nu C_{2,(2)} \rb, \nn \\
	\tilde\o_{q(0)} &= - 2\vq C_{2,(1)}, \qquad
	\tilde\o_{q(1)} = 2\vq \lb\frac{Q}{R} C_{2,(1)}- C_{2,(2)}\rb.
}

\paragraph{Odd Dimensional Galilean Fluids:}
Using reduction of data enlisted in \cref{tab:dataredn}, we can read out the odd dimensional ($d=2n$) constitutive relations from \cref{recap},
\bea{
	\pi^{\mu\nu} &=
	- \eta \s^{\mu\nu} 
	- \sum_{r=0}^{n-1} \binom{n-1}{r} \tilde\eta_{(r)} l_{(r)}^{\r(\mu} \s^{\nu)\s} p_{\r\s}
	- p^{\mu\nu} \lb
		\z \Q
		+ \sum_{r=0}^{n} \binom{n}{r} \tilde\z_{(r)} l_{(r)}
	\rb, \nn\\
	\vs_q^\mu &=
	\lb
		p^{\mu\nu} \k_{q}
		+ \sum_{r=0}^{n-1} \binom{n-1}{r}  l^{\mu\nu}_{(r)}\tilde\k_{q(r)} 
	\rb \dow_\nu \vq
	+ \lb
		p^{\mu\nu} \s_{q}
		+ \sum_{r=0}^{n-1} \binom{n-1}{r}  l^{\mu\nu}_{(r)} \tilde\s_{q(r)} 
	\rb \lb \cF_{\nu\r}u^\r - \vq \dow_\nu \nu \rb \nn\\
	&\qquad 
	+ \vq n  \sum_{r=0}^{n-1} \binom{n-1}{r}  l^{\mu\nu}_{(r)} \lb 
		\frac{Q}{R} \dow_\nu \cS_{2,(r)} - \dow_\nu \cS_{2,(r+1)} \rb, \nn\\
%
%	\hat\e_{tot} &= E + \half R v^\mu v_\mu, \nn\\
%
	\vs_\e^\mu &= 
	\lb
		p^{\mu\nu} \k_{\e}
		+ \sum_{r=0}^{n-1} \binom{n-1}{r}  l^{\mu\nu}_{(r)} \tilde\k_{\e(r)} 
	\rb\dow_\nu \vq 
	+ \vq \lb
		p^{\mu\nu} \k_q
		- \sum_{r=0}^{n-1} \binom{n-1}{r}  l^{\mu\nu}_{(r)} \tilde\k_{q(r)} 
	\rb \lb \cF_{\nu\r}u^\r - \vq \dow_\nu \nu \rb  \nn\\
	&\qquad + \vq n \sum_{r=0}^{n-1} \binom{n-1}{r}  l^{\mu\nu}_{(r)} 
		\lb \vq \dow_\nu \cS_{1,(r)} + \frac{E + P - \vq\vp R}{R} \dow_\nu \cS_{2,(r)} - \vq\nu \dow_\nu \cS_{2,(r+1)} \rb.
}
The transport coefficients in parity even sector are same as before; however parity-odd transport coefficients $\tilde\eta_{(r)}$ (Hall viscosity), $\tilde\k_{\e(r)}$ (thermal Hall conductivity), $\tilde\k_{q(r)}$ (thermo-electric Hall coefficient), $\tilde\s_{q(r)}$ (electric Hall conductivity), $\cS_{1,(r)}$ and $\cS_{2,(r)}$ are some arbitrary functions of $\vq,\vp,\nu$. Finally $\tilde\z_{(r)}$ is determined as,
\bee{
	\tilde\z_{(r)} = - \lB\vq^2 \frac{\dow P}{\dow E} \frac{\dow}{\dow\vq} + \frac{\dow P}{\dow R} \frac{\dow}{\dow\vp} + \frac{\dow P}{\dow Q} \frac{\dow}{\dow\nu} \rB \cS_{2,(r)}.
}
As a special case we would like to write down the 3 dimensional results,
\bea{\label{E:NC_odd_CR}
%	j_\r^\mu &= \hat p^\mu = R v^\mu, \nn\\
	\pi^{\mu\nu} &= 
	- \eta \s^{\mu\nu} 
	- \tilde\eta \ve^{\r(\mu} \s^{\nu)\s}p_{\r\s}
	- p^{\mu\nu} \lb
		\z \Q
		+ \tilde\z_{\o} \o
		+ \tilde\z_{B} B
	\rb, \nn\\
	\vs_q^\mu &=
	\lb
		p^{\mu\nu} \k_{q}
		+ \ve^{\mu\nu} \tilde\k_q
	\rb \dow_\nu \vq 
	+ \lb
		p^{\mu\nu} \s_{q}
		+ \ve^{\mu\nu} \tilde\s_{q} 
	\rb \lb \cF_{\nu\r}u^\r - \vq \dow_\nu \nu \rb \nn\\
	&\qquad + \ve^{\mu\nu} \lb
		\frac{Q}{R} \vq\dow_\nu \cS_{2,(0)} 
		- \vq\dow_\nu \cS_{2,(1)}
	\rb, \nn\\
%
%	\hat\e_{tot} &= E + \half R v^\mu v_\mu, \nn\\
	\vs_\e^\mu &= 
	\lb
		p^{\mu\nu} \k_{\e}
		+ \ve^{\mu\nu} \tilde\k_{\e} 
	\rb \dow_\nu \vq 
	+ \vq \lb
		p^{\mu\nu} \k_q 
		- \ve^{\mu\nu} \tilde\k_{q} 
	\rb \lb \cF_{\nu\r}u^\r - \vq \dow_\nu \nu \rb \nn\\
	&\qquad + \vq\ve^{\mu\nu} \lb \vq \dow_\nu \cS_{1,(0)} + \frac{E + P - \vq\vp R}{R} \dow_\nu \cS_{2,(0)} - \vq\nu \dow_\nu \cS_{2,(1)} \rb,
}
where,
\bee{
	\o = l_{(0)} = \half \ve^{\mu\nu} \o_{\mu\nu}, \qquad
	B = l_{(1)} = \half\ve^{\mu\nu} \cF_{\mu\nu},
}
are again the vorticity and gauge magnetic fields and we have renamed $\tilde\s_{q}  = \tilde\s_{q(0)} $, $\tilde\k_{\e}  = \tilde\k_{\e(0)} $, $\tilde\k_{q} = \tilde\k_{q(0)}$, $\tilde\eta = \tilde\eta_{(0)}$, $\tilde\z_{\o} = \tilde\z_{(0)}$, $\tilde\z_{B} = \tilde\z_{(1)}$. The $3$ dimensional Galilean fluid was also studied by \cite{Geracie:2015xfa}, however we find certain discrepancies in their and our results. A detailed comparison has been provided in \cref{Geracie:2015xfa}.

Before closing this discussion we would like to note that \cite{Jensen:2014ama} also constructed an
equilibrium partition function and entropy current for an uncharged $3$ and $4$ dimensional Galilean
fluid, and used it to constraint the respective constitutive relations. By switching off the charge
sector and setting $d=2,3$ we see that we trivially recover their results. 

This finishes our discussion of (non-anomalous) constitutive relations of a Galilean fluid upto leading order in derivatives in arbitrary number of dimensions, obtained by light cone reduction of a null fluid. Unlike the hydrodynamic reductions before this work \cite{Rangamani:2008gi,Banerjee:2014mka}, there is no non-trivial mapping between the relativistic (null) fluid and the Galilean fluid. In fact term by term, null fluid constitutive relations are same as Galilean constitutive relations. Same is true for thermodynamics, entropy current and the equilibrium partition function as well. We deduce that we can see null fluid as Galilean fluid written in extended space representation. Many aspects of it are already hinted by extended space construction of \cite{{Geracie:2015xfa}}. In next section we extend this approach to study effect of $U(1)$ anomaly on fluid transport.

\section{Anomalies} \label{anomalies}

%\hl{Put $\L_V = 0$}

Upto this point we have studied hydrodynamics on non-anomalous null/Galilean backgrounds. In this
section we want to explore if the null background construction can also be used to introduce $U(1)$
anomaly in Galilean theories\footnote{Galilean anomalies considered in \cite{Jensen:2014hqa} are
  different than what we are considering in this paper, because our background field content does
  not match that of \cite{Jensen:2014hqa} after reduction (we have chosen $\cA_- = 0$). A detailed
  comparison of these issues along with an extension to non-abelian and gravitational anomalies will
  shortly appear in a companion paper \cite{akash}.}. Later we will find how constraints of Galilean fluid modify in presence of anomalies. We use the anomaly inflow mechanism of usual relativistic backgrounds to achieve this goal, with appropriate modifications due to the null structure of the background.

Consider a \emph{bulk} manifold $\cB_{(d+3)}$, on whose boundary $\cM_{(d+2)}$
our theory of interest, i.e. null fluid lives. Indices on $\cB_{(d+3)}$ are denoted with a bar $\bar M,\bar N\ldots$. We define $\cB_{(d+3)}$ also as a null background, with respective fields $\cA_{\bar M}$, $G_{\bar M \bar N}$ and a compatible null isometry\footnote{We would like to mention that this construction only seems to work when we set $\L_{(V)} = 0$. We give more reasoning in this regard in a companion paper.} $\p_V = \{V = V^{\bar M} \dow_{\bar M}, \L_{(V)}=0 \}$, such that transverse components of all these fields vanish at boundary. We can define respective fields on $\cM_{(d+2)}$ by pulling back the bulk fields, which gives it a null background structure.

We start with the assumption that full theory on $\cB_{(d+3)} \cup \cM_{(d+2)}$ described by a partition function $\cW$ is gauge invariant. Most generic such partition function can be decomposed into a bulk and a boundary piece,
\bee{
	\cW = W [\cM_{(d+2)}] + W_{bulk} [\cB_{(d+3)}],
}
which individually are not gauge invariant. Here $W$ is the partition function of the boundary null theory which is \emph{anomalous}, i.e. is not gauge invariant. $W_{bulk}$ on the other hand is a pure bulk piece whose gauge variation must be a boundary term. While constructing $W_{bulk}$ out of background fields, we can let go of any terms which are gauge invariant upto a total derivative (we can always redefine $W$ to absorb this total derivative term at the boundary), as they will not induce any anomalies in the boundary theory. Hence allowed $W_{bulk}$ can be written as integration of a full rank form,
\bee{
	W_{bulk} = \int_{\cB_{(d+3)}} \bI^{(d+3)},
}
such that $\bI^{(d+3)}$ has an exact gauge variation $\d_{\L_{(\xi)}}\bI^{(d+3)} = \df G_{(\xi)}$, and it must not be symmetry invariant upto an exact form\footnote{ $ \bI \neq \bI' + \df \bX $, for some gauge invariant $ \bI'$.}. 
%It is generally more tractable to work with anomaly polynomial $\bcP^{(d+4)} = \df \bI^{(d+3)}$, which contains all non-trivial information about anomalies. The requirements of $\bI^{(d+3)}$ implies that $\bcP^{(d+4)}$ must be closed (obviously), symmetry invariant, and must not be expressible as exterior derivative of a symmetry invariant form.
In usual relativistic theories, $\bI^{(d+3)}$ can only be written in odd bulk dimensions $(d=2n-2)$, and is given by the \emph{Chern-Simons} form $\bI^{(2n+1)}_{CS}$,
\bee{
	\bI^{(2n+1)}_{CS} = C^{(2n)}\bcA \wedge \bcF^{\wedge n}.
}
However for null backgrounds, this term identically vanishes, as it is a full rank form but does not have any component along $V$. We are therefore forced to modify $\bI^{(d+3)}$, by adding some term which has non-vanishing component along $V$. We do it by choosing an arbitrary time-field $T$ and use it to define a conjugate null field $\bar V_{(T)}$. Now we can define an analogue of Chern-Simons form, but in even bulk dimensions $(d=2n-1)$,
\bee{
	\bI^{(2n+2)} = - C^{(2n)} \bm{{\bar V}}_{(T)} \wedge \bcA \wedge \bcF^{\wedge n}.
}
We need to check if it fits our requirements. We will leave it for the readers to convince themselves that this expression cannot be transformed into a symmetry invariant term by adding an exact form. For the other criteria we need to compute its gauge variation,
\bee{
	\d_{\L_{(\xi)}}\bI^{(2n+2)} = 
	C^{(2n)} \df \L_{(\xi)} \wedge  \bm{{\bar V}}_{(T)} \wedge \bcF^{\wedge n}
	= \df \lb C^{(2n)}  \L_{(\xi)} \bm{{\bar V}}_{(T)} \wedge \bcF^{\wedge n} \rb
	- C^{(2n)}  \L_{(\xi)} \df \bm{{\bar V}}_{(T)} \wedge \bcF^{\wedge n}.
}
%Corresponding anomaly polynomial is given by,
%\bee{
%	\bcP^{(2n+3)} = - C^{(2n)}\df \bm{{\bar V}} \wedge \bcA \wedge \bcF^{\wedge n} + C^{(2n)}\bm{{\bar V}} \wedge \bcF^{\wedge (n+1)}.
%}
Last term vanishes as it has again no component along $V$, thus we verify that gauge variation of $\bI^{(2n+2)}$ is a boundary term. 
%It is obviously closed, and one can check that it cannot be written as exterior derivative of a gauge invariant form. 
It is important to note that while we have used $\bar V_{(T)}$ to define $\bI^{(2n+2)}$, it is invariant under $T$-redefinition. One can check there does not exist any other term which meets these criteria. Hence contrary to usual relativistic backgrounds, here we can only define $\bI^{(2n+2)}$ in even bulk dimensions. It means that only odd dimensional null backgrounds (the one at the boundary) and hence even dimensional Galilean backgrounds (that we get by reducing the boundary null theory) can be anomalous, which is what we expect.

\subsection{Anomalous Ward Identities}

In this section, we present the modified Ward identities in presence of $U(1)$ anomaly.
In presence of anomalies, variation of boundary partition function $W$ generates \emph{consistent currents} $T^{MN}_{cons}$ and $J^{M}_{cons}$ which are not gauge invariant. Varying $\bI^{(2n+2)}$ we can now write down variation of the full partition function,
\bee{
	\d\cW
	=
	\int_{\cB_{(2n+2)}} (n+1) C^{(2n)} \d\bcA \wedge \bm{{\bar V}}_{(T)} \wedge \bcF^{\wedge n}
	+ \int \lbr \df x^M \rbr \sqrt{\rmG} \lB
		\half T^{MN}\d \rmG_{MN}
		+ J^M \d\cA_M
	\rB,
}
where we have defined the covariant currents,
\bee{\label{E:cov_current}
	T^{MN} = T^{MN}_{cons}, \qquad
	J^M = J^M_{cons} + n C^{(2n)} \star\lB \bm{{\bar V}}_{(T)} \wedge \bcA \wedge \bcF^{\wedge (n-1)}\rB^M.
}
Since the full partition function is gauge invariant, and we see that the bulk piece is manifestly gauge invariant, therefore covariant currents must also be gauge invariant. Demanding $\cW$ to be gauge invariant we can get the Ward identities in the boundary theory,
\bee{
	\N_M T^{MN} = \cF^{NM} J_M, \qquad
	\N_M J^M = - (n+1) C^{(2n)} \star \lB \bm{{\bar V}}_{(T)} \wedge \bcF^{\wedge n} \rB.
}
We observe that the system exhibits $U(1)$ anomaly.

\subsection{Anomalous Equilibrium Partition Function}

In our earlier discussion on equilibrium in \cref{relEQB}, we wrote the most generic equilibrium partition function as a gauge invariant scalar. Now we need to modify this partition function appropriately with a gauge non-invariant piece to account for anomaly. We decompose the equilibrium partition function into,
\bee{
	W^{eqb} = W^{eqb}_{cons} + W^{eqb}_{anom}.
}
Here $W^{eqb}_{cons}$ is the most generic gauge invariant partition function which can be written out of background fields, which has been discussed thoroughly in preceding sections. $W^{eqb}_{anom}$ on the other hand is completely determined in terms of anomaly coefficient $C^{(2n)}$.  We suggest its explicit form to be,
\bee{
	W^{eqb}_{anom} 
	= - \int_{\cM_{(2n+1)}} n C^{(2n)} \vq_o\nu_o \bV \wedge \bm{{\bar V}}_{(K)} \wedge \bcA \wedge  \lb 
		\bcF^{\wedge (n-1)}
		+ \half (n-1) \df (\vq_o\nu_o \bV) \wedge \bcF^{\wedge (n-2)}
	\rb,
}
which generates correct anomalies. Recall that we are allowed to use any arbitrary time field $T$ to specify anomalies; in equilibrium configuration $T=K$ is the most natural choice. Performing on shell variation ($\df V = 0$) of $W^{eqb}_{anom}$,
\bem{
	\d W^{eqb}_{anom} 
	= 
	- \int_{\cM_{(2n+1)}} \half n (n+1) C^{(2n)} \vq_o^2\nu_o^2 \d \bV \wedge \bV \wedge \bm{{\bar V}}_{(K)} \wedge \bcF^{\wedge (n-1)} \\
	- \int_{\cM_{(2n+1)}} n C^{(2n)} \d \bcA \wedge \lbr 
		(n+1) \vq_o\nu_o \bV \wedge \bm{{\bar V}}_{(K)} \wedge \bcF^{\wedge (n-1)} 
		- \bm{{\bar V}}_{(K)} \wedge \bcA \wedge \bcF^{\wedge (n-1)}
	\rbr,
}
and using \cref{E:cov_current}, we can find find the covariant anomalous currents at equilibrium,
\bea{
	J_{o,anom}^M
	&= n(n+1)C^{(2n)} \vq_o\nu_o \star\lB \bV \wedge \bm{{\bar V}}_{(K)} \wedge \bcF^{\wedge (n-1)} \rB^M, \nn\\
	T^{MN}_{o,anom} &= n(n+1)  C^{(2n)} \vq_o^2 \nu_o^{2} \star\lB \bV \wedge \bm{{\bar V}}_{(K)} \wedge \bcF^{\wedge (n-1)} \rB^{(M} V^{N)}. 
	%\qquad q^M_{anom}  = \half n(n+1)  C^{(2n)} \vq_o^2 \nu_o^{2} \star\lB \bV \wedge \bm{{\bar V}}_{(K)} \wedge \bcF^{\wedge (n-1)} \rB^M.
}
One can check that these currents identically satisfy the anomalous conservation equations. Note that these currents are also to be supplemented with the non-anomalous pieces discussed in previous sections.
In the local rest of reference frame $K$, the equilibrium partition function can be expressed in Kaluza-Klein notation,
\bee{
	W^{eqb}_{anom} 
	= \int_{\cM_{(2n-1)}} n C^{(2n)} \nu_o \bA \wedge  \lb 
		(\df \bA)^{\wedge (n-1)}
		+ \half (n-1) \nu_o \tilde\vq \df \bm a \wedge (\df \bA)^{\wedge (n-2)}
	\rb.
}
Above we have left more than one powers of $\df\bm a$, as they do not contribute on-shell. Varying it we can find the anomalous contribution to Galilean currents; only non-trivial contributions are given by,
\bee{
	j_{q,anom}^i
	= n(n+1)C^{(2n)} \vq_o\nu_o \ast\lB (\df \bA)^{\wedge (n-1)} \rB^i, \quad 
	j_{\e,anom}^i  = \half n(n+1)  C^{(2n)} \vq_o^2 \nu_o^{2} \ast\lB (\df \bA)^{\wedge (n-1)} \rB^i.
}
When generating constitutive relations of a Galilean fluid using equilibrium partition function, above results are naturally written in equilibrium hydrodynamic frame. Interestingly mass current does not get any anomalous correction, hence these results are automatically written in mass frame as well. Correspondingly the constraints of parity-odd sector in odd spatial dimensions ($d=2n-1$) \cref{E:constraints_PF_leading_odd} modify to include contribution from anomalies,
\bea{\label{E:constraints_PF_leading_anom}
	\tilde\o_{\e(n-1)}
	&=
	\vq n \lb
		\vq C_{1,(n-1)}
		+ \frac{E + P - \vq\vp R}{R} C_{2,(n-1)}
		- \vq\nu C_{2,(n)}
		+ \half (n+1) \vq\nu^{2}  C^{(2n)}
	\rb, \nn \\
	\tilde\o_{q(n-1)}
	&= 
	\vq n \lb
		\frac{Q}{R} C_{2,(n-1)}
		- C_{2,(n)}
		+ (n+1) \nu C^{(2n)}
	\rb.
}
Note that only $r=n-1$ component of \cref{E:constraints_PF_leading_odd} is modified, while other constraints remain unchanged.

\subsection{Anomalous Entropy Current}

In this section, we shall try to get the anomalous contribution to constitutive relations found in last sub-section, using second law constraint. In presence of anomaly, the canonical entropy current divergence \cref{E:entropycurrent_div} will get modified to,
\bem{
	\vq\N_M J_{s(can)}^M 
	=
	- \Pi^{MN} \N_M u_N
	- \frac{1}{\vq}  \cE^{M}\dow_M \vq
	+ \U^M \lb
		\cF_{MN}u^N
		- \vq\dow_M \nu
	\rb \\
	+ \vq\nu (n+1) C^{(2n)} \star \lB \bm{{\bar V}}_{(T)} \wedge \bcF^{\wedge n} \rB.
}
Using leading order parity-odd ($d=2n-1$) constitutive relations \cref{leading_cons_odd_1} we can evaluate it to get,
\bem{
	\vq\N_M J_{s(can)}^M 
	=
	\sum_{r=0}^{n-1} \binom{n-1}{r} l^M_{(r)} 
	\lB
	- \tilde\o_{\e(r)} \frac{1}{\vq} \dow_M \vq
	+ \tilde\o_{q(r)} \lb
		\cF_{MN}u^N
		- \vq\dow_M \nu
	\rb
	\rB \\
	- \vq\nu n(n+1) C^{(2n)} l^M_{(n-1)} \cF_{MN} u^N.
}
Clearly it will modify the constraints \cref{E:constraints_PF_leading_anom} only for $r=n-1$. Plugging in the expression for $\N_M \U^M_s$ from section \cref{const_EC}, we will reproduce expression for $\tilde\o_{q(n-1)}$ in \cref{E:constraints_PF_leading_anom}, and get the following differential equations for $\tilde\o_{\e(n-1)}$,
\bee{\nn
	\tilde\o_{\e(n-1)} = \vq n \lb \vq \frac{\dow}{\dow\vq} \tilde\o_{s(n-1)} + \frac{E+P - \vq\vp R}{R} C_{2,(n-1)} - \vq\nu C_{2,(n)} \rb,
}
\bee{
	\frac{\dow}{\dow\vp} \tilde\o_{s(n-1)} = 0, \qquad
	\frac{\dow}{\dow\nu} \tilde\o_{s(n-1)} = n(n+1) \vq^2\nu C^{(2n)}.
}
Last equation will imply,
\bee{
	\tilde\o_{s(n-1)} = \half n(n+1) \vq\nu^2 C^{(2n)} + f(\vq), \qquad
	\frac{\dow}{\dow\vq} \tilde\o_{s(n-1)} = \half n(n+1) \nu^2 C^{(2n)} + C_{1,(n-1)}(\vq).
}
This gives the remaining $\tilde\o_{\e(n-1)}$ constraint in \cref{E:constraints_PF_leading_anom}, except that $C_{1,(n-1)}$ is an arbitrary function of $\vq$ similar to what we saw in \cref{const_EC}. This can be remedied by putting in torsion to relax $\cH_{MN} = 0$ condition, as we shall present in the \cref{MinTor}.

\section{Discussion}

% \hl{Left for Suvankar.}

In this work we have proposed an innovative and interesting approach to construct the equilibrium
partition function and constitutive relations of a Galilean fluid in arbitrary dimensions, starting
from a relativistic system, namely {\it null fluid}. The basic idea of this construction has already
been presented in a previous paper \cite{Banerjee:2015uta}; here we have generalized it to include a
global (anomalous) $U(1)$ current. The beauty and importance of our approach lies in the
construction of {\it null fluids}. We have showed that the symmetries and background field content
of a Galilean theory is exactly captured by a theory defined on null background. This motivates us
to define a theory of hydrodynamics on null backgrounds (i.e. null fluid) from scratch, and use it
to derive constitutive relations of a Galilean fluid.

Although, the main aim of this construction has been to write down leading order constitutive
relations of a Galilean fluid (in presence of anomaly), but in the process we learnt that null fluids can be
considered as a robust stage to study properties of the most generic Galilean fluid. We found an
exact one to one correspondence (not corrected order by order in derivatives) between all aspects of
a null fluid and a Galilean fluid, but more than that, the actual map of this correspondence is
essentially trivial. Our approach has been to study the null fluid itself as an independent theory,
and later exploit the triviality of this map to say something useful about the Galilean fluids.
% Light cone reduction merely facilitates setting up this dictionary precisely.

The triviality of this map is an important feature. Works in past in this direction, including our
own in \cite{Banerjee:2014mka}, have at best found a mapping between usual relativistic fluids and
(a subset of) Galilean fluids, which has to be corrected order by order in the derivative expansion
(not to forget the faulty thermodynamics it endows to the Galilean fluid), and not much useful could
be said thereof. Note that in the current work however, our mapping is exact to all orders in the
derivative expansion, which enables us to directly use much sophisticated relativistic machinery to
study non-relativistic fluids and hence is very interesting in realistic scenarios.

% As we have mentioned in the main work, our way of performing light cone reduction is different
% from what we did along with Roychowdhury in \cite{Banerjee:2014mka}, which failed to give
% `correct' Galilean fluid after reduction. However, not only \cite{Banerjee:2014mka} but all the
% works in this direction preceding it were plagued with the same problems, i.e. did not reproduce
% correct thermodynamics after reduction. Our current null fluid construction allows us to resolve
% this problem by performing the reduction in a different way, as has been described in the paper.

We also found that null backgrounds allow us to introduce $U(1)$ anomalies in an odd-dimensional
null theory (i.e. an even dimensional Galilean theory) and forbid them in an even dimensional null
theory (i.e. an odd dimensional Galilean theory). This is in sharp contrast with the usual
relativistic results, where it is the other way round. However from the perspective of Galilean
fluids it is pretty natural.
% In a previous work \cite{Banerjee:2014mka}, we showed that if we start with an anomalous
% relativistic fluid (in even dimensions obviously), after light cone reduction the lower
% dimensional Galilean fluid becomes anomaly free.  This is because all the physical fields have
% zero component along the compact direction. H However, presence of null isometry changes the
% scenario completely. In this work we explain how one can construct an even dimensional Galilean
% fluid with $U(1)$ anomaly.
A generalization of this construction to include non-abelian and gravitational anomalies will be
presented in a companion paper \cite{akash}.
% Needless to mention that these anomalous terms in the constitutive relations are parity-odd.

Apart from these anomalous terms, we also have other parity-odd terms in the constitutive relations,
for both, even and odd dimensional Galilean fluid. The study of parity-odd non-relativistic
hydrodynamics has become a very fascinating topic in recent years. Fluid consisting of chiral
molecules breaks parity at the microscopic level. This kind of fluid plays an important role in many
biochemical processes, where only the molecule with right chirality can fit into a
protein. Therefore to model such a fluid, we would be forced to add parity-odd terms in the
constitutive relations. Our construction gives a consistent way to obtain the possible parity-odd
terms in a Galilean fluid, at any desired order in the derivative expansion. It would be very
interesting and important to understand the effect of these terms in some practical examples.

Finally, we would like to make some comments on physical aspects of the null fluid. Although we
construct the null fluid dynamics and show that it is in one to one correspondence to a lower
dimensional Galilean fluid; the physical significance of this null fluid itself is not yet clear.
Presence of an extra background field, i.e. the null Killing vector, has allowed us to introduce a
set of new transport coefficients (e.g. $R$) in the null fluid constitutive relations, as compared
to a usual relativistic fluid.
% Usually a relativistic fluid is described by constitutive relations written order by order in
% derivatives of fluid variables (a normalized timelike velocity vector, and thermodynamic variables
% like temperature and chemical potentials) and background fields (e.g. curved metric or
% electromagnetic fields), upto some undetermined transport coefficients.  and some undetermined
% transport coefficients. If we place the system in some background fields (for example
% electromagnetic fields) then the field variables are also incorporated in the system and we write
% constitutive relations as order by order derivative expansion of all these variables.  In the case
% of a null fluid, we have considered the null Killing vector also as a background field and
% incorporated it in the constitutive relation. This enabled us to introduce a new set of transport
% coefficients (like $\cR$) appearing with the null vector.  The dynamics of these constitutive
% relations can be constrained either by demanding the of an equilibrium partition function or by
% constructing an entropy current for the null system and demanding its positive divergence.
The physical meaning of these new transport coefficients, which come coupled to the null vector,
becomes clear only once we identify the dynamics of null fluid with that of a Galilean fluid living
in one lower dimension. Hence, at this stage, the correct physical interpretation of the null fluid
appears to be that it is a particular embedding of the Galilean fluid in a spacetime with one higher
dimension. This approach is more in lines with the axiomatic approach of defining a Galilean fluid,
but has the benefit that we have all the well-developed machinery of relativistic physics at our disposal.

\section*{Acknowledgments}

We would like to thank Felix Haehl for various helpful discussions. The work of NB is supported by DST Ramanujan Fellowship. AJ would like to thank Durham Doctoral Scholarship for financial support, and hospitality of IISER Bhopal where part of this project was done. NB and SD would like to thank the people of India for their generous support 
 to basic science research.

\appendix

\section{Minimal Torsion Model} \label{MinTor}

In the main text we alluded at more that one instances that entropy current fails to capture all the constraints a null/Galilean fluid should follow on torsionless backgrounds.
In this appendix we provide heuristic arguments for consistency between entropy current and equilibrium partition function constraints in parity-odd sector by introducing `minimal torsion'. We introduce a torsion in the connection, $\G^R_{\ [MN]} = - \half u^R \cH_{MN}$, so the full connection becomes,
\bee{
	\G^R_{\ MS} = \half \rmG^{RN} \lb \dow_M \rmG_{NS} + \dow_S \rmG_{NM} - \dow_N \rmG_{MS} \rb - \half \lb u^R \cH_{MS} - u_M \cH_{S}^{\ \ R} - u_S \cH_{M}^{\ \ R} \rb.
}
One can check that with this connection, $\N_M V^N = 0$ does not require $\cH_{MN}$ to be zero. In fact, for our purposes it suffices to choose $\bcH = \df \bV = \bV\wedge \bX$ where $X^M$ is a projected vector such that $\df \bX = 0$. We call this \emph{minimal torsion}. For spinless theories, constitutive relations remain the same except $\N_M \ra \underline{\N}_M = \N_M - X_M$. We can define an entropy current as before, whose canonical part will now have divergence,
\bee{\label{E:torr_EC_div}
	\vq\underline{\N}_M J_{s(can)}^M 
	=
	- \Pi^{MN} \N_M u_N
	+ \cE^{M} \lb X_M - \frac{1}{\vq} \dow_M \vq \rb
	+ \U^M \lb
		\cF_{MN}u^N
		- \vq\dow_M \nu
	\rb.
}
We will now write the most generic constitutive relations and entropy current corrections in presence of minimal torsion, and compute constraints on hydrodynamic transport. In the following we only consider parity-odd sector; in parity-even sector we did not find any discrepancy between entropy current and equilibrium partition function to start with, and moreover calculation with torsion turns out to be trivially equivalent to what was done without torsion.

\paragraph{Odd Dimensions (d=2n-1):}

In odd dimensions introduction of $X^M$ does not lead to any new data; hence constitutive relations \cref{leading_cons_odd_1} do not modify, neither does the entropy current correction \cref{E:odd_dim_EC}. However divergence of entropy current does modify,
\bem{
	- \N_M \U^M_s 
	= 
	l^M_{(0)}  n\lB 
		- \dow_M \tilde\o_{s(0)} + C_{2,(1)} \vq\nu \lb X_M - \frac{1}{\vq} \dow_M \vq \rb
		- C_{2,(1)} \lb \cF_{MN} u^N - \vq \dow_M \nu \rb
	\rB \\
	+ \sum_{r=1}^{n-1} \binom{n-1}{r} l^M_{(r)}
	n \lB 
		- \dow_M \tilde\o_{s(r)} 
		+ \tilde\o_{s(r)} \frac{1}{\vq} \dow_M \vq
		+ \lb \frac{Q}{R} C_{2,(r)} - C_{2,(r+1)} \rb \lb \cF_{MN} u^N - \vq \dow_M \nu \rb \dbrk
		+ \lb \tilde\o_{s(r)} + \frac{E+P - \vq\vp R}{R} C_{2,(r)} - \vq\nu C_{2,(r+1)} \rb \lb X_M - \frac{1}{\vq} \dow_M \vq \rb
	\rB.
}
Plugging in the canonical part of entropy current \cref{E:torr_EC_div} and demanding $\underline{\N}_M J_s^M \geq 0$, we will find a consistency condition on entropy current,
\bee{
	\frac{\dow}{\dow\vq} \tilde\o_{s(r)} = \frac{\tilde\o_{s(r)}}{\vq}, \qquad
	\frac{\dow}{\dow\vp} \tilde\o_{s(r)} = 0, \qquad
	\frac{\dow}{\dow\nu} \tilde\o_{s(r)} = 0 \quad\Ra\quad
	\tilde\o_{s(r)} = C_{1,(r)} \vq,
}
where $C_{1,(r)}$ is a constant. This is the missing constraint, which we did not find through entropy current in absence of torsion. Using this we can check that we get all the constraints \cref{E:constraints_PF_leading_odd} which we got from equilibrium partition function. One can now take a torsionless limit, which is trivial as there is no $X_M$ dependence in the constitutive relations or entropy current.

\paragraph{Even Dimensions (d=2n):} The even dimensional case is more interesting. Constitutive relations \cref{leading_cons_odd_2} modify as,
\bea{
	\tilde\Pi_{(n)}^{MN} &= 
	- P^{MN} \sum_{r=0}^{n} \binom{n}{r} \tilde\z_{(r)} l_{(r)}
	- \sum_{r=0}^{n-1} \binom{n-1}{r} \tilde\eta_{(r)} l_{(r)}^{R(M} \s^{N)}{}_{R}, \nn\\
	\tilde\cE_{(n)}^{M} &= 
	\sum_{r=0}^{n-1} \binom{n-1}{r} l^{MN}_{(r)} 
		\lB \tilde\l_{\e\vp(r)} \dow_N \vp + \tilde\l_{\e\nu(r,s)} \dow_N \nu 
			+ \tilde\k_{\e(r)} \dow_N \vq 
			+ \tilde\l_{\e X(r)} X_N
			+ \tilde\s_{\e(r)} \lb \cF_{NR}u^R - \vq \dow_N \nu \rb 
		\rB, \nn\\
	\tilde\U_{(n)}^{M} &= \sum_{r=0}^{n-1} \binom{n-1}{r} l^{MN}_{(r)} 
		\lB \tilde\l_{q\vp(r)} \dow_N \vp 
			+ \tilde\l_{q\nu(r)} \dow_N \nu 
			+ \tilde\k_{q(r)} \dow_N \vq 
			+ \tilde\l_{q X(r)} X_N
			+ \tilde\s_{q(r)} \lb \cF_{NR}u^R - \vq \dow_N \nu \rb 
		\rB.
}
On the other hand, most generic entropy current correction (omitting terms that will give pure derivative terms in the divergence) will be given as,
\bem{
	\U^M_s = 
	\sum_{r=0}^{n} \binom{n}{r}
	\cS_{2,(r)} \star\lB \bm u \wedge \bm{{\hat\cF}}^{\wedge r} \wedge \bm{{\hat \O}}^{\wedge (n-r-1)} \rB^M \\
	+ \vq n\sum_{r=0}^{n-1} \binom{n-1}{r} l_{(r)}^{MN}
	\lB \tilde\l_{s\vp(r)} \dow_N \vp + \tilde\l_{s\nu(r)} \dow_N \nu + \tilde\l_{s\vq(r)} \dow_N \vq + \tilde\l_{sX(r)} \lb X_N - \frac{1}{\vq}\dow_N \vq\rb \rB.
}
Its divergence can be computed to be,
\bem{
	\underline{\N}_M\U^M_s = 
	- \sum_{r=0}^{n} \binom{n}{r} l_{(r)} u^M\dow_M \cS_{2,(r)} \\
	+ n\sum_{r=0}^{n-1} \binom{n-1}{r} l_{(r)}^{MN} \lB
		\lb \frac{Q}{R} \dow_M\cS_{2,(r)} - \dow_M\cS_{2,(r+1)} \rb \lb \cF_{NR} u^R - \vq \dow_N \nu \rb \dbrk
		+ \lb I_{(r)M} \vq 
			+ \frac{E+P - \vq\vp R}{R} \dow_M\cS_{2,(r)}
			- \vq\nu \dow_M\cS_{2,(r+1)}
		\rb \lb X_N - \frac{1}{\vq} \dow_N \vq \rb \dbrk
		\vq \lb \frac{\dow}{\dow \vq} \tilde\l_{s\vp(r)} - \frac{\dow}{\dow \vp} \tilde\l_{s\vq(r)} \rb \dow_M\vq \dow_N\vp \dbrk
		+ \vq \lb \frac{\dow}{\dow \nu} \tilde\l_{s\vq(r)} - \frac{\dow}{\dow\vq} \tilde\l_{s\nu(r)} \rb \dow_M\nu \dow_N\vq
		+ \vq \lb \frac{\dow}{\dow\vp} \tilde\l_{s\nu(r)} - \frac{\dow}{\dow \nu} \tilde\l_{s\vp(r)} \rb \dow_M\vp \dow_N\nu
	\rB,
}
where,
\bee{
	I_{(r)} = \df \tilde\l_{sX(r)} + \tilde\l_{s\vp(r)} \df \vp + \tilde\l_{s\nu(r)} \df \nu + \tilde\l_{s\vq(r)} \df \vq.
}
The divergence of canonical piece on the other hand is given as:
\bem{
	\vq\underline{\N}_M J_{s(can)}^M =
	- \Q \sum_{r=0}^{n} \binom{n}{r} \tilde\z_{(r)} l_{(r)} \\
	- \sum_{r=0}^{n-1} \binom{n-1}{r} l^{MN}_{(r)} 
		\lB 
			\lbr \tilde\l_{\e\vp(r)} \dow_M \vp + \tilde\l_{\e\nu(r)} \dow_M \nu + \lb\tilde\k_{\e(r)} + \frac{1}{\vq} \tilde\l_{\e X(r)}\rb \dow_M \vq \rbr \lb X_N - \frac{1}{\vq} \dow_N \vq \rb \dbrk
			+ \lb \tilde\s_{\e(r)} - \tilde\l_{\e X(r)} \rb \lb \cF_{MR}u^R - \vq \dow_M \nu \rb \lb X_N - \frac{1}{\vq} \dow_N \vq \rb \dbrk
			+ \lbr \tilde\l_{q\vp(r)} \dow_M \vp + \tilde\l_{q\nu(r)} \dow_M \nu + \lb\tilde\k_{q(r)} + \frac{1}{\vq} \tilde\l_{qX(r)}\rb \dow_M \vq \rbr\lb \cF_{NR}u^R - \vq\dow_N \nu\rb
		\rB.
}
Comparing the two we will get three consistency conditions on the entropy current,
\bee{
	\frac{\dow}{\dow \vq} \tilde\l_{s\vp(r)} = \frac{\dow}{\dow \vp} \tilde\l_{s\vq(r)}, \qquad
	\frac{\dow}{\dow \nu} \tilde\l_{s\vq(r)} = \frac{\dow}{\dow\vq} \tilde\l_{s\nu(r)}, \qquad
	\frac{\dow}{\dow\vp} \tilde\l_{s\nu(r)} = \frac{\dow}{\dow \nu} \tilde\l_{s\vp(r)},
}
which will have most generic solution,
\bea{
	\l_{s\vp(r)} &= \frac{\dow}{\dow\vp} f_{1(r)}(\vq,\vp,\nu), \nn\\
	\l_{s\nu(r)} &= \frac{\dow}{\dow\nu} f_{1(r)}(\vq,\vp,\nu) + \frac{\dow}{\dow\nu} f_{2(r)}(\vq,\nu) \nn\\
	\l_{s\vq(r)} &= \frac{\dow}{\dow\vq} f_{1(r)}(\vq,\vp,\nu) + \frac{\dow}{\dow\vq} f_{2(r)}(\vq,\nu) + \frac{\dow}{\dow\vq} f_{3(r)}(\vq),
}
for some arbitrary functions $f_{1(r)}(\vq,\vp,\nu)$, $f_{2(r)}(\vp,\nu)$, $f_{3(r)}(\nu)$. We define,
\bee{
	\cS_{1,(r)} = \tilde\l_{sX(r)} + f_{1(r)} + f_{2(r)} + f_{3(r)}.
}
One can check that $I_{(r)} = \df \cS_{1,(r)}$. Demanding entropy current divergence to be non-negative one can check that (in torsionless limit) we get all the equilibrium partition function constraints, as well as the additional entropy current constraints which we found before.

Hence we have established the agreement of equilibrium partition function and entropy current in arbitrary number of dimensions. These results upon reduction, agree with the entropy current calculation for 2 spatial dimensional fluid in \cite{Geracie:2015xfa}, except for a few discrepancies. We provide a detailed comparison with their results in \cref{Geracie:2015xfa}.

\section{Non-Covariant Results} \label{non_cov}

{\renewcommand{\arraystretch}{1.5}
\mktbl{t}{tab:dataredn_noncov}{Leading Derivative Order Data for Galilean Fluid (Non-Covariant)}{} {|l|l|} { 
	
	\hline
	\textbf{Newton-Cartan Data} & \textbf{Non-Covariant Data} \\
	\hline\hline
	\multicolumn{2}{|c|}{\textbf{Parity Even}} \\ \hline
	
	$\Ndot_\mu u^\mu$ &
	$\Nsp_{i} \bar u^{i} - \half (\E{\F} - a_i \bar u^i) g_{ij} \dow_t g^{ij} - a_i \dow_t \bar u^{i}$ \\ \hline
	$p^{\mu\nu}\dow_\nu \vq$, $p^{\mu\nu}\dow_\nu \vp$, $p^{\mu\nu}\dow_\nu \nu$
	& $\Nsp^i\vq - a^i\dow_t \vq$, $\Nsp^i\vp - a^i\dow_t \vp$, $\Nsp^i\nu - a^i\dow_t \nu$ \\ \hline
	$p^{\mu\nu}\lb \cF_{\nu\r} u^\r - \vq \dow_\nu \nu \rb$
	& $ e^i + \b^{ij}\bar u_j - \vq \Nsp^i\nu + \vq a^i\dow_t \nu$ \\ \hline
	$\s^{\mu\nu} = 2 p^{\s(\mu} \Ndot_{\s} u^{\nu)} - \frac{2}{d} p^{\mu\nu} \Ndot_\mu u^\mu $ &
	\begin{minipage}{8.5cm}\raggedright$ \s^{ij} = 2\Nsp^{(i} \bar u^{j)} - \frac{2}{d} g^{ij} \Nsp_{i} \bar u^{i}$
	$- 2a^{(i} \dow_t \bar u^{j)} + \frac{2}{d} g^{ij} a_i \dow_t \bar u^{i}$ \\
	$\qquad - (\E{\F} - a_i \bar u^i) \lb \dow_t g^{ij} 
	+ \frac{1}{d} g^{ij} g_{kl} \dow_t g^{kl} \rb $\end{minipage} \\ \hline\hline
	
	\multicolumn{2}{|c|}{\textbf{Parity Odd -- Odd Dimensions $(d=2n-1)$}} \\ \hline
	$l^\mu_{(r)} \big\vert_{r=0}^{n-1} \equiv \ast \lB \bm\b^{\wedge r} \wedge \bm\o^{\wedge (n-r-1)} \rB^\mu$ &
	$l^i_{(r)} \big\vert_{r=0}^{n-1} \equiv \ast \lB \bm\b^{\wedge r} \wedge \bm\o^{\wedge (n-r-1)} \rB^i$ \\ \hline\hline
	
	\multicolumn{2}{|c|}{\textbf{Parity Odd -- Even Dimensions $(d=2n)$}} \\ \hline
	$l_{(r)} \big\vert_{r=0}^{n} \equiv \ast \lB \bm\b^{\wedge r} \wedge \bm\o^{\wedge (n-r)} \rB$ &
	$l_{(r)} \big\vert_{r=0}^{n} \equiv \ast \lB \bm\b^{\wedge r} \wedge \bm\o^{\wedge (n-r)} \rB$ \\ \hline
	$l_{(r)}^{\mu\nu}\dow_\nu \vq$, $l_{(r)}^{\mu\nu}\dow_\nu \vp$, $l_{(r)}^{\mu\nu}\dow_\nu \nu$ &
	\begin{minipage}{7cm}\raggedright $l_{(r)}^{ij}\lb\dow_j \vq - a_i \dow_t \vq \rb$, $l_{(r)}^{ij}\lb\dow_j \vp - a_i \dow_t \vp \rb$, $l_{(r)}^{ij}\lb\dow_j \nu - a_i \dow_t \nu \rb$
	\end{minipage} \\ \hline
	$l_{(r)}^{\mu\nu}\lb \cF_{\nu\r} u^\nu - \vq \dow_\nu \nu \rb$ &
	$l_{(r)}^{ij}\lb e_j + \b_{jk} \bar u^k - \vq \dow_j \nu + \vq a_j \dow_t \nu \rb$ \\
\hline
	$l_{(r)}^{\r(\mu} \s^{\nu)\s}p_{\r\s}$ &
	$l_{(r)}^{k(i} \s^{j)l} g_{kl}$ \\ \hline
	\multicolumn{2}{|l|}{where,} \\ 
	$l^{\mu\nu}_{(r)} \big\vert_{r=0}^{n-1} \equiv \ast \lB \bm\b^{\wedge r} \wedge \bm\o^{\wedge (n-r-1)} \rB^{\mu\nu}$ &
	$l^{\mu\nu}_{(r)} \big\vert_{r=0}^{n-1} \equiv \ast \lB \bm\b^{\wedge r} \wedge \bm\o^{\wedge (n-r-1)} \rB^{ij}$ \\ \hline
}
}

We mentioned in the main text that vector field $T$ defines a reference frame. We can go to \emph{local rest} of one such frame by choosing a basis $x^M = \{ x^-, t, x^i \}$ such that $V = \dow_-$ and $T = \dow_t$. This essentially amounts to setting $v^i = 0$ in the Newton-Cartan construction. For example in the equilibrium configuration discussed in \cref{relEQB}, we have studied the system in local rest of frame defined by isometry $K^M$. Using the same field decomposition as given in \cref{relEQB}, partition function \cref{NC_PF} reduces to:
\bem{\label{E:dKKreducedPF}
	\d W = \tilde R \int \lbr \df x^\mu \rbr \sqrt{-G} \bigg[
		\E{-\F} \e_{tot} \d \F
		- \E{-2\F} \lb
			j_\e^i 
			- j_\r^i \E{\F} \cB_t
			- j_q^i \E{\F} \cA_t
		\rb \d a_i 
		+ \E{-\F} \half t^{ij} \d g_{ij} \\
		+ \lb \r \d \cB_t + \E{-\F} j_\r^i \d B_i \rb
		+ \lb q \d \cA_t + \E{-\F} j_q^i \d A_i \rb
	\bigg].
}
It is exactly same as \cref{E:PF_eqb_LCR}, except that the observables are now defined with respect to frame $T$ and are not independent of $t$. Note that choosing a frame fixes the Milne invariance in the partition function, and hence Milne Ward identity $p^i = j_\r^i$ goes on-shell. Under Milne boost background fields transform as,
\bee{
	\cB_t \ra \cB_t - \E{-\F} \half \bar\p^2, \qquad
	B_i \ra B_i + \bar\p_{i}. 
}
On the other hand various densities and currents transform as,
\bee{\nn
	\r \ra \r, \qquad
	j^i_\r \ra j^i_\r - \r\bar\p^i, \qquad
	t^{ij} \ra t^{ij} - 2 j^{(i}_\r \bar\p^{j)} + \r \bar\p^2,
}
\bee{\nn
	\e_{tot} \ra \e_{tot} - j^i_\r \bar\p_i + \half \r \bar\p^2, \qquad
	j^i_\e \ra
	\lb j^i_\e - \e_{tot} \bar\p^i \rb
		- \lb t^{ij} - \bar\p^i j^j_\r \rb \bar\p_j 
		+ \half \bar\p^2 \lb j^i_\r - \r \bar\p^i \rb.
}
\bee{
	q \ra q, \qquad
	j^i_q \ra j^i_q - q \bar\p^i.
}
The conservation equations in non-covariant basis becomes,
\bea{\label{E:inertial_CR}
	\frac{1}{\sqrt{g}} \dow_t \lb \sqrt{g} q_{\text{nc}}\rb + \Nsp_i \lb \E{-\F}  j_q^i \rb &= 0, \nn\\
	\frac{1}{\sqrt{g}} \dow_t \lb \sqrt{g} \r_{\text{nc}} \rb + \Nsp_i \lb \E{-\F}  j_\r^i \rb &= 0, \nn\\
	\frac{1}{\sqrt{g}}\dow_t \lb \sqrt{g} \e_{tot,\text{nc}}\rb + \Nsp_i \lb \E{-\F}  j_\e^i \rb
	&= - \half t^{ij} \dow_t g_{ij} + \E{-\F} e_{i}j^i_q - \E{-\F} \a_{(T)i} j_\r^i \nn\\
	\frac{1}{\sqrt{g}}\dow_t \lb\sqrt{g} p_{\text{nc},i} \rb
	+ \Nsp_j \lb \E{-\F} t^{j}_{\ i} \rb
	&= - \half \E{-\F} a_i t^{jk} \dow_t g_{jk}
	+ \E{-\F} \lb q e_i + \b_{ij} j^j_{q} \rb
	+ \E{-\F} \lb - \r \a_{(T)i} + \o_{(T)ij} j^{j}_\r \rb,
}
where non-covariant densities gets a contribution from temporal curvature,
\bee{
	\r_{\text{nc}} = \r - \E{-\F}  j_\r^i a_i, \quad
	\e_{tot,\text{nc}} = \e_{tot} - \E{-\F}  j_\e^i a_i, \quad
	q_{\text{nc}} = q - \E{-\F}  j_q^i a_i, \quad
	p^i_{\text{nc}} = j_\r^i - \E{-\F}  t^{ij} a_j.
}
These are just the usual Galilean conservation equations, generalized to curved space-time. Constitutive relations of a Galilean fluid written in mass frame can be found as,
\bee{\nn
	j_\r^i = p^i = R \bar u^i, \qquad
	t^{ij} = R \bar u^i \bar u^j + P g^{ij} + \pi^{ij}, \qquad
	j_q^i = Q \bar u^i + \vs_q^i,
}
\bee{\nn
	\e_{tot} = E + \half R \bar u^2, \qquad
	j_\e^i = \lb E + P + \half R \bar u^2 \rb \bar u^i + \vs_\e^i + \pi^{ij}\bar u_j.
}
\bee{
	\hat s = S - \U_{s-}, \qquad
	j_{s}^i = S \bar u^i + \frac{1}{\vq} \vs_\e^{i} - \nu \vs_q^i + \U^i_s - \U_{s-} \bar u^i.
}
They follow conservation laws \cref{E:inertial_CR}. Finally we can explicitly obtain the constitutive relations upto leading order in derivative expansion for by reducing results of \cref{nonrel_hydro} in mentioned basis. Reduction of various data down to non-covariant basis is given in \cref{tab:dataredn_noncov}. In the following we present results for a special case when time is flat $(a_i = \F = 0)$, space is time independent $(\dow_t g_{ij} = 0)$ and reference frame is inertial $\a_i = \o_{ij} = 0$ for simplicity.

\paragraph*{Odd Spatial Dimensions:} We first present constitutive relations for a fluid living in odd spatial dimensions $(d=2n-1)$,
\bea{
	\pi^{ij} &= - \eta \s^{ij} - g^{ij}\z \Nsp_k \bar u^k, \nn\\
	\vs_\e^i &= \k_{\e} \Nsp^i \vq + \vq \k_q \lb e^i + \b^{ij} \bar u_j - \vq \Nsp^i \nu \rb + \sum_{r=0}^{n-1} \binom{n-1}{r}  \tilde\o_{\e(r)} l^i_{(r)}, \nn\\
	\vs_q^i &= \k_{q} \Nsp^i \vq + \s_{q} \lb e^i + \b^{ij} \bar u_j - \vq \Nsp^i \nu \rb + \sum_{r=0}^{n-1} \binom{n-1}{r}  \tilde\o_{q(r)} l^i_{(r)}.
}
In the special case of $3$ spatial dimensions we will get the well known results \cite{landau1959fluid},
\bea{
	\pi^{ij} &= 
	- \eta \s^{ij} - \z g^{ij} \Nsp_k \bar u^k, \nn\\
	\vs_\e^i &= 
	\k_{\e}\Nsp^i \vq + \s_{\e} \lb e^i + (\bar u \times B)^i - \vq \Nsp^i \nu \rb + \tilde\o_{\e(0)} \o^i + \tilde\o_{\e(1)} B^i, \nn\\
	\vs_q^i &= \k_{q} \Nsp^i \vq + \s_{q} \lb e^i + (\bar u \times B)^i - \vq \Nsp^i \nu \rb 
	+ \tilde\o_{q(0)} \o^i + \tilde\o_{q(1)} B^i,
}
where $B^i = \half \ve^{ijk}\b_{jk}$, $\o^i = \ve^{ijk}\dow_j v_k$, $(v \times B)^i = \ve^{ijk} v_j B_k = \b^{ij}v_j$.

\paragraph*{Even Spatial Dimensions:} Similarly in even spatial dimensions $(d=2n)$ we can get,
\bea{
	\pi^{ij} &= 
	- \eta \s^{ij} 
	- \sum_{r=0}^{n-1} \binom{n-1}{r} \tilde\eta_{(r)} l_{(r)}^{k(i} \s^{j)l} g_{kl}
	- g^{ij} \lb
		\z \Nsp_i \bar u^i
		+ \sum_{r=0}^{n} \binom{n}{r} \tilde\z_{(r)} l_{(r)}
	\rb, \nn\\
	\vs_\e^i &=
	\lb
		g^{ij} \k_{\e}
		+ \sum_{r=0}^{n-1} \binom{n-1}{r}  l^{ij}_{(r)} \tilde\k_{\e(r)} 
	\rb\dow_j \vq 
	+ \vq \lb
		g^{ij} \k_q
		- \sum_{r=0}^{n-1} \binom{n-1}{r}  l^{ij}_{(r)} \tilde\k_{q(r)} 
	\rb \lb e_j + \b_{jk} \bar u^k - \vq \dow_j \nu \rb  \nn\\
	&\qquad + \vq n \sum_{r=0}^{n-1} \binom{n-1}{r}  l^{ij}_{(r)} 
		\lb \vq \dow_j \cS_{1,(r)} + \frac{E + P - \vq\vp R}{R} \dow_j \cS_{2,(r)} - \vq\nu \dow_j \cS_{2,(r+1)} \rb, \nn\\
	\vs_q^i &=
	\lb
		g^{ij} \k_{q}
		+ \sum_{r=0}^{n-1} \binom{n-1}{r}  l^{ij}_{(r)}\tilde\k_{q(r)} 
	\rb \dow_j \vq
	+ \lb
		g^{ij} \s_{q}
		+ \sum_{r=0}^{n-1} \binom{n-1}{r}  l^{ij}_{(r)} \tilde\s_{q(r)} 
	\rb \lb e_j + \b_{jk} \bar u^k - \vq \dow_j \nu \rb \nn\\
	&\qquad 
	+ \vq n  \sum_{r=0}^{n-1} \binom{n-1}{r}  l^{ij}_{(r)} \lb 
		\frac{Q}{R} \dow_j \cS_{2,(r)} - \dow_j \cS_{2,(r+1)} \rb.
}
These results for sure are messy, but take a cleaner form in 2 spatial dimensions,
\bea{
	\pi^{ij} &=
	- \eta \s^{ij}
	- \tilde\eta \ve^{k(i} \s^{j)l} g_{kl}
	- g^{ij} \lb \z \Nsp_k \bar u^k + \tilde\z_{\o} \o + \tilde\z_{B} B \rb, \nn\\
	\vs_\e^i &= 
	\lb g^{ij} \k_{\e} + \ve^{ij} \tilde\k_{\e} \rb \dow_j \vq
	+ \vq \lb g^{ij} \k_q - \ve^{ij} \tilde\k_{q} \rb \lb e_j + \e_{jk} \bar u^k B - \vq \dow_j \nu \rb\nn\\
	&\qquad + \vq \ve^{ij}
		\lb \vq \dow_j \cS_{1,(0)} + \frac{E + P - \vq\vp R}{R} \dow_j \cS_{2,(0)} - \vq\nu \dow_j \cS_{2,(1)} \rb, \nn\\
	\vs_q^i &= 
	\lb g^{ij} \k_{q} + \e^{ij} \tilde\k_{q} \rb \dow_j \vq
	+ \lb g^{ij} \s_q + \e^{ij} \tilde\s_{q} \rb \lb e_j + \e_{jk} \bar u^k B - \vq \dow_j \nu \rb \nn\\
	&\qquad + \vq \ve^{ij} \lb \frac{Q}{R} \dow_j \cS_{2,(0)} - \dow_j \cS_{2,(1)}\rb,
}
where $B = \half \ve^{ij}\b_{ij}$, $\o = \ve^{ij}\dow_i v_j$, $\ve^{ij} v_j B = \b^{ij} v_j$.

Under the assumption of flat time, time independent space and reference frame being inertial, these constitutive relations follow very simplified and familiar conservation laws,
\bee{\nn
	\dow_t Q + \Nsp_i j_q^i = 0, \qquad
	\dow_t R + \Nsp_i (R v^i) = 0,
}
\bee{
	\dow_t \lb E + \half R v^i v_i \rb + \Nsp_i j_\e^i = j^i_q e_{i}, \qquad
	\dow_t (R v^i)
	+ \Nsp_j t^{ji}
	= 
	Q e^i - j_{q j}\b^{ji}.
}
Last term can be seen as $(j_{q}\times B)^i$ or $\ve^{ij}j_{qj} B$ depending on the number of dimensions. 

\section{Comparison with Geracie et al. \cite{Geracie:2015xfa}} \label{Geracie:2015xfa}

Our null fluid construction is computationally similar to \cite{Geracie:2015xfa}, but has a
different essence to it. Authors in \cite{Geracie:2015xfa} considered an extended $(d+2)$-dim
representation of the Galilean group, and realized it with the help of a $(d+2)$-dim flat
space. They then defined an extended Vielbein which connects this $(d+2)$-dim space to $(d+1)$-dim
Newton-Cartan manifold, and used this formalism to write Ward identities and constitutive relations
of the Galilean fluid in a covariant manner. In this work however, we do hydrodynamics on the
$(d+2)$-dim curved manifold (null background) to start with, and later perform light cone reduction
to get a Galilean fluid. As we mentioned, fluid on null background (null fluid) is essentially
equivalent to the Galilean fluid, so we can expect computational similarities with the construction of \cite{Geracie:2015xfa}.

\cite{Geracie:2015xfa} also studied torsional Galilean fluid in 2 spatial dimensions. We should be able to reproduce their results restricted to torsion-less case. In two spatial dimensions $(d=2)$, the hydrodynamic frame invariants in parity-odd sector \cref{E:NC_odd_CR} become,
\bea{\label{E:Geracie_comp}
	\pi^{\mu\nu} &= - \eta \s^{\mu\nu} - \tilde\eta \e^{\r(\mu} \s^{\nu)\s}p_{\r\s} - p^{\mu\nu}\lb \z \Q + \tilde f_\o \o + \tilde f_B B \rb, \nn\\
	\vs_\e^{\mu} &= 
	\vq \s_T p^{\mu\nu}\lb \cF_{\nu\r}v^\r - \vq \dow_\nu \nu \rb
	+ \k_{T} p^{\mu\nu}\dow_\nu \vq 
	- \vq\tilde\s_T \e^{\mu\nu} \lb \cF_{\nu\r}u^\r - \vq \dow_\nu \nu \rb
	+ \tilde\k_T \e^{\mu\nu} \dow_\nu \vq \nn\\ 
	&\qquad 
	- \tilde m \e^{\mu\nu}  \cF_{\nu\r}u^\r
	+ \frac{E + P - \vq\vp R}{R} \vq \e^{\mu\nu}  \dow_\nu \tilde n
	+ \dow_\nu \e^{\mu\nu} \tilde m_\e, \nn\\
	\vs_q^{\mu} &= 
	\s_{E} p^{\mu\nu}\lb \cF_{\nu\r}u^\r - \vq \dow_\nu \nu \rb + \s_{T} p^{\mu\nu}\dow_\nu \vq 
	+ \tilde\s_{E} \e^{\mu\nu}\lb \cF_{\nu\r}v^\r - \vq \dow_\nu \nu \rb 
	+ \e^{\mu\nu}\tilde\s_{T} \dow_\nu \vq  \nn\\
	&\qquad + \e^{\mu\nu} \dow_\nu \tilde m + \frac{Q}{R} \vq \e^{\mu\nu} \dow_\nu \tilde n,
}
where we have made some redefinitions to make results look similar to \cite{Geracie:2015xfa},
\bee{\nn
	\tilde m = - \vq\cS_{2,(1)}, \quad
	\tilde n = \cS_{2,(0)}, \quad
	\tilde\s_T = \tilde\k_{q} - \frac{1}{\vq} \tilde m, \quad
	\tilde m_\e = \vq^2 \cS_{1,(0)} + \vq\nu \tilde m, \quad
	\tilde\k_T = \tilde\k_{\e} - \frac{2}{\vq} \tilde m_\e,
}
\bee{
	\tilde\s_{E} = \tilde\s_{q}, \quad
	\s_E = \s_q, \quad
	\s_{T} = \k_{q}, \quad
	\k_T = \k_\e,
}
\bee{\nn
	\tilde f_\o = \tilde\z_\o = - \lB\vq^2 \frac{\dow P}{\dow E} \frac{\dow}{\dow\vq} + \frac{\dow P}{\dow R} \frac{\dow}{\dow\vp} + \frac{\dow P}{\dow Q} \frac{\dow}{\dow\nu} \rB \tilde n,
}
\bee{
	\tilde f_B = \tilde\z_B = \lB\vq^2 \frac{\dow P}{\dow E} \frac{\dow}{\dow\vq} + \frac{\dow P}{\dow R} \frac{\dow}{\dow\vp} + \frac{\dow P}{\dow Q} \frac{\dow}{\dow\nu} \rB \frac{\tilde m}{\vq}.
}
Relations \cref{E:Geracie_comp} are same as eqn. (6.64) of \cite{Geracie:2015xfa}, except few subtle
points. We do not have the coefficient $\tilde\k_Q$ while they don't have $\tilde n$ and $\tilde
f_\o$. We would like to mention few points which might explain this discrepancy. In the following we use notation as used in their paper.

Just before eqn. (6.52) of \cite{Geracie:2015xfa}, authors dropped the $\tilde\z_\o \o u^\mu$ term in the entropy current as it gives rise to a `genuine second order data' $\dot\o$. However one can show that just like magnetic field, there exists an independent combination,
\bee{
	\tilde\z_\o \lb \o u^\mu + \tilde\a_\nu \rb,
}
which has composite divergence. Here $\a^\nu = u^\mu\N_\mu u^\nu$ and $\tilde V^\mu = \e^{\mu\nu} V_\nu$ is the duality operation. In other words $\dot\o = u^\mu \dow_\mu \o$ is not an independent genuine data, and can be decomposed using first order Ward identities,
\bea{
	\dot\o 
	&= - \o \Q - \e^{\mu\nu} \N_\mu \a_\nu \nn\\
	&= 
	\lb \frac{q}{\r} B - \o \rb \Q 
	- \e^{\mu\nu} \N_\mu e_\nu
	+ \frac{q}{\r} \dot B
	- \tilde\N^\mu \lb \frac{q}{\r} \rb E_\mu
	- \tilde\N^\mu \lb \frac{\e+p}{\r} \rb G_\mu
	- \frac{1}{\r^2} \tilde\N^\mu \r \N_\nu p.
}
Therefore \cite{Geracie:2015xfa} missed the $\tilde n$ and its dependent $\tilde f_\o$
coefficients. For other discrepancy we note that they have a term $-\frac{1}{T}(\tilde\s_G -
\tilde\k_E) \tilde E^\mu G_\mu$ in entropy production eqn. (6.55), which implies $\tilde\s_G =
\tilde\k_E$. 
%Authors missed this while noting the constraints in eqn. (6.56). 
This will give rise to another consistency condition, which can be read out directly from eqn. (6.62),
\bee{
	\tilde c_Q + T \lb \dow_Q \tilde c_T - \dow_T \tilde c_Q \rb = 0 \quad\Ra\quad
	\tilde c_Q = T \lB \dow_Q f_1(T,\nu_Q) + \dow_Q f_2(\nu_Q) \rB, \quad
	\tilde c_T = T \dow_T f_1(T,\nu_Q),
}
for some arbitrary functions $f_1(T,\nu_Q)$, $f_2(\nu_Q)$. One can in turn absorb these functions by redefining,
\bee{
	\tilde g \ra \tilde g - f_1 - f_2.
}
As an effect, in eqn. (6.62) of \cite{Geracie:2015xfa} all the $c$'s drop out. Consequently by eqn. (6.70) all $c$ dependent coefficients $\tilde\s_\F$, $\tilde\k_Q$, $\tilde\k_\F$ vanish (or can be absorbed in definition of $\tilde m_\e$). Barring these modifications, we find our results to be in exact correspondence with \cite{Geracie:2015xfa} for torsionless fluids.

\section{Conventions of Differential Forms} \label{forms}

In this appendix we will recollect some results about differential forms, and will set notations and conventions used throughout this work. An $m$-rank differential form $\bm\mu^{(m)}$ on $\cM_{(d+2)}$, can be written in a coordinate basis as,
\bee{
	\bm\mu^{(m)} = \frac{1}{m!}\mu_{M_1M_2\ldots M_m} \df x^{M_1} \wedge \df x^{M_2} \wedge \ldots \wedge \df x^{M_m},
}
where $\mu$ is a completely antisymmetric tensor. On $\cM_{(d+2)}$, volume element is given by a full rank form, 
\bee{
	\bm\e^{(d+2)} = \frac{1}{(d+2)!}\e_{M_1M_2\ldots M_{d+2}} \df x^{M_1} \wedge \df x^{M_2} \wedge \ldots \wedge \df x^{M_{d+2}},
}
where $\e$ is the totally antisymmetric Levi-Civita symbol with value $\e_{0,1,2,\ldots,d+1} = \sqrt{|G|}$ and $G =\det{G_{MN}}$. Using it,  Hodge dual is defined to be a map from $m$-rank differential forms to $(d+2-m)$-rank differential forms,
\bee{
	\star[\bm\mu^{(m)}] = \frac{1}{(d+2-m)!} \lb\frac{1}{m!}\mu^{M_1\ldots M_m} \e_{M_1\ldots M_{m}N_{1}\ldots N_{d+2-m}} \rb 
	\df x^{N_{1}} \wedge \ldots \wedge \df x^{N_{d+2-m}}.
}
One can check that $\star\star\bm\mu^{(m)} = \sgn(G) (-)^{m(d-m)}$. For us obviously $\sgn(G) = -1$ due to Minkowski signature of the metric, but we tag along this factor for clarity.
The exterior product of a differential form is defined to be,
\bee{
	\df \bm\mu^{(m)} = \frac{1}{(m+1)!} \lB (m+1) \dow_{[M_1} \mu_{M_2\ldots M_{m+1}]} \rB \df x^{M_1} \wedge \ldots \wedge \df x^{M_{p+1}}.
}
%One can check a useful relation,
%%\bee{
%%	\star \df \bm\mu^{(m)} = (-)^{m}\i_{\N} \star \lB\bm\mu^{(m)}\rB, \qquad
%%	\df \star \lB\bm\mu^{(m)}\rB = (-)^{m-1} \star \i_\N \bm\mu^{(m)}.
%%}
%%Here $\i_{\N}$ is interior product with respect to operator $\N^M$. As a special case, for a one form,
%\bee{
%	\star \df \bm\mu^{(d+1)} = (-)^{d+1} \underline{\N}_M \star \lB\bm\mu^{(d+1)}\rB^M, \qquad
%	\df \star \lB\bm\mu^{(1)}\rB = \star \underline{\N}_M \mu^M.
%}
%The lie derivative of a differential form satisfies,
%\bee{
%	\lie_X \bm\mu^{(m)} = \i_X \df \bm\mu^{(m)} + \df \lb \i_X\bm\mu^{(m)} \rb.
%}
Integration of a full rank form is defined as,
\bea{
	\int_{\cM_{(d+2)}} \bm\mu^{(d+2)} 
	&= \sgn(G)\int \lbr \df x^M \rbr \sqrt{|G|} \ \star[\bm\mu^{(d+2)}]  \nn\\
	&= \sgn(G)\int \lbr \df x^M \rbr \sqrt{|G|} \ \frac{1}{(d+2)!}\e^{M_1\ldots M_{d+2}}\mu_{M_1\ldots M_{d+2}}.
}
Here the raised Levi-Civita symbol has value $\e^{0,1,2,\ldots,d+1} = \sgn(G)/\sqrt{|G|}$. Integration of an exact full rank form is given by integration on the boundary,
\bee{
	\int_{\cM_{(d+2)}} \df\bm\mu^{(d+1)} 
	= \int_{\dow\cM_{(d+2)}} \bm\mu^{(d+1)},
}
where given a unit vector $\rmN$ normal to boundary, volume element on the boundary is defined as $\i_\rmN \bm\e^{(d+2)} = \star \mathbf N$.

\subsection*{Newton-Cartan Differential Forms}

A Newton-Cartan differential form is a differential form on $\cM_{(d+2)}$ which does not have a leg along $V$, i.e. $\i_V \bm\mu^{(m)}$. Such a form can be expanded as,
\bee{
	\bm\mu^{(m)} = \frac{1}{m!}\mu_{\mu_1\mu_2\ldots \mu_m} \df x^{\mu_1} \wedge \df x^{\mu_2} \wedge \ldots \wedge \df x^{\mu_m}.
}
Volume element of NC manifold is defined as,
\bee{
	\bm\ve^{(d+1)}_\downarrow = \star\bV = \frac{1}{(d+1)!} \lb V^{M} \e_{M \mu_{1}\ldots \mu_{d+1}} \rb 
	\df x^{\mu_{1}} \wedge \ldots \wedge \df x^{\mu_{d+1}}.
}
Since there is no degenerate metric on NC manifold we cannot define a Hodge dual. Hodge dual can however be defined if we chose a frame $T$. We can hence define spatial differential forms with the requirement that they should not have any leg along $V$ and $\bar V_{(T)}$. For these forms, indices can be raised and lowered using $p^{\mu\nu}$ and $p_{\mu\nu}$. We can define a spatial volume element,
\bee{
	\bm \ve^{(d)} = \star[\bV \wedge \bm{{\bar V}}] = \frac{1}{d!} \lb V^{M} \bar V^N \e_{MN \mu_{1}\ldots \mu_{d}} \rb 
	\df x^{\mu_{1}} \wedge \ldots \wedge \df x^{\mu_{d}},
}
and corresponding to it a Hodge duality operation,
\bee{
	*\lB\bm\mu^{(m)}\rB 
	= \star \lB \bV \wedge \bm u \wedge \bm\mu^{(m)} \rB
	= \frac{1}{(d-m)!} 
	\lb\frac{1}{m!} \mu^{\mu_1\ldots \mu_m} \ve_{\mu_1\ldots \mu_m \nu_1\ldots \nu_{d-m}} \rb
	\df x^{\nu_{1}} \wedge \ldots \wedge \df x^{\nu_{d-m}}.
}
One can check that $**= - \sgn(G) (-)^{m(d-m)} $.
%Finally we need to define integration for NC full rank forms and contra-forms,
%\bea{
%	\int_{\cM_{(d+1)}} \bm\mu^{(d+1)} 
%	&= \sgn(G)\int_{\cM_{(d+2)}} \bm{{\bar V}} \wedge \bm\mu^{(d+1)} 
%	= \sgn(\g)\int \lbr \df x^\mu \rbr \sqrt{|\g|} \ *_\uparrow\lB\bm\mu^{(d+1)}\rB, \nn\\
%	\int_{\cM_{(d+1)}} \bm\mu^{[d+1]} 
%	&= \sgn(G)\int_{\cM_{(d+2)}} \bV \wedge \bm\mu^{\flat(d+1)} 
%	= \sgn(\g)\int \lbr \df x^\mu \rbr \sqrt{|\g|} \ *^\downarrow\lB\bm\mu^{[d+1]}\rB,
%}
%where $\g_{\mu\nu} = p_{\mu\nu} + n_\mu n_\nu$ and $\g = \det \g_{\mu\nu} = - G$. Obviously a full rank spatial form would be zero. Rest of the notations and conventions follow from our relativistic discussion.

\subsection*{Non-Covariant Differential Forms}

%Choosing a non-covariant basis given in \cref{non_cov}, a vector and a one-form can be decomposed as,
%\bea{
%	\cX^M \dow_M &= 
%	- \E{\F} \lb \cX_t + \cB_t \cX_\sim \rb \dow_\sim 
%	- \E{\F} \cX_\sim \lb
%		\cB_t \dow_\sim 
%		+ \dow_t 
%	\rb
%	+  \cX^i \lb 
%		\dow_i 
%		- a_i \dow_t
%		+ \lb\cB_i - a_i \cB_t \rb \dow_\sim
%	\rb, \nn\\
%	\cY_M \df x^M &= \cY_\sim \lb \df x^\sim - \cB_\mu \df x^\mu \rb 
%	+ \lb
%		\cY_\sim \cB_t
%		+ \cY_t 
%	\rb \lb \df t + a_i \df x^i\rb
%	+ g_{ij} \cY^j \df x^i.
%}
Going to the local rest of frame $T$ used to define the Newton-Cartan spatial forms, we can check that the spatial forms behave covariantly on the spatial slice, i.e. can be expressed as,
\bee{
	\bm\mu^{(m)} = \frac{1}{m!}\mu_{i_1i_2\ldots i_m} \df x^{i_1} \wedge \df x^{i_2} \wedge \ldots \wedge \df x^{i_m}.
}
One can check that the volume element $\bm\ve^{(d)}$ defined before is indeed a full rank form on the spatial slice and can be written in this setting as,
\bee{
	\bm \ve^{(d)} = \frac{1}{d!} \lb V^{M} \bar V^N \e_{MN i_{1}\ldots i_{d}} \rb 
	\df x^{i_{1}} \wedge \ldots \wedge \df x^{i_{d}}.
}
The Hodge dual $\ast$ associated with it serves as Hodge dual operation on the spatial slice, 
\bee{
	*\lB\bm\mu^{(m)}\rB 
	= \frac{1}{(d-m)!} 
	\lb\frac{1}{m!} \mu^{i_1\ldots i_m} \ve_{i_1\ldots i_m j_1\ldots j_{d-m}} \rb
	\df x^{j_{1}} \wedge \ldots \wedge \df x^{j_{d-m}}.
}
Finally a full rank spatial form can be integrated on a spatial slice,
\bee{
	\int_{\cM_{(d)}} \bm\mu^{(d)} 
	= \sgn(G)\int_{\cM_{(d+2)}} \E{\F} \bm{{V}} \wedge \bm{{\bar V}} \wedge \bm\mu^{(d)} 
	= \sgn(g)\int \lbr \df x^\mu \rbr \sqrt{|g|} \ *[\bm\mu^{(d)}].
}
Here $g = \det g_{ij} = \E{2\F} \g = - \E{2\F} G$. Since $g_{ij}$ is a spatial metric $\sgn(g) = +1$. Other conventions and notations are same as relativistic case.

\bibliographystyle{utcaps.bst}
\bibliography{aj-bib}

\end{document}